\documentclass[12pt]{article}
\pdfoutput=1
\usepackage{amssymb}
\usepackage{amsfonts}
\usepackage{amsmath}
\usepackage{amsthm}
\usepackage{cancel}
\usepackage{slashed}
\usepackage{fullpage}
\usepackage{color}
\usepackage{braket}
\usepackage{graphicx}
\usepackage{multirow}
\usepackage{verbatim}
\usepackage{cite}
\usepackage{bigints}
\usepackage{simplewick}
\usepackage{authblk}
\usepackage{empheq}
\usepackage{bbm}
\usepackage{braket}
\usepackage{tcolorbox}
\usepackage{relsize}
\usepackage{simplewick}
\usepackage[T1]{fontenc}

\usepackage{feynmp}
\usepackage{feynmp-auto}
\usepackage{tikz}

\usepackage{caption}
\usepackage{subcaption}
\usepackage{dsfont}
\usepackage{latexsym,amsthm,amssymb,amsmath,amscd,euscript,graphicx,fontenc,eufrak,float, url,hyperref,makecell}
\usepackage{fullpage}

\newcommand{\eq}[1]{\begin{equation}\begin{aligned}#1\end{aligned}\end{equation}}
\newcommand{\eqbreak}[1]{\allowdisplaybreaks \begin{align}#1\end{align}}

\newcommand{\deltaD}[2]{\left(2\pi\right)^{#1} \delta^{#1} \left(#2\right)}

\newcommand{\momintFields}[1]{\int \frac{d^3 #1}{\left(2 \pi \right)^3 \sqrt{2 \omega_{#1}}}}

\usetikzlibrary{patterns}

\usetikzlibrary{shapes.misc}
\tikzset{point/.style={
    draw=black,
    cross out,
    inner sep=0pt,
    minimum width=4pt,
    minimum height=4pt,
}}
\tikzset{plussign/.style={
    draw=black,
    cross out,
    inner sep=0pt,
    rotate = 45,
    minimum width=4pt,
    minimum height=4pt,
}}

\newcommand{\beq}{\begin{equation}}
\newcommand{\eeq}{\end{equation}}
\newcommand{\be}{\begin{equation}}
\newcommand{\ee}{\end{equation}}
\newcommand{\beqa}{\begin{eqnarray}}
\newcommand{\eeqa}{\end{eqnarray}}
\newcommand{\beqar}{\begin{eqnarray*}}
\newcommand{\eeqar}{\end{eqnarray*}}
\newcommand{\bea}{\begin{eqnarray}}
\newcommand{\eea}{\end{eqnarray}}

\newcommand{\mc}[1]{\mathcal{#1}}

\newcommand{\cL}{{\cal L}}

\newcommand{\eps}{\varepsilon}

\allowdisplaybreaks

\definecolor{darkred}{rgb}{0.5,0.0,0.0}
\definecolor{darkblue}{rgb}{0.0,0.0,0.9}
\definecolor{darkerblue}{rgb}{0.0,0.0,0.5}
\definecolor{darkgreen}{rgb}{0.0,0.5,0.0}
\definecolor{black}{rgb}{0.0,0.0,0.0}
\definecolor{brown}{rgb}{0.6,0.4,0.2}

\def\be{\begin{equation}}
\def\ee{\end{equation}}
\def\cL{\mathcal{L}}
\def\cN{\mathcal{N}}

\def\msbar{\overline{\text{MS}}}

\usepackage{MnSymbol}

\renewcommand{\be}{\begin{equation}}
\renewcommand{\ee}{\end{equation}}

\definecolor{darkgreen}{rgb}{0.0,0.5,0.0}
\usepackage{hyperref}
\hypersetup{
    linktocpage,
     colorlinks,
     citecolor=darkgreen,
     linkcolor= darkgreen,
     urlcolor=darkgreen
}

\renewcommand{\eps}{\varepsilon}

\newcommand{\cM}{\mathcal M}

\newcommand{\cO}{\mathcal O}
\newcommand{\cS}{\mathcal S}

\newcommand{\bbone} { {\mathds 1}}
\newcommand{\euv}{\epsilon_{\text{UV}}}
\newcommand{\eir}{\epsilon_{\text{IR}}}
\newcommand{\Li}{\text{Li}}
\newcommand{\cdt}{ \! \cdot \!}

\newcommand{\omax}{\omega^{\text{max}}}
\newcommand{\tmax}{\theta^{\text{max}}}

\newcommand{\Has}{H_{\text{as}}}

\title{An $S$-Matrix for Massless Particles}
\author{Holmfridur Hannesdottir}
\author{Matthew D. Schwartz}
\affil{\small \emph{Department of Physics, Harvard University, Cambridge, MA 02138, USA}}
 
\date{}
\begin{document}

\begin{fmffile}{feyngraph}
\unitlength = 0.4mm
\maketitle
\thispagestyle{empty}

\begin{abstract}
The traditional $S$-matrix does not exist for theories with massless particles, such as quantum electrodynamics. The difficulty in isolating asymptotic states manifests itself as infrared divergences at each order in perturbation theory. Building on insights from the literature on coherent states and factorization, we construct an $S$-matrix that is free of singularities order-by-order in perturbation theory. Factorization guarantees that the asymptotic evolution in gauge theories is universal, i.e.\ independent of the hard process. Although the hard $S$-matrix element is computed between well-defined few particle Fock states, dressed/coherent states can be seen to form as intermediate states in the calculation of hard $S$-matrix elements. We present a framework
for the perturbative calculation of hard $S$-matrix elements combining Lorentz-covariant Feynman rules for the dressed-state scattering with time-ordered perturbation theory for the asymptotic evolution. With hard cutoffs on the asymptotic Hamiltonian, the cancellation of divergences can be seen explicitly. In dimensional regularization,
where the hard cutoffs are replaced by a renormalization scale, the contribution from the asymptotic evolution produces scaleless integrals that vanish. A number of illustrative examples are given in QED, QCD, and $\cN=4$ super Yang-Mills theory.
\end{abstract}

\newpage
 \tableofcontents
\newpage
\pagenumbering{arabic}

\section{Introduction}
The scattering matrix, or $S$-matrix, is a fundamental object in physics.
Intuitively, the $S$-matrix is meant to transform an ``in'' state $| \psi_{\text{in}} \rangle$ at $t = - \infty$ into an ``out'' state $\langle \psi_{\text{out}} |$ at $t = + \infty$. Unfortunately, constructing an operator in quantum field theory which achieves this projection is far from trivial. To begin, one might imagine that $S= \lim_{t\to\infty} e^{-i H t}$. However, this operator does not exist, even in a free theory. For example, acting on states with energies $E_i$, matrix elements of this operator would be infinitely oscillating phases.
 The proper resolution in quantum mechanics was first understood by Wheeler~\cite{Wheeler}, who defined the $S$-matrix to project from a basis of metastable asymptotic states $|\psi_\text{in}\rangle$ (a nucleus) to other states (other nuclei) $|\psi_\text{out}\rangle$. This idea was expanded for use in quantum field theory by Heisenberg, Feynman, and Dyson~\cite{Heisenberg,Feynman,Dyson1} for calculations in quantum electrodynamics (QED).  In modern language,  one must factor out the evolution due to the free Hamiltonian $H_0$ to make $S$ well-defined.

In the Wheeler-Heisenberg-Feynman-Dyson (henceforth ``traditional'') approach, one assumes that in the far past, the ``in'' state is well-approximated with a freely evolving state, i.e.\ a state that evolves with the free Hamiltonian $H_0$: $e^{-i H t} | \psi \rangle \rightarrow e^{- i H_0 t} |\psi_{\text{in}} \rangle$ as $t \rightarrow - \infty$. The interaction is assumed to occur during some finite time interval so that in the far future, the time evolution is again  nearly free: $e^{-i H t} | \psi \rangle \rightarrow e^{- i H_0 t} | \psi_{\text{out}} \rangle$ as $t \rightarrow + \infty$.   The state $|\psi\rangle$ is then related to the in and out states by M\o ller operators
\be
   \Omega_\pm = \lim_{t \to \pm\infty}  e^{i Ht} e^{-i H_0 t} \label{Mollerdef}
\ee
as $|\psi\rangle  = \Omega_+ |\psi_\text{out}\rangle = \Omega_- |\psi_\text{in}\rangle$ and so
 $| \psi_\text{out} \rangle =S |\psi_\text{in}\rangle$ where the traditional $S$-matrix is defined as
\be
S = \Omega_+^\dagger \Omega_- \label{Sdef}
\ee
Unfortunately, this textbook approach has problems too: bare $S$-matrix elements computed this way are both ultraviolet (UV) and infrared (IR) divergent.\footnote{In this paper, we use ``IR divergences'' to refer to any divergence that is not of short-distance origin. So IR divergences come from both soft and collinear regions.}
Ultraviolet divergences are by now completely understood: they are an artifact of computing $S$-matrix elements using unphysical fields in terms of unphysical (bare) parameters. When $S$-matrix elements are computed with physical, renormalized, fields in terms of physical observable parameters, the UV divergences disappear. IR divergences, however, are not as well understood and remain an active area of research. In theories with massless charged particles, such as QCD, $S$-matrix elements have IR divergences of both soft and collinear origin. Historically, three approaches have been explored to ameliorate the problem: the ``cross section method'', the ``dressed-state method''  and the ``modification-of-$S$ method''.

The first way of dealing with IR divergences, referred to as the \textit{cross section method} (following~\cite{nelson1981origin,Contopanagos:1991yb}) is the most common. It argues that $S$-matrix elements themselves are not physical; only cross sections, determined by the squares of $S$-matrix elements integrated over sufficiently inclusive phase space regions, correspond to observables. Importantly, in this method, IR divergences cancel between virtual contributions and real emission contributions to {\it different} final states.
The cancellation in QED was demonstrated definitively by Bloch and Nordsieck~\cite{Bloch:1937pw} in 1937. They showed that cross sections in QED (with massive fermions) are IR finite order-by-order in perturbation theory when processes with all possible numbers of final state photons with energies less than some cutoff $\delta$ are summed over. The proof of Bloch-Nordsieck cancellation~\cite{Yennie:1961ad,Weinberg:1965nx,Grammer:1973db} crucially relies on Abelian exponentiation~\cite{Yennie:1961ad}: the soft singularities at any given order in $\alpha$ in QED are given by the exponential of the 1-loop soft-singularities. For theories with massless charged particles, such as QCD, Bloch-Nordsieck fails~\cite{Doria:1980ak}.

In non-Abelian gauge theories, the theorem of Kinoshita, Lee and Nauenberg (KLN) ~\cite{Kinoshita:1962ur,Lee:1964is} is often invoked to establish IR finiteness. The KLN theorem states that for any given process a finite cross section can be obtained by summing over all possible initial and final states for processes whose energy $E$ lies within some compact energy window around a reference energy $E_0$, i.e.\ $\vert E-E_0\vert < \delta$ for a given $\delta$. In fact, the KLN theorem is weaker and its proof more complicated than required. First of all, energy is conserved, so the cancellation must occur without the energy window. Second of all, one does not need to sum over initial and final states; the sum over only final states for a fixed initial state will do, as will the sum over initial states for a fixed final state. This stronger version of the KLN theorem was proven recently by Frye et al.~\cite{Frye:2018xjj}. The proof is one line: for a given initial state, the probability of it becoming {\it anything} is 1, which is finite to all orders in perturbation theory.
Importantly, both the KLN theorem and its stronger version by Frye et al.\ generically require the sum of diagrams to include the forward scattering contribution, which is usually excluded from a cross section definition. Unless they happen to be IR finite on their own, the forward scattering diagrams are crucial to achieve IR finiteness. Multiple illustrative examples can be found in~\cite{Frye:2018xjj}. If one wants the cross section to be finite when summing over only a restricted set of final states, insights beyond Block-Nordsieck, KLN, and Frye et al.\ are required, such as those coming from factorization~(e.g.
\cite{Collins:1988ig,Collins:1989gx,CATANI1989323,Bauer:2000ew,Bauer:2000yr,Beneke:2002ph,Beneke:2002ni,Feige:2013zla,Feige:2014wja}).

In the second approach to remedy IR divergences, the \textit{dressed-state method}, the $S$-matrix is defined in the traditional way, but it is evaluated between states $|\psi^d\rangle$ that are not the usual few-particle Fock states $|p_1, \cdots, p_n\rangle$. One of the first proposals in this direction was by Chung~\cite{Chung:1965zza} (see also~\cite{Greco:1967zza}), who argued that in QED one should replace single-particle electron states $|p\rangle$ wifth dressed states of the form $|p^d\rangle = e^{R} |p\rangle$ with $R$ defined as
\be
   R \ket{p} = e\sum_{j=1}^2 \int \frac{d^{d-1}k}{\left(2\pi\right)^{d-1} \sqrt{2 \omega_k}} \frac{p \cdot \epsilon_j(k)}{p \cdot k} a^{j\,\dagger}_k \ket{p}
   \label{RChung}
\ee
where $\epsilon_j$ is a photon polarization vector and $a_k^{j\,\dagger}$ is its corresponding creation operator.
The idea behind this dressing is that the eikonal factors $\frac{p\, \cdot \,\epsilon}{p\, \cdot \,k}$ give the real emission amplitude in the singular (soft) limit, which is then canceled by virtual contributions, so that
$\langle p_3^d \cdots p_n^d | S | p_1^d p_2^d \rangle$ is IR finite. The exponentiation of the eikonal interaction is the same mechanism (Abelian exponentiation) as invoked in the Bloch-Nordsieck cancellation.
Indeed, the proof of the IR finiteness of these dressed states in QED is essentially the same as in the proof of the Bloch-Nordsieck theorem.
 This cloud of photons in the dressing has the same form as Glauber's coherent states~\cite{glauber1963coherent} used in quantum optics~\cite{frantz1965compton,Ilderton:2017xbj} (these are, roughly speaking, $e^R|0\rangle$), and so the dressed states in this case are commonly called {\it coherent states}.

While the coherent state approach is in some ways appealing, it has drawbacks. The main problem is that the IR divergences are just moved from the amplitudes to the states. That is, the coherent states themselves are IR divergent and therefore not normalizable elements of a Fock space (although they may be understood as living in a non-separable von Neumann space, as explained in a series of papers by Kibble~\cite{Kibble:1968sfb,Kibble:1969ip,Kibble:1969ep,Kibble:1969kd}). The IR divergence problem is therefore still present in this construction; it has merely been moved from the $S$-matrix elements to the states of the theory. Additionally, generalizing beyond massive QED to theories like QCD with collinear divergences and color factors has remained elusive~\cite{Catani:1985xt,Gonzo:2019fai}. In particular, no prescription is given for how to go beyond the singular points (zero energy or exactly collinear). For example, the coherent states are sums over particles with different momenta, so they do not have well-defined momenta themselves. Is momentum then conserved by the $S$-matrix in the coherent-state basis? How does one integrate over coherent states to produce an observable cross section? These problems are not commonly discussed in the literature. As far as we know, no one has explicitly computed an $S$-matrix element between coherent states. This defect gives the coherent-state literature a rather formal aspect.

The third approach to removing the IR divergences in scattering theory is to redefine the $S$-matrix rather than the states. That the traditional $S$-matrix inaccurately captures the asymptotic dynamics arises already in non-relativistic scattering of a charged particle off a Coulomb potential in non-relativistic quantum mechanics. The standard assumption that particles move freely at asymptotic times is not justified for non-square-integrable potentials, like the $\frac{1}{r}$ Coulomb potential, and leads to ill-defined $S$-matrix elements.
In modern language, the $S$-matrix element for non-relativistic Coulomb scattering has the form
\be
\langle \vec{p}_f | S | \vec{p}_i \rangle \sim \frac{\alpha}{(\vec{p}_i-\vec{p}_f)^2}
e^{-i \alpha \frac{m}{|\vec{p}_i-\vec{p}_f|}\frac{1}{2\eir}}
\ee
We see that the leading term of order $\alpha$, corresponding to the first Born approximation, is not problematic: except in the exactly forward limit, there are no divergences in the tree-level scattering process. The logarithmic IR divergence (showing up as a $\frac{1}{\eir}$ pole in $d=4-2\epsilon$ dimensions) first appears in the second Born approximation, where it is seen to be purely imaginary. Moreover, the IR divergent part exponentiates (as do all IR divergences in QED), into the {\it Coulomb phase}. Thus, in non-relativistic quantum mechanics, one can apply the cross section ideology even without the inclusive phase space integrals: the cross section for the scattering of a single electron off a Coulomb potential is well-defined. However, the $S$-matrix is not.

One of the first attempts to define an $S$-matrix for potentials that are not square-integrable was made by Dollard~\cite{dollard1971quantum} in 1971.  He noted that when incoming momentum eigenstates are evolved to late times with the Coulomb interaction $H=H_0 + \frac{\alpha}{r}$, there is a residual logarithmic time dependence for large $t$:
\be
 e^{-i \int^t H(t') dt' }|p\rangle \cong e^{-i \left( \frac{p^2}{2m} t + \frac{m \alpha}{|p|} \ln t \right)}|p \rangle \label{dollardtime}
\ee
The intuition for this form is that at large $t$, the particle moves approximately on a classical trajectory with $r = \frac{p t}{m}$, which gives the logarithmic dependence on $t$ when integrated up to infinity. While the $e^{-i \frac{p^2}{2m} t}$ is removed by Wheeler's $e^{i H_0 t}$ factor, the other term is not and persists to generate the $\frac{1}{\eir}$ divergences in the $S$-matrix. Dollard then proposed to replace to the $e^{i H_0 t}$ factor with a $e^{i \Has(t)}$ factor, with $\Has(t)$ defined with exactly the logarithmic time dependence needed to cancel the time dependence in Eq.~\eqref{dollardtime}. He then showed that a modified $S$-matrix, defined with his asymptotic Hamiltonian replacing $H_0$, exists for Coulomb scattering.

When the electron is relativistic, the IR divergence in the second Born approximation has a real part that does not cancel at the cross section level. So first-quantized quantum mechanics is insufficient to produce an IR-finite cross section: QED is needed.
Faddeev and Kulish~\cite{Kulish:1970ut} combined the aforementioned work of Chung in QED and Dollard's  in non-relativistic quantum mechanics. They observed that in QED, infrared divergences have both a real part (as Chung observed) and an imaginary part (the relativistic generalization of the Coulomb phase). These can be combined into a modified $S$-matrix of the form
\be
    S_{\text{FK}} = \lim_{t_{\pm} \to \pm \infty} e^{-R(t_+)} e^{-i \Phi(t_+)} S  e^{-i \Phi(t_-)}e^{R(t_-)} \label{FKform}
\ee
where
\be
\Phi(t) = \frac{\alpha}{2} \int \frac{d^3p}{(2\pi)^3} \frac{d^3 q}{(2\pi)^3} : \rho(p) \rho(q): \frac{p \cdot q}{\sqrt{(p \cdot q)^2 - m^4}} \ln |t|
\ee
corresponds to the Coulomb phase (compare to the $\ln t$ dependence in Dollard's form, Eq.~\eqref{dollardtime}).
The factor $R$ is similar to Chung's in Eq.~\eqref{RChung} but with a power-expanded phase, and annihilation operators included as well:
\be
    R(t) =  e\sum_{j=1}^2 \int \frac{d^3p}{ (2\pi)^3}
    \frac{d^3 k}{(2\pi)^3\sqrt{2 \omega_k} }
    \left[\frac{p\cdot \epsilon^\star_j(k)}{p \cdot k} a^{j \, \dagger}_k e^{i \frac{p \cdot k}{\omega_p}t} - \frac{p\cdot \epsilon_j(k)}{p \cdot k} a^j_k e^{-i \frac{p \cdot k}{\omega_p}t}  \right] \rho (\vec{p}) \label{FKR}
\ee
where
\be
    \rho \left(p\right) = \sum_s \left( a^{s \, \dagger}_p a^s_p - b^{s\,\dagger}_p b^s_p \right) \label{FKrho}
\ee
is the electron-number operator. Acting on states, it pulls out the direction $p$ of each fermion and multiplies the contribution by 1 for electrons or -1 for positrons:
$\rho(p) |q_1 \cdots q_n\rangle = \sum \pm \left(2\pi\right)^3 \delta^3\left(\vec{p} - \vec{q}_j\right)|q_1 \cdots q_n\rangle $.
 Faddeev and Kulish proceed to argue that $S_\text{FK}$ has finite matrix
elements between coherent states in QED. They argued that one should include the phase factors in a redefinition of the $S$-matrix while including the $e^R$ factors in dressing the states.
Although there are some suspicious orders-of-limit and signs in Faddeev and Kulish's paper (see~\cite{Contopanagos:1991yb}), we believe their construction is essentially valid. Indeed, one goal of our paper is to translate this classic work in QED to modern language. As we will show in Section~\ref{sec:softWilson}, both the real and imaginary parts in the factor $e^{i \Phi(t_-)}e^{R}$ are reproduced by the action of a single Wilson line.

In the 50 odd years since Faddeev and Kulish's work, there has been intermittent progress on generalizing the coherent state construction from QED to non-Abelian theories.
Early work~\cite{Catani:1985xt,Butler:1978rd,Nelson:1980qs} focused on trying to use coherent states to salvage the Bloch-Nordsieck theorem, following the QCD counterexamples given by Doria et al.~\cite{Doria:1980ak,Andrasi:1980qw}.
Although soft divergences in QCD do not exponentiate into a compact form as they do in QED~\cite{Gatheral:1983cz,Frenkel:1984pz}, they still have a universal
form and factorize off of the hard scattering~\cite{Collins:1988ig,Feige:2014wja}. Using this
observation, it has been argued using a frequency-ordered formalism that soft-finite dressed
states can be constructed between $S$-matrix
elements in QCD~\cite{Catani:1985xt,Giavarini:1987ts}.
Collinear divergences and the soft-collinear overlap in gauge theories were explored in~\cite{curci1978mass,Havemann:1985ra,DelDuca:1989jt,Contopanagos:1991yb}.
An explicit check of the dressed formalism was performed by
Forde and Signer~\cite{Forde:2003jt} who used explicit cutoffs
to separate the regions and showed that the cross section for $e^+e^- \to$ jets can be reproduced at leading power at order $\alpha_s$ through finite $S$-matrix elements.
Ref. \cite{DelDuca:1989jt} argued that if soft-collinear factorization holds in QCD, then the dressed state formalism should allow one to construct a finite $S$-matrix in QCD to all orders. Collinear factorization was proven diagrammatically at large $N$ a decade later~\cite{Kosower:1999xi} and a full proof of collinear factorization and soft/collinear factorization for QCD to all orders in perturbation theory was given in ~\cite{Feige:2013zla,Feige:2014wja}, inspired by~\cite{Collins:1988ig,Collins:1989gx,CATANI1989323,Bauer:2000yr,Bauer:2002nz,Beneke:2002ph}. One goal of the current paper is to combine these various insights to provide, for the first time, an explicit construction of an IR-finite $S$-matrix for QCD.

In all of this literature, there are a number of unresolved issues. First, there are essentially no results about the finite parts of a finite $S$-matrix. Showing the cancellation of the IR singularities is one thing, but to evaluate $S$ one needs to deal with complications of momentum conservation, cutoffs, UV divergences, and to actually be able to compute the resulting integrals.
A prescription to determine the finite parts of the modified $S$-matrix is required if we are to explore the $S$-matrix's properties. While some authors have suggested criteria such as that the dressed states should be gauge~\cite{Bagan:1999jf} or BRST invariant~\cite{catani1987gauge}, or have asymptotic charges~\cite{Kapec:2017tkm,Strominger:2017zoo}, or be compatible with decoherence~\cite{Carney:2017jut,Carney:2018ygh,Gomez:2018war}, the necessity of these choices is unclear. Certainly nothing goes wrong at the level of cross sections if we proceed using the cross section method. After the finite part is fixed, one must further explain how to relate modified $S$-matrix elements to observables: what is the measure for integration over momenta in the von Neumann space of dressed states (if one goes that route)? To agree with data, the predictions had better reduce to what one calculates using the IR-divergent $S$, but how that will happen in any of the approaches to dressed states is rarely discussed.
In this paper, we attempt to raise the bar for constructing a finite $S$-matrix by providing a motivated, calculable scheme, and give explicit expression for $S$-matrix elements and observables in a number of cases in QED, QCD, and $\cN=4$ super Yang-Mills theory.

The organization of this paper is as follows. We start by motivating and defining a ``hard'' $S$-matrix in Section \ref{sec:def_SH}. We show how to get finite answers, and connect to the previous work on QED using dressed states in Section~\ref{SHandd}. In Section~\ref{sec:observables}, we discuss how to compute observables and show that the same predictions for infrared-safe differential cross sections results from $S_H$ as from the traditional $S$. In Section~\ref{sec:softWilson} we connect our construction to the  expressions of Faddeev and Kulish in QED. We then proceed to explicit calculations, working out the Feynman rules and some toy examples in Section \ref{sec:compute}. In Section \ref{sec:DIS} we demonstrate IR finiteness in the process $\gamma^\star e^- \to e^-$ in QED using cutoffs, and illustrate the relative simplicity when pure dimensional regularization is invoked. In Section \ref{sec:Zee} we discuss  $Z \to e^+e^-$ including the connection to the Coulomb phase and the Glauber operator as well as an explicit calculation of the thrust distribution, both exactly at NLO and to the leading logarithmic level using the asymptotic interactions. Section~\ref{sec:thrust} makes explicit some of the general observations about exclusive measurements from Section~\ref{sec:observables}.
Section~\ref{sec:Nis4} gives some examples in  $\cN=4$ super Yang-Mills theory, connecting to observations about remainder functions, renormalization and subtractions schemes. Concluding remarks and a summary of our main results are given in Section \ref{sec:summary}.

\section{The hard $S$-matrix}
\label{sec:def_SH}
The intuition behind scattering is that one starts with some initial state, usually well-approximated as a superposition of momentum eigenstates, which then evolves with time into a region of spacetime where it interacts, and then a new state emerges. The $S$-matrix is meant to be a projection of this emergent final state onto a basis of momentum eigenstates. For scattering off a local (square-integrable) potential, this picture works fine. The $S$-matrix is then defined as $ S = \Omega_+^\dagger \Omega_-$ as in Eq.~\eqref{Sdef} with the M\o ller operators $\Omega_\pm$ defined in Eq.~\eqref{Mollerdef}.
However, when the interactions cannot be confined to a finite-volume interaction region, as in Coulomb scattering or in a quantum field theory with massless particles, this picture breaks down: the states at early and late times continue to interact, so the momentum-eigenstate approximation is no longer valid.

As mentioned in the introduction, the simplest example with the traditional definition of $S$ breaks down is for non-relativistic scattering off a Coulomb potential. In this case, the M\o ller operators acting on momentum eigenstates generate an infrared divergent ``Coulomb'' phase. While the infrared divergence is a problem for a formal definition of the $S$-matrix, it is
not a problem for cross section calculations that depend only on squares of $S$-matrix elements. In relativistic Coulomb scattering, or in QED, $S$ has both an infrared divergent Coulomb phase and an infrared divergent real part. A convenient feature (Abelian exponentiation~\cite{Yennie:1961ad}) of QED is that a closed form expression is known for the IR-divergent contribution to all orders in perturbation theory for any process. Indeed, the 1-loop divergences are given by $S \sim \frac{\gamma_\text{cusp}}{\eir}$ where the cusp-anomalous dimension is (see~\cite{Chien:2011wz})
\be
\gamma_\text{cusp} =-\frac{\alpha}{\pi} \left[(\beta-i \pi) \coth \beta -1\right]
\ee
with the cusp angle defined by $\cosh \beta =\frac{v_1 \cdot v_2}{|v_1| |v_2|}$ and $v_1^\mu = \frac{p_1^\mu}{E_1}$  and $v_2^\mu = \frac{p_2^\mu}{E_2}$ are the 4-velocities of the incoming and outgoing electrons. To all orders, the IR divergences exponentiate as $S \sim \exp\frac{-\gamma_\text{cusp}}{2\eir}$~\cite{Korchemsky:1987wg}.
Thus, it is possible to factor out IR-divergent parts from the $S$-matrix and redefine a new $S$-matrix that is IR-finite order-by-order. This was done by Chung and Faddeev and Kulish, as discussed in the introduction.  Note that the non-relativistic limit corresponds to $\beta \to 0$ in which case $\gamma_\text{cusp} =i \alpha \frac{1}{\beta}$ becomes the purely imaginary Coulomb phase.

When the charged particles are also massless, as in QED with $m_e = 0$, new IR divergences appear associated with collinear divergences. Soft-collinear divergences appear as double IR-poles. Indeed, in the $m_e \to 0$ limit, $v^\mu_i$ becomes lightlike, so $\beta \to \infty$. At large $\beta$ in the cusp angle $\gamma_\text{cusp} \sim -\frac{\alpha}{\pi} \beta$ diverges linearly with $\beta$, so the $S$-matrix now has double, $\frac{1}{\eir^2}$ poles. In QCD, or other non-Abelian theories, the cusp angle gets corrections beyond one loop and the IR divergences do not exponentiate into a closed form expression~\cite{Gatheral:1983cz,Frenkel:1984pz,Gardi:2013ita}. These complications have made it difficult to come up with a complete formulation of an IR-finite $S$-matrix in general quantum field theories~\cite{DelDuca:1989jt,Contopanagos:1991yb,Forde:2003jt}.

The approach we take in this paper is to construct an $S$-matrix that is IR finite by replacing the free Hamiltonian $H_0$ in the definition of the traditional $S$-matrix
with an appropriate asymptotic Hamiltonian  $\Has$.
That is, we can define new {\bf hard M\o ller operators}
\be
   \Omega_{\pm}^H = \lim_{t_{\pm} \to \pm \infty} e^{i H t_\pm} e^{- i H_{\text{as}} t_\pm}
   \label{OmegaHdef}
\ee
and a \textbf{hard $S$-matrix} as
\be
S_H  = \Omega^{H\dag}_+\Omega_-^H \label{SHdef}
\ee
Ideally, we would want to choose $\Has$ so that the hard M\o ller operators exist, as unitary operators on the Hilbert space. Proving their existence is challenging, as even in a mass-gapped theory, where we can take $\Has=H_0$, they do not exist by Haag's theorem~\cite{Haag:1955ev}. From a practical point of view, we can be less ambitious and aim to choose $\Has$ so that the hard $S$-matrix is free of IR divergences at each order in perturbation theory. If this was our only criteria, we could choose $\Has=H$, so that $S_H=\bbone$.
%

A better criteria for defining $\Has$ is that, in addition to capturing long-distance interactions, the asymptotic Hamiltonian should be defined so that the asymptotic evolution of the states is independent of how they scatter. It is possible to define $\Has$ this way due to universality of infrared divergences in gauge theories. Using factorization~\cite{Collins:1988ig,Collins:1989gx,CATANI1989323,Bauer:2000ew,Bauer:2000yr,Beneke:2002ph,Beneke:2002ni,Feige:2013zla,Feige:2014wja}, the soft and collinear interactions can be separated from the hard scattering process: Any $S$-matrix element in gauge theories can be reproduced by the product of a hard factor, collinear factors for each relevant direction, and a single soft factor. See~\cite{Feige:2014wja} for a concise statement of factorization at the amplitude level.

In order to exploit factorization, we employ methods developed in Soft-Collinear Effective Theory (SCET). The theory provides a systematic power expansion of the QED or QCD Lagrangian, and reproduces all infrared effects. The leading power Lagrangian in SCET is~\cite{stewart2013lectures,Becher:2014oda}
\begin{multline}
    \mathcal{L}_{\text{SCET}} = - \frac{1}{4} (F_{\mu \nu}^s)^2 + \sum_n -\frac{1}{4} (F_{\!\mu \nu}^{c, n})^2 \\
+ \sum_n    \bar{\psi}_n^c \frac{\slashed{\bar{n}}}{2}  \left[
  i n \cdot D_c+g n \cdt A_{s}^a(x_{-})T^a
+ i \slashed{D}_{c \perp} \frac{1}{i \bar{n} \cdt D_c} i
  \slashed{D}_{c \perp} \right] \psi_n^c
  + \cL_\text{Glauber}
  \label{lscet}
\end{multline}
where $s$ and $c,n$ are soft and collinear labels respectively and the collinear covariant derivative is
\be
 i D_{\mu}^{c} =i \partial_{\mu}+g A_{\mu}^{c, a} T^{a} \\
\ee
The last term $\cL_\text{Glauber}$ describes Coulomb or Glauber gluon interactions~\cite{Rothstein:2016bsq} (see also~\cite{Schwartz:2017nmr}). Pedagogical introductions to SCET can be found in~\cite{Becher:2014oda, stewart2013lectures, Schwartz:2013pla}.
 
We define the asymptotic Hamiltonian $\Has$ to be the SCET Hamiltonian
appended with free Hamiltonians for massive particles. The hard $S$-matrix is then defined in terms of $\Has$ using Eqs.~\eqref{OmegaHdef} and \eqref{SHdef}.

Although the SCET Lagrangian looks complicated and non-local, much of the complication comes from being careful
to include only leading-power interactions. In principle, for a theory to be valid at leading power, one could include any
subleading power interactions one wants. Exploiting this flexibility, the collinear interactions in $\cL_\text{SCET}$ can be replaced simply with the full interactions of QCD: $i\bar{\psi}_n^c \slashed D_c \psi_n^c$. The soft interactions, from the
$  \bar{\psi}_n^c \frac{\slashed{\bar{n}}}{2} n \cdt A_{s}^a(x_{-}) \psi_n^c$ term, are also not that complicated: they are equivalent to treating the collinear fermions as being infinitely energetic, with no recoil. That is, the fermions act as classical sources for radiation moving in a straight line along the $n^\mu$ direction. This leads to an alternative representation of the soft interactions as coming from Wilson lines. This connection is made more precise in Section~\ref{sec:softWilson}.

In practice, when computing $S_H$ elements we will not use the explicit and cumbersome interactions in $\cL_\text{SCET}$. Instead, we will
take the method-of-regions approach~\cite{Beneke:1997zp, Becher:2014oda}. We start with a particular Feynman diagram and then expand to leading power based on the collinear or soft scaling associated with particles involved. In a sense, this is the most straightforward and foolproof way to compute $S_H$ amplitudes. Numerous examples are given in subsequent sections.

We also, in accord with the general principles of the method of regions, do not impose any hard cutoffs on the momenta of the soft and collinear particles that interact through $\Has$. Imposing cutoffs is helpful for demonstrating explicit IR-divergence cancellation, and some examples are provided in Section~\ref{sec:cutoffs}. However, cutoffs generally lead to  very difficult integrals, and moreover they break symmetries like gauge-invariance that we would like $S_H$ to respect. More precisely, it is only the finite, cutoff-dependent remainder terms that may depend on gauge -- the IR divergence cancellation mechanism is gauge-independent. Since the cutoff-dependent finite parts are unphysical anyway, it is not a problem that they are also gauge-dependent. In general, however, the whole framework with cutoffs is rather unwieldy.

When using pure dimensional regularization, the diagrams involving $\Has$ interactions will lead to scaleless integrals. These integrals are both UV and IR divergent. The IR divergences cancel in other contributions to $S_H$ (as we will provide ample demonstration), but the UV divergences must be removed through renormalization.
As a consequence, in pure dimensional regularization, $S_H$-matrix elements are not guaranteed to be independent of renormalization scheme. Indeed, they are generally complex and will depend on the scale $\mu$ at which renormalization is performed. The $S_H$-matrix is {\it not} scale independent: $\frac{d}{d\mu} S_H \ne 0$, in contrast to $S$ which does satisfy the Callan-Symanzik equation $\frac{d}{d\mu } S=0$. This is unsatisfying, but not unsettling, as $S_H$ elements are not themselves observable. (To be fair, if $S$-matrix elements are IR divergent, it is not clear what it means to say they are scale-independent). In any case, one should think of $S_H(\mu)$ like one thinks about the strong coupling constant $\alpha_s(\mu)$ in $\msbar$. While $\alpha_s(\mu)$ is not observable, it is still an extraordinarily useful concept. The running coupling indeed encodes qualitatively and quantitatively a lot of important physics, such as unification and confinement. As with $\alpha_s(\mu)$, when $S_H(\mu)$ is used to compute an observable, the scale dependence will cancel. We demonstrate that in general in Section~\ref{sec:observables}, and provide an explicit example in Section~\ref{sec:Zee}.

%
%

\subsection{$S_H$ and dressed states \label{SHandd}}
The usual way of calculating $S$-matrix elements in perturbation theory is to work in the interaction picture,
where one expands the interactions in terms of freely evolving fields. The propagators for free fields have a relatively simple form, and $S$-matrix elements then become integrals over these propagators.
One might try to work out Feynman rules for $S_H$ analogously, in an asymptotic interaction picture. Then propagators would correspond to non-perturbative Green's functions for the soft and collinear fields in $\mathcal{L}_{\text{SCET}}$, including all of their interactions. Unfortunately, finding a closed-form expression for these propagators is not possible. In any case, it is not necessary, since if we want to work perturbatively in the coupling constants, we must do so consistently in both $H$ and $\Has$.

To proceed, we note that the hard S-matrix can be written suggestively as
\eq{
    S_H =\Omega_+^{H\dag}\,  \Omega^H_-  = \Omega_+^{\text{as}} \, \Omega_+^{\dag} \, \Omega_- \, \Omega^{\text{as}\dag}_-
    =  \Omega_+^{\text{as}}\,S\,\Omega^{\text{as}\dag}_-
}
where
\be
 \Omega^\text{as}_{\pm} = \lim_{t\to \pm\infty} e^{i H_{\text{as}} t}e^{-i H_0 t}
\ee
 are asymptotic M\o ller operators and $\Omega_{\pm} = \lim_{t\to \pm\infty} e^{ i H t}e^{-i H_0 t}$ are the usual M\o ller operators. Inserting complete sets of states lets us write hard $S$-matrix elements between a Heisenberg picture out-state $| \psi_{\text{out}} \rangle$ and a Heisenberg picture in-state $| \psi_{\text{in}} \rangle$ as
\eq{
    \langle \psi_{\text{out}} | S_H | \psi_{\text{in}} \rangle  =
    \int d\Pi_{\psi_{\text{out}}'}
    \int d\Pi_{\psi_{\text{in}}'}
    \langle \psi_{\text{out}} | \Omega_{+}^{\text{as}}
    \ket{\psi_{\text{out}}'}\bra{\psi_{\text{out}}'}
    S
    \ket{\psi_{\text{in}}'}\bra{\psi_{\text{in}}'}
    \Omega^{\text{as}\dag}_-
    | \psi_{\text{in}} \rangle
    \label{eq:SH_terms}
}
Here the integral is over complete sets of Fock-space states $\ket{\psi_{\text{in}}'}$ and $\ket{\psi_{\text{out}}'}$.
The hard scattering matrix elements are written as a product of three terms. The middle term is the traditional $S$-matrix and the outer terms correspond to evolution with the asymptotic M\o ller operators. The Feynman rules for these contributions closely resemble those of time-ordered perturbation theory and are derived in Section~\ref{sec:asymrules} below.

\begin{figure}[t]
\centering
{{
\begin{tikzpicture}
\draw [|->,line width=1,black] (0,0.1) to (-2,0.1);
\draw [<-,line width=1,darkblue] (0,-0.5) to [out = 180, in = 0] (-2,0);
\draw [|->,line width=1,darkblue] (0,-0.5) to [out = 0, in = 180] (2,-1);
\draw [|<-,line width=1,black] (0,-1-0.1) to (2,-1-0.1);
\node[above] at (0,0.1) {$|\psi_{\text{in}}\rangle$};
\node[above,black] at (-1,0.1) {$e^{i H_0 t}$};
\node[] at (0,-0.25) {$|\psi\rangle$};
\node[above,darkblue] at (0.8,-0.5) {$e^{i H t}$};
\node[below] at (0,-1.1) {$|\psi_{\text{out}}\rangle$};
\node[below,black] at (1,-1.1) {$e^{i H_0 t}$};
\node[darkred] at (-1.5,1) {${}^{t=-\infty}$};
\node[darkred] at (0,1)  {${}^{t=0}$};
\node[darkred] at (1.5,1)  {${}^{t=\infty}$};
\draw [->, line width=0.5,darkred] (-2,1.3) to (2,1.3);
\node[darkred] at (-1.5,1.5)  {${}^{\text{time}}$};
\node[] at (-1.5,-1.2) {\large $\boxed  S$};
\node[] at (-1.5,-2.1) {};
\end{tikzpicture}
\hspace{30pt}
\begin{tikzpicture}
\draw [|->,line width=1,darkgreen] (0,1) to [out = 180, in = 0] (-2,0.3);
\draw [<-,line width=1,black] (0,0.2) to (-2,0.2);
\draw [|->,line width=1,black] (0,0.1) to (-2,0.1);
\draw [<-,line width=1,darkblue] (0,-0.5) to [out = 180, in = 0] (-2,0);
\draw [|->,line width=1,darkblue] (0,-0.5) to [out = 0, in = 180] (2,-1);
\draw [|<-,line width=1,black] (0,-1.1) to (2,-1.1);
\draw [->,line width=1,black] (0,-1.2) to (2,-1.2);
\draw [|<-,line width=1,darkgreen] (0,-2) to [out = 0, in = 190] (2,-1.3);
\node[above] at (0.4,0.75) {$|\psi_{\text{in}}\rangle$};
\node[above] at (0.3,0.1) {$|\psi_\text{in}^\text{d}\rangle$};
\node[below] at (0,-1.9) {$|\psi_{\text{out}}\rangle$};
\node[below] at (-0.3,-1) {$|\psi_\text{out}^\text{d}\rangle$};
\node[above,darkgreen] at (-1,0.7) {$e^{i H_\text{sc} t}$};
\node[above,black] at (-0.5,0.2) {$e^{i H_0 t}$};
\node[above,darkblue] at (0.8,-0.5) {$e^{i H t}$};
\node[below,black] at (0.9,-1.15) {$e^{i H_0 t}$};
\node[below,darkgreen] at (1.2,-1.6) {$e^{i H_\text{sc} t}$};
\draw [->, line width=0.5,darkred] (-2,1.5) to (2,1.5);
\node[darkred] at (-1.6,1.6) {${}^{t=-\infty}$};
\node[darkred] at (0,1.6)  {${}^{t=0}$};
\node[darkred] at (1.6,1.6)  {${}^{t=\infty}$};
\node[] at (-1.5,-1)  {\large ${\boxed {S_H}}$};
\end{tikzpicture}
}}
  \caption{(Left) The traditional $S$-matrix is
  computed from Fock states evolved using $H_0$ and $H$.
   (Right) The hard $S$-matrix is computed either using Fock states evolved with $H_\text{sc}$ and $H$  or using dressed states evolved with $H_0$ and $H$.
\label{fig:SH}}
\end{figure}
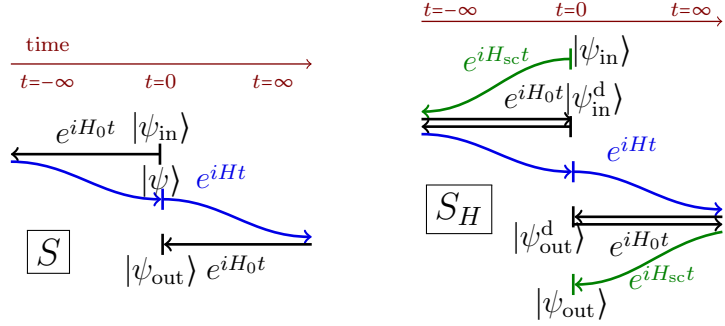

Another interpretation of the hard matrix elements can be obtained by defining dressed states as
\eq{
| \psi_{\text{in}}^d \rangle & \equiv
      \Omega_-^{\text{as}\dag} |\psi_\text{in}\rangle \\
| \psi_{\text{out}}^d \rangle & \equiv
    \Omega_+^{\text{as}\dag} |\psi_\text{out}\rangle
}
Then,
\eq{
    \langle \psi_{\text{out}} | S_H | \psi_{\text{in}} \rangle  = \langle \psi_{\text{out}}^d | S | \psi_{\text{in}}^d \rangle
}
i.e.\ the matrix elements of the hard $S$-matrix are equivalent to matrix elements of the traditional $S$-matrix between dressed states. This connection was made in the context of QED in~\cite{Contopanagos:1991yb}.
The role of the asymptotic evolution can then be viewed as transforming the in-state defined at $t=0$ into a dressed state at $t=-\infty$ that scatters in the traditional way (with $S$). The role of dressed states is illustrated in Figure~\ref{fig:SH}.

The dressed states $|\psi^d_\text{in}\rangle$ and  $|\psi^d_\text{out}\rangle$ are not normalizable elements of the Fock space that $|\psi_{\text{in}}\rangle$ and  $|\psi_{\text{out}}\rangle$ live in. Indeed, if we expand them perturbatively their coefficients in the Fock space basis contain infrared divergent integrals.
 For example, starting with an $|e^+e^-\rangle$  state
\eq{
    | \psi_{\text{in}} \rangle =|\bar v_s(p_1)   u_{s'}(p_2)\rangle
    = \sqrt{2 \omega_{p_1}} b^{s\,\dagger}_{p_1} \sqrt{2 \omega_{p_2}} a^{s'\, \dagger}_{p_2} | 0 \rangle
}
in QED, the asymptotic M\o ller operator can add or remove soft photons with each factor of the coupling $e$. Up to order $\mc{O}(e^2)$ the dressed state will be a superposition of the leading order $| e^+  e^- \rangle$ state,  $| e^+ e^- \gamma \rangle$ states
and $| e^+ e^- \gamma \gamma \rangle$ Fock states. Explicitly,
\eq{
|\psi_\text{in}^d\rangle & =
|\bar v(p_1)   u(p_2) \rangle  \\
&- e \int \frac{d^3 k}{(2\pi)^3} \frac{1}{2\omega_k} \left[
 \frac{ p_1 \cdot \epsilon}{p_1 \cdot k} \Big|\bar v(p_1-k)   u(p_2) \epsilon(k) \Big\rangle
-
\frac{p_2 \cdot \epsilon}{p_2 \cdot k} \Big|\bar v(p_1) u(p_2-k) \epsilon(k) \Big\rangle  \right]
\\
&
+ \frac{e^2}{2}
\int \frac{d^3 k_1}{(2\pi)^3} \frac{1}{2\omega_{k_1}}
\int \frac{d^3 k_2}{(2\pi)^3} \frac{1}{2\omega_{k_2}}
\\
&\hspace{-10mm}\times \left[
\frac{ p_1 \cdt \epsilon_1}{p_1 \cdt k_1}\frac{ p_1 \cdt \epsilon_2}{p_1 \cdt k_2} \Big|\bar v(p_1-k_1-k_2) u(p_2) \epsilon_1(k_1) \epsilon_2(k_2) \Big\rangle
+  \frac{ p_2 \cdt \epsilon_1}{p_2 \cdt k_1} \frac{ p_2 \cdt \epsilon_2}{p_2 \cdot k_2}\Big|\bar v(p_1) u(p_2-k_1-k_2) \epsilon_1(k_1) \epsilon_2(k_2) \Big\rangle
\right. \\
&\hspace{-8mm}
- \frac{ p_1 \cdt \epsilon_1}{p_1 \cdt k_1} \frac{ p_2 \cdt \epsilon_2}{p_2 \cdt k_2}\Big|\bar v(p_1-k_1)   u(p_2-k_2) \epsilon_1(k_1) \epsilon_2(k_2) \Big\rangle
\left. -\frac{ p_1 \cdt \epsilon_2}{p_1 \cdt k_2} \frac{ p_2 \cdt \epsilon_1}{p_2 \cdt k_1}\Big|\bar v(p_1-k_2)   u(p_2-k_1) \epsilon_1(k_1) \epsilon_2(k_2) \Big\rangle \right]
\\ &
- e^2 \int \frac{d^3 k}{(2\pi)^3}\frac{1}{2\omega_k}
\frac{p_1\cdt p_2}{p_1\cdt k \, p_2\cdt k} \Big|\bar v(p_1-k)   u(p_2+k)\Big\rangle  + \cdots
}
Let us make a few observations about these dressed states.
First, note that the Fock states being added have different 3-momenta.
When $k$ has exactly zero momentum (the case almost exclusively considered in the literature), momentum is conserved. But if one really wants to take these dressed states seriously, $k$ must be allowed to have finite energy too, and then $|\psi_\text{in}^d\rangle$ is not a momentum eigenstate.

Second, the coefficient at order $e^2$ is a UV and IR divergent integral. The IR divergence is expected; it is exactly the IR divergence that cancels the IR divergence in elements of $S$ to make elements of $S_H$ IR finite. Nevertheless, it makes $|\psi_\text{in}^d\rangle$ hard to deal with as a state. The divergence requires an excursion from the traditional Fock space to a von Neumann space~\cite{Kibble:1968sfb,Kibble:1969ip,Kibble:1969ep,Kibble:1969kd}. The UV divergence is due to the fact a soft momentum is not sensitive to any hard scale in the problem, so there is no natural cutoff on the $k$ integrals. One could, of course, put in explicit hard cutoffs on the soft momenta, however, it is easier to simply renormalize the UV divergence by rescaling $|\psi_\text{in}^d\rangle$.

Third, it is not each separate electron that is being dressed. Rather it is the combination. Indeed, the IR divergence in the example above comes from loops connecting the two electrons.  These loops are critical to cancelling the IR divergences in $S_H$.  In
Chung's original formulation (cf. Eq.~\eqref{RChung}), a picture can be sketched for a coherent state as an electron moving with a cloud of photons around it. But this picture is too naive: the cloud depends on all the charged particles. This is even clearer in QCD,
where the soft factors come with non-Abelian color matrices so one cannot rely on the crutch of Abelian exponentiation to move the dressing factors from state to state at will.  A discussion of additional complications in QCD and the failure of Bloch-Nordsieck mechanism, can be found in~\cite{Catani:1985xt}.

In conclusion, although the dressed state picture fits in naturally with the construction of $S_H$ we have presented,  we doubt
that thinking of the dressed states as physical states will ultimately be profitable.


%
%
%
%
We emphasize that for the purpose of having finite matrix elements, neither the  in- and out-states $| \psi_{\text{in}} \rangle$ and $| \psi_{\text{out}} \rangle$, nor the dressed states $| \psi^d_{\text{in}} \rangle$ and $| \psi^d_{\text{out}} \rangle$, need to be eigenstates of the asymptotic Hamiltonian. In the examples to follow we will take $| \psi_{\text{in}} \rangle$ and $| \psi_{\text{out}} \rangle$ to be eigenstates of the free momentum operator $P_0^\mu$ with a finite number of particles, but in principle they can be taken to be any sensible linear combination of states in the relevant Hilbert space, i.e.\ with finite coefficients, in contrast to the usual coherent states which are an infinite linear superposition of Fock state elements. The $S_H$-matrix elements between any such states are always finite.

\subsection{Computing observables using $S_H$ \label{sec:observables}}
To compute an observable using $S_H$, one must specify what is to be included in the measurement and what is not. As a concrete example, consider computing the inclusive decay rate of the $Z$ boson in perturbation theory. Since the $Z$ does not couple to massless gauge bosons, it has no interactions in $\Has$
and therefore $\Omega_\pm^\text{as} |Z \rangle = |Z\rangle$.
The rate is then (up to kinematic factors)
\be
\Gamma_Z
\propto \sum_{X \ne Z} |\langle X | S_H | Z \rangle|^2
=   \sum_{X \ne Z} \langle Z | \Omega_-^{\text{as}}\,S^\dag\,\Omega^{\text{as}\dag}_+ | X \rangle
\langle X | \Omega_+^{\text{as}}\,S\,\Omega^{\text{as}\dag}_- | Z \rangle
\ee
The sum is over all states in the theory except the $Z$ itself, since $Z\to Z$ does not contribute to the rate and includes
an implicit integral over the phase space for $|X\rangle$.
Now we write $ \sum_{X\ne Z}| X \rangle
\langle X |  =\bbone - |Z\rangle \langle Z|$ to get
\be
\Gamma_Z
\propto\langle Z | Z \rangle -
\langle Z | S^\dag\,\Omega^{\text{as}\dag}_+ |Z \rangle \langle Z| \Omega_+^{\text{as}} S | Z\rangle
= \sum_{X\ne Z} |\langle X | S | Z \rangle|^2
\ee
where $\Omega_+^{\text{as}\dag} |Z \rangle = |Z\rangle$ was used in the last step. So the sum over final states gives the same decay rate using $S_H$ as it would using $S$. The key here was that there are no asymptotic interactions for $Z$. If there were, then the derivation would not hold. But in that case, the $Z\to Z$ forward scattering amplitude would be infrared divergent using $S$ so it is not clear what physical result we should expect.

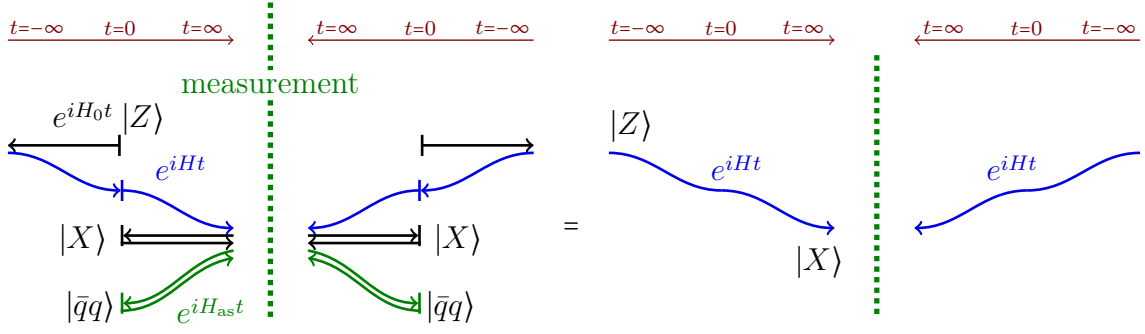
\begin{figure}[t]
\centering
{
\hspace{0pt}
\begin{tikzpicture}
\draw [|->,line width=1,black] (0,0.1) to (-1.5,0.1);
\draw [<-,line width=1,darkblue] (0,-0.5) to [out = 180, in = 0] (-1.5,0);
\draw [|->,line width=1,darkblue] (0,-0.5) to [out = 0, in = 180] (1.5,-1);
\draw [|<-,line width=1,black] (0,-1.1) to (1.5,-1.1);
\draw [->,line width=1,black] (0,-1.2) to (1.5,-1.2);
\draw [|<-,line width=1,darkgreen] (0,-2) to [out = 0, in = 190] (1.5,-1.3);
\draw [->,line width=1,darkgreen] (0,-2.1) to [out = 0, in = 190] (1.5,-1.4);
\node[above] at (0.3,0.1) {$|Z\rangle$};
\node[below] at (-0.4,-1.7) {$|\bar{q}q\rangle$};
\node[below] at (-0.5,-0.8) {$|X \rangle $};
\node[above,black] at (-0.5,0.2) {$e^{i H_0 t}$};
\node[above,darkblue] at (0.8,-0.5) {$e^{i H t}$};
\node[below,darkgreen] at (1.2,-1.8) {$e^{i \Has t}$};
\draw [->, line width=0.5,darkred] (-1.5,1.5) to (1.5,1.5);
\node[darkred] at (-1.1,1.6) {${}^{t=-\infty}$};
\node[darkred] at (0,1.6)  {${}^{t=0}$};
\node[darkred] at (1.1,1.6)  {${}^{t=\infty}$};
\end{tikzpicture}
\hspace{-40pt}
\begin{tikzpicture}
\draw[dotted,darkgreen,line width = 2] (-2.0,2) to (-2.0,1.1);
\draw[dotted,darkgreen,line width = 2] (-2.0,0.7) to (-2.0,-2.2);
\node[below,darkgreen] at (-2.0,1.2) {$\text{measurement}$};
\draw [|->,line width=1,black] (0,0.1) to (1.5,0.1);
\draw [<-,line width=1,darkblue] (0,-0.5) to [out = 180-180, in = 180-0] (1.5,0);
\draw [|->,line width=1,darkblue] (0,-0.5) to [out =180- 0, in = 180-180] (-1.5,-1);
\draw [|<-,line width=1,black] (0,-1.1) to (-1.5,-1.1);
\draw [->,line width=1,black] (0,-1.2) to (-1.5,-1.2);
\draw [|<-,line width=1,darkgreen] (0,-2) to [out = 180-0, in = 180-190] (-1.5,-1.3);
\draw [->,line width=1,darkgreen] (0,-2.1) to [out = 180-0, in = 180-190] (-1.5,-1.4);
\node[below] at (0.4,-1.7) {$|\bar{q}q\rangle$};
\node[below] at (0.5,-0.8) {$|X \rangle $};
\draw [->, line width=0.5,darkred] (1.5,1.5) to (-1.5,1.5);
\node[darkred] at (1.1,1.6) {${}^{t=-\infty}$};
\node[darkred] at (0,1.6)  {${}^{t=0}$};
\node[darkred] at (-1.1,1.6)  {${}^{t=\infty}$};
\end{tikzpicture}
%
%
%
\begin{tikzpicture}
\node at (-2,-1) {$=$};
\draw [-,line width=1,darkblue] (0,-0.5) to [out = 180, in = 0] (-1.5,0);
\draw [->,line width=1,darkblue] (0,-0.5) to [out = 0, in = 180] (1.5,-1);
\draw [|<-,line width=1,white] (0,-2) to [out = 0, in = 190] (1.5,-1.3);
\draw [->,line width=1,white] (0,-2.1) to [out = 0, in = 190] (1.5,-1.4);
\node[above] at (-1.2,0) {$|Z\rangle$};
\node[above] at (1.3,-1.8) {$|X\rangle$};
\node[below,white] at (-0.4,-1.7) {$|\bar{q}q\rangle$};
\node[above,darkblue] at (0.2,-0.5) {$e^{i H t}$};
\draw [->, line width=0.5,darkred] (-1.5,1.5) to (1.5,1.5);
\node[darkred] at (-1.1,1.6) {${}^{t=-\infty}$};
\node[darkred] at (0,1.6)  {${}^{t=0}$};
\node[darkred] at (1.1,1.6)  {${}^{t=\infty}$};
\end{tikzpicture}
\hspace{0pt}
\begin{tikzpicture}
\draw[dotted,darkgreen,line width = 2] (-2.0,1.3) to (-2.0,-2.2);
\draw [-,line width=1,darkblue] (0,-0.5) to [out = 180-180, in = 180-0] (1.5,0);
\draw [->,line width=1,darkblue] (0,-0.5) to [out = 180-0, in = 180-180] (-1.5,-1);
\draw [|<-,line width=1,white] (0,-2) to [out =180- 0, in = 180-190] (-1.5,-1.3);
\draw [->,line width=1,white] (0,-2.1) to [out =180- 0, in = 180-190] (-1.5,-1.4);
\node[below,white] at (0.4,-1.7) {$|\bar{q}q\rangle$};
\node[above,darkblue] at (-0.2,-0.5) {$e^{i H t}$};
\draw [->, line width=0.5,darkred] (1.5,1.5) to (-1.5,1.5);
\node[darkred] at (1.1,1.6) {${}^{t=-\infty}$};
\node[darkred] at (0,1.6)  {${}^{t=0}$};
\node[darkred] at (-1.1,1.6)  {${}^{t=\infty}$};
\end{tikzpicture}
}
\caption{
An observable is computed by integrating the square of an amplitude against a measurement function, inserted at $t=\infty$.
In computing an exclusive observable sensitive to the asymptotic dynamics,
 one must evolve the dressed states to $+\infty$ using the asymptotic Hamiltonian. The example $Z \to \text{jets}$ is illustrated on the left.
The result is equivalent to evolving the initial state $|Z\rangle$ at $t=-\infty$ with the full Hamiltonian
to the set of states $|X\rangle$ on which the measurement is  performed at $t=+\infty$ (right).
\label{fig:jets}
}
\end{figure}

Suppose we wanted to compute something less inclusive than the total decay rate. The observable has to be infrared safe. For example, we could consider a 2-jet rate in $e^+e^- \to \text{hadrons}$. Such a rate depends on the jet definition, which depends on exactly how the soft and collinear momenta are handled. In other words, it depends not only on the hard process, which is roughly speaking the jet-production amplitude, but also on the evolution of the jets after the hard scattering occurs. For this evolution, we need to include the dynamics induced by $e^{- i\Has t_+} \equiv \lim_{t\to\infty}e^{- i\Has t}$, as the state evolves from $t=0$ to $t=\infty$ after the hard scattering. That is,
we should define our exclusive cross section as
\be
\sigma_{\text{2-jet}}
= \sum_X \sum_{Y} |\langle X |e^{-i  \Has t_+} | Y \rangle \langle Y | S_H | Z\rangle|^2 \delta\,\Big[N_\text{jets}(X) - 2\Big]
\ee
Here $N_\text{jets}(X)$ is the measurement function which takes as input the momenta of the particles in the final state $X$ and  returns the number of jets according to some jet definition.
The factor  $ \langle Y | S_H | Z\rangle$ gives the amplitude to produce the jets and $\langle X |e^{-i  \Has t_+}|Y \rangle$ gives the amplitude
for those jets to evolve into a state with the particles in $|X\rangle$ at asymptotic times. The sum over $Y$ can be as restrictive as desired. For example, if $Y$ is taken to be only $|q \bar{q}\rangle$ quark-antiquark states, the distribution will be valid to leading power. To get the jet mass distribution exactly right, including subleading power effects, one should extend the sum from over $|\bar{q} q\rangle$ states to anything that could possibly evolve into a state $X$ with $N_\text{jets}(X) = 2$. For example,  $|\bar{q} q g\rangle$ should be included. If all states are allowed then one can replace $\sum_{Y}  | Y \rangle \langle Y|$ with $\bbone$. In that case, the rate reduces to
\be
\sigma_{\text{2-jet}}  = \sum_X  |\langle X |e^{i H_0 t_+} S | Z\rangle|^2 \delta\,\Big[N_\text{jets}(X) - 2\Big]
\ee
The $e^{iH_0 t_+}$ factor generates a phase $e^{i E_X t}$ which is constant for all $X$ by energy conservation and therefore drops out of the absolute value. Thereby the exclusive cross section reduces to the same thing one would compute using $S$ (in agreement with a century of theory/experiment comparisons).
A cartoon of the reduction of the cross section to the one computed with $S$ for this process is shown in Fig.~\ref{fig:jets}.

Just because one {\it can} reduce cross section calculations using $S_H$ to those using $S$, does not mean one should. Additional physical insight is gained by maintaining the separation into a calculation of $S_H$ first and then of the evolution using  $e^{-i  \Has t_+}$ or equivalently $\Omega_\text{as}^+$. In particular, since $\Has$ is independent of the hard scattering, the separation leads to the physical picture of a short-distance amplitude for jet production followed by an evolution from short-to-long distances where the jets are resolved into their constituents. For example, in the computation of thrust in $e^+e^-$ events, when the events comprise pencil-like jets, the structure of the distribution is almost completely determined by the asymptotic evolution alone. This example, and the utility of the separation will be discussed more in Section~\ref{sec:Zee}.

The above discussion of observables also helps clarify how one should think of assigning hard or soft/collinear labels to the particles in the states. Consider, for example, the process $Z \to \bar{q} q g$. In what circumstances should one consider the gluon momentum to be collinear to the quark or antiquark momenta, or soft?

On the one hand, if one declares the gluon momentum to be soft or collinear, then there are necessarily interactions in $\Has$ that can produce the gluon through a real emission. Due to factorization, the amplitude for this emission from $\Has$ will approach that from $H$, but with an opposite sign. So the two will cancel in the exact soft/collinear limits. In other words, if the gluon momentum is soft/collinear, then the hard matrix element $\langle \bar q q g| S_H |Z\rangle$ will vanish in soft/collinear limits. In this case, there is also a contribution to a $\bar{q} q g$ final state from the hard $\bar q q$ production $\langle \bar{q}q| S_H |Z\rangle$ and then an emission of $g$ though the asymptotic interactions.  This additional contribution is not power suppressed and adds to the $\langle \bar q q g| S_H |Z\rangle$ amplitude to produce the full distribution, in agreement with  $\langle \bar q q g| S |Z\rangle$. Such a deconstruction corresponds to the picture of matching onto a 2-jet operator $C_2 \cO_2$ and then matching on to a 3-jet operator $C_3 \cO_3$ in SCET.~\cite{Bauer:2006mk,Bauer:2006qp}.  In such matching, the Wilson coefficient $C_3$ vanishes in soft and collinear limits.

On the other hand, it does not really make sense to compute $\langle \bar{q} q g| S_H |Z\rangle$ when the gluon is soft or collinear. The hard $S$-matrix is meant to give amplitudes for production of hard particles. The evolution of those hard particles into jets with soft/collinear substructure is subsequently determined by $\Has$. Thus, a more sensible convention is to  consider only matrix elements  $\langle\bar q q g| S_H |Z\rangle$  when all 3 final state particles are considered hard. In this case, these particles have no interactions with each other in $\Has$ and there are no contributions to $\langle  \bar{q} q  g| S_H |Z\rangle$ that have real emissions from the asymptotic region. Thus, all the contributions to $S_H$ involving the asymptotic region are virtual (and give scaleless integrals in pure dimensional regularization).  In other words, if one is interested in 3-jet production, one should study  $\langle  \bar{q} q g| S_H |Z\rangle$ and if one is interested in 2-jet production, one should study  $\langle \bar{q} q | S_H |Z\rangle$. Although the final predictions for IR-safe differential cross sections are independent of what convention we take for assigning labels to the final state particles (and always agree with the result from $S$), the hard $S$-matrix should always be thought of as giving the amplitudes for producing hard particles. With this convention $\langle \bar{q} q g | S_H |Z\rangle$ no longer vanishes in soft or collinear limits. Instead in these limits, it factorizes into
$\langle \bar q q g | e^{-i \Has t_+} | \bar{q}{q} \rangle \langle \bar{q} q | S_H |Z\rangle$. Since the splitting amplitudes $\langle \bar q q g | e^{-i \Has t_+} | \bar{q}{q} \rangle$
are universal~\cite{Kosower:1999xi,Feige:2013zla,Feige:2014wja}, this restricts the possible form that $\langle \bar{q} q g | S_H |Z\rangle$ could have. Implications of these restrictions have
been discussed extensively
(see~\cite{Becher:2009qa, Gardi:2009qi}) and are one instance of the deep structure present in $S_H$-matrix elements.

In summary, one has two choices
\begin{itemize}
    \item Allow states in which $S_H$ matrix elements are taken to have soft or collinear momenta. Observables computed this way will only be valid to leading power, but can be computed efficiently exploiting factorization.
    \item Insist that all states in which $S_H$ matrix elements are taken have only hard momenta. Then all the contributions from the asymptotic regions are virtual, and scaleless in dimensional regularization. Observables agree {\it exactly} with their computation using $S$.
\end{itemize}
We emphasize that with either choice, $S_H$ matrix elements are IR finite.
The general observations in this section are backed up with explicit calculations in Section~\ref{sec:thrust}.

\subsection{Soft Wilson lines \label{sec:softWilson}}
To connect our framework to previous work, we consider the QED case with massive electrons. In this case, there are only soft interactions in the asymptotic Hamiltonian.
The  interaction in the SCET Hamiltonian between soft photons and collinear fermions has the form (see Eq.~\eqref{lscet})
\be
  H_{\text{soft}}^\text{int}(t) = e \sum_n \int d^3 x \, n  \cdt A (x_-) \bar{\xi}_n(x) \frac{\slashed{\bar n}}{2} \xi_n (x)
\ee
where $n^\mu$ is a lightlike 4-vector labeling the fermion, $\bar n^\mu$ is the direction backwards to $n^\mu$, and  $x_- = \bar n \cdot x$.
For simplicity, we take $n^\mu = (1,0,0,1)$ so $\bar n^\mu = (1,0,0,-1)$ and $x_- = t+z$.
The dependence of the interaction only on $x_-$ follows from the multipole expansion\footnote{A collinear momentum scales as
$(p^-,p^+,p_\perp)\sim(\lambda^2,1,\lambda)$ so $x$ scales like $(x_-,x_+,x^\perp)\sim (1,\lambda^{-2},\lambda^{-1})$. Then since a soft momentum scales
 homogeneously like $k\sim\lambda^2$, only the $k^+ x_-$ component is relevant at leading power. See~\cite{Becher:2014oda} for more details.}. The collinear
fields have only half the degrees of freedom of fields in QED: they only describe electrons in this case, as pair-creation is power-suppressed. So we can write
\be
\xi_n(x) = \int \frac{d^3p}{(2\pi)^3} \frac{1}{\sqrt{2\omega_p}} u(p) a_{p} e^{-i p x}, \qquad
\bar \xi_n(x) = \int \frac{d^3 q}{(2\pi)^3} \frac{1}{\sqrt{2\omega_q}} \bar u(q) a_{q}^\dagger  e^{ i q x}
\ee
The field expansion for the soft photon is as usual, but the phase is power expanded
\be
A_{\mu}(x_-)=\sum_{j=1}^{2}\int \frac{d^{3} k}{(2 \pi)^{3}} \frac{1}{\sqrt{2 \omega_{k}}} \left[\epsilon_{\mu}^{j}(k) a_{k}^j e^{-i \frac{1}{2} k^+ x_-}+\epsilon_{\mu}^{j *}(k) a_{k}^{j\,\dagger} e^{i \frac{1}{2} k^+ x_-}\right]
\label{fieldexpansions}
\ee
Inserting these field expansions and integrating over $d^3 x$ gives
\begin{multline}
  H_{\text{soft}}^\text{int}(t)  = e  \sum_n \int \frac{d^3p}{(2\pi)^3 \sqrt{2 \omega_{p}}} \frac{d^3q}{(2\pi)^3\sqrt{2 \omega_{q}}} \frac{d^3k}{(2\pi)^3 \sqrt{2 \omega_{k}}} (2\pi)^3\delta^2(\vec p_\perp - \vec q_\perp) \overline{u}(q) \frac{\slashed{\bar n}}{2} u(p) a_{q}^\dagger  a_{p}  \\
   \times
  \sum_{j=1}^{2}
  \left[n \cdot \epsilon^{j}(k) a_{k}^j \delta\left(q^z - p^z - \frac{1}{2} k^+\right)e^{i (\omega_q - \omega_p - \frac{1}{2} k^+)t}+n \cdot \epsilon^{j *}(k) a_{k}^{j\,\dagger}  \delta\left(q^z - p^z + \frac{1}{2} k^+\right)e^{i (\omega_q - \omega_p + \frac{1}{2} k^+)t}\right]
 \end{multline}
Since $k^+ \ll p^z$ after doing the $q$ integral, we can replace $a_{q}^\dagger \cong a_{p}^\dagger$ at leading power and write
\eq{\frac{1}{\sqrt{2 \omega_p}} \frac{1}{\sqrt{2 \omega_q}} \overline{u}(q) \frac{\slashed{\bar n}}{2} u(p) \cong \frac{1}{2 \omega_p} p \cdot \bar n \cong 1
}
Power expanding the energy $\omega_q$ gives
\eq{
    \omega_q = \sqrt{\vec{p}^2_\perp+\left(p^z \pm \frac{1}{2} k^+\right)^2} \cong \omega_p \pm \frac{p^z}{2\omega_p} k^+
    \cong \omega_p \mp \frac{1}{2} k^+
}
and hence the argument of the exponential becomes $i (\omega_q-\omega_p \mp \frac{1}{2} k^+) t \cong \mp i k^+ t$.
So we get
\be
 H_{\text{soft}}^\text{int}(t) = e \sum_n A_\mu ( t n^\mu)  \int\frac{d^3p}{(2\pi)^3} a_{p}^\dagger a_{p} \label{rhosoft}
\ee
Then we find that the asymptotic M\o ller operator acting on a single electron state gives
\be
    \Omega^\text{soft}_+ |p\rangle = T     \left\{ \text{exp} \left[     -i  \int_0^\infty dt \,     H_{\text{soft}}^\text{int} (t)     \right] \right\} |p \rangle =
    P   \left\{ \text{exp} \left[-ie \int_0^\infty ds \, n\! \cdot \!A(s n^\mu)     \right] \right\} |p \rangle
\ee
with $P$ a path-ordered product.  The path ordering is actually superfluous in QED, but is important in the non-Abelian case. The soft Wilson line in QED is defined as
\be
Y_n^\dagger = \text{exp} \left[     -i  e  \int_0^\infty ds \, n \cdt A(s n^\mu )e^{-\varepsilon s}     \right]
\label{eq:wilsonQED}
\ee
where the factor $e^{-\varepsilon s}$ ensures convergence near $s=\infty$.
Then, the action of the asymptotic soft M\o ller operator is the same as that of a product of soft Wilson lines
\be
   \Omega^\text{soft}_+ |p_1 \cdots p_j \rangle = T \Big\{ Y_{n_1}^\dagger \cdots Y_{n_j}^\dagger \Big\} |p_1 \cdots p_j \rangle
   \label{eq:softWilson}
\ee
For antiparticles, one would have $Y_n$ factors instead, and for incoming particles, one would have factors of
$\overline {Y_n}$, defined as $Y_n^\dagger$ but with an integral from $-\infty$ to $0$~\cite{Feige:2013zla}.

We can combine the time-ordered product of exponential into a single exponential using the Magnus expansion~\cite{Magnus:1954zz},
\begin{multline}
    T\left\{ \text{exp} \left[ \int_0^\infty d t\, \cO (t)  \right] \right\}
    =
    \text{exp} \left\{ \int_0^\infty dt\, \cO (t)
    + \frac{1}{2} \int_0^\infty dt \int_t^\infty ds \left[ \cO (s), \cO (t) \right] \right. \\
    \left.
    \frac{1}{6} \int_0^\infty dt \int_t^\infty ds \int_s^\infty du \,
    \Big( \Big[\cO(u),[\cO(s),\cO(t)] \Big]+\Big[\cO(t),[\cO(s), \cO(u)] \Big] \Big)
    + \cdots \right\}
    \label{eq:magnus}
\end{multline}
where the higher order terms are sums of nested commutators.
The commutators of two fields in Feynman gauge can be computed directly from the field expansions in Eq.~\eqref{fieldexpansions}
\be
\Big[n_1 \cdt A ( s n_1^\mu), n_2 \cdt A (t n_2^\mu)\Big]
=- \int \frac{d^3 k}{(2\pi)^3} \frac{n_1\cdt n_2}{2\omega_k}  \left[ e^{-i (s n_1^\mu - t n_2^\mu) k_\mu} - e^{i (s n_1^\mu - t n_2^\mu) k_\mu} \right] \label{comm}
\ee
Since the commutator in Eq.~\eqref{comm} is a $c$-number, additional commutators vanish. This is the essence of Abelian exponentiation. Then,
we can combine all the time-ordered exponentials into a single exponential:
\be
 T \left\{ Y_{n_1}^\dagger \cdots Y_{n_j}^\dagger \right\} =  \text{exp} \left[-i e \sum_j \int_0^\infty ds \, n_j  \cdt A(s n_j^\mu) \right] \text{exp} \left[ i \sum_{ij} \Phi_{ij}    \right]  \label{Ycombine}
\ee
where
\be
i \Phi_{ij} \equiv -e^2 \, \frac{1}{2} \int_0^\infty dt \int_t^\infty d s \Big[n_i \cdt A ( s n_i^\mu), n_j \cdt A (t n_j^\mu)\Big]e^{-\varepsilon(s+t)}
\label{stplus}
\ee
When acting on states with electrons, this combination is exactly of the form $e^R e^{i \Phi}$ that Faddeev and Kulish write (see Eq.~\eqref{FKform}), with $R$ the expression in Eq.~\eqref{FKR}. The electron-number operator $\rho(\vec{p})$ from Eq.~\eqref{FKrho} is of the same origin as the $a^\dagger_p a_p$ in Eq.~\eqref{rhosoft}.

Consider the case of an outgoing electron and positron in QED, where we want to simplify the time-ordered product of two
Wilson lines $T\{Y_{n_1}^\dagger Y_{n_2}\}$. Then
\be
\cO(t) = -i e \left[n_1 \cdt A(t n_1^\mu) - n_2 \cdt A(t n_2^\mu)\right]
\ee
To see the connection to the Coulomb phase, let us do the integrations over $s$ and $t$ in Eq.~\eqref{stplus} using
Eq.~\eqref{comm}
\be
i \Phi_{ij} = i e^2 \int \frac{d^3 k}{(2\pi)^3}\frac{1}{2\omega_k}
 \text{Im} \frac{n_1\cdt n_2}{( n_1\cdt k - i \varepsilon)( \left(n_1-n_2\right)\cdt k - 2i \varepsilon)}
\ee
Taking $n_1 = (1,0,0,1)$ and $n_2 = (1,0,0,-1)$ we can simplify this to
\be
\Phi = e^2 \int \frac{d^3 k}{(2\pi)^3}\frac{1}{2\omega_k}
 \text{Im}\frac{2}{\left( \omega_k-k_z - i \varepsilon\right) \left(-2 k_z - 2 i \varepsilon \right)}  = - \frac{e^2}{16\pi^2} \int \frac{d^2 k_\perp}{k_\perp^2}
\ee
This is the usual divergent integral appearing in the Coulomb phase (cf. Eq.~\eqref{eq:glaub}). When one of the electrons is incoming, the
$ \int_0^\infty ds$ gets replaced with $ \int_{-\infty}^0 ds$ in Eq.~\eqref{stplus} and we get
\be
\Phi = - e^2 \int \frac{d^3 k}{(2\pi)^3}\frac{1}{2\omega_k}
 \text{Im}\frac{2}{( \omega_k + k_z + i \varepsilon)(\omega_k-k_z - i \varepsilon)}  =  0
\ee
which is consistent with the Coulomb phase vanishing for timelike kinematics.

In this way, we have shown
that our framework agrees with previous work in the case of QED, where there are soft but not collinear singularities and the gauge boson is Abelian. Note that both the Coulomb phase and the real part of the exponent emerge from the single soft-collinear interaction
in $\Has$.

 In the non-Abelian case, one cannot combine the path-ordered exponentials into the  exponential of a single closed-form expression as in Eq.~\eqref{Ycombine}: the gauge generators do not commute.
There is an analog of Abelian exponentiation, called non-Abelian exponentiation~\cite{Gatheral:1983cz,Frenkel:1984pz,Gardi:2013ita} but one must include higher order commutators, and no closed form expression is known.  Thus, a Faddeev-Kulish type formulation of the dressed states is impossible for QCD. The Wilson-line description of the soft interactions is still valid, however, and the soft interactions in QCD still factorize off of the scattering operator into soft Wilson lines.

\section{Computing the hard $S$-matrix \label{sec:compute}}
\begin{figure}[t]
\centering
{
    \begin{tikzpicture}
    \node[darkred] at (-1.5,1.5) {${}^{t=0}$};
    \draw [<-, line width=0.5,darkred] (-2,1.7) to (2,1.7);
    \node[darkred] at (-1.5,1.9)  {${}^{\text{time}}$};
    \draw[dashed, line width=1] (1.5,-2) to (0,-0.5);
    \draw[dashed, line width=1] (1.5,1) to (0,-0.5);
    \draw[dashed, line width=1] (0,-0.5) to (-1.5,-0.5);
    \draw[fill=white, radius=0.5] (0,-0.5) circle;
    \draw[pattern=horizontal lines, radius=0.5] (0,-0.5) circle;
    \node at (2.2,-0.5)  {$\ket{\psi^d_{\text{in}}}$};
    \node at (-2.2,-0.5)  {$\ket{\psi_{\text{in}}}$};
    \draw[dashed,darkred,thick] (2.2,2.2) -- (2.2,0);
   \node[above,darkred] at (2.2,2.0) {${}^{t=-\infty}$};
    \draw[dashed,darkred,thick] (2.2,-1) -- (2.2,-2);
    \node[above,darkgreen]  at  (-1,2) {$\text{asymptotic region}$};
    \end{tikzpicture}
    \hspace{-20pt}
    \begin{tikzpicture}
    \draw [->, line width=0.5,darkred] (-2,1.7) to (2,1.7);
    \draw[dashed, line width=1] (1.5,-2) to (0,-0.5);
    \draw[dashed, line width=1] (1.5,1) to (0,-0.5);
    \draw[dashed, line width=1] (-1.5,-2) to (0,-0.5);
    \draw[dashed, line width=1] (-1.5,1) to (0,-0.5);
    \draw[fill=white, radius=0.5] (0,-0.5) circle;
    \draw[pattern=horizontal lines, radius=0.5] (0,-0.5) circle;
    \draw[dashed,darkred,thick] (2.2,2.2) -- (2.2,0);
   \node[above,darkred] at (2.2,2.0) {${}^{t=\infty}$};
    \draw[dashed,darkred,thick] (2.2,-1) -- (2.2,-2);
    \node[above,darkblue]  at  (0,2) {$\text{central region}$};
    \end{tikzpicture}
    \hspace{-40pt}
    \begin{tikzpicture}
    \node[darkred] at (1.5,1.5)  {${}^{t=0}$};
    \draw [<-, line width=0.5,darkred] (-2,1.7) to (2,1.7);
    \draw[dashed, line width=1] (-1.5,-2) to (0,-0.5);
    \draw[dashed, line width=1] (-1.5,1) to (0,-0.5);
    \draw[dashed, line width=1] (0,-0.5) to (1.5,-0.5);
    \draw[fill=white, radius=0.5] (0,-0.5) circle;
    \draw[pattern=horizontal lines, radius=0.5] (0,-0.5) circle;
    \node at (2.2,-0.5)  {$\ket{\psi_{\text{out}}}$};
    \node at (-2.2,-0.5)  {$\ket{\psi^d_{\text{out}}}$};
    \node[above,darkgreen]  at  (1,2) {$\text{asymptotic region}$};
    \end{tikzpicture}
    \caption{In order to facilitate calculations in perturbation theory, we divide the matrix elements of $S_H$ into three parts. In the two outer parts, the asymptotic evolution M\o ller operators $ \Omega^\text{as}_{\pm}$ work to dress the in- and out-states. The middle part corresponds to a calculation of traditional $S$-matrix elements between dressed states.
        \label{fig:SH3regions}}
}
\end{figure}
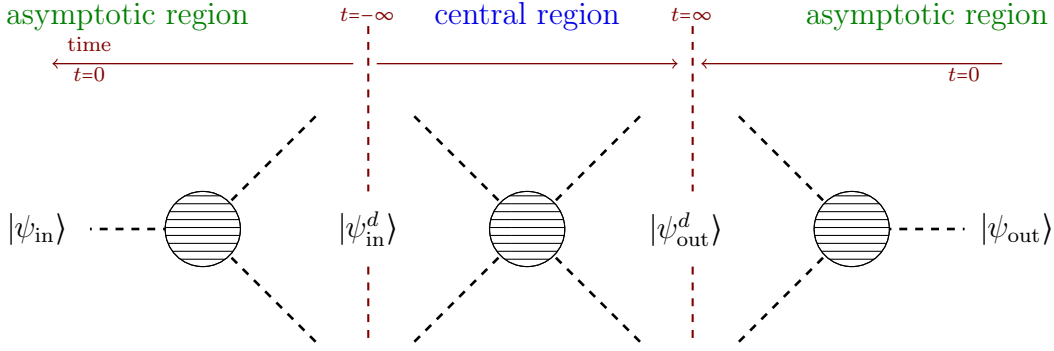

In this section, we show how to compute $S_H$-matrix elements perturbatively. We will use the formula in Eq.~\eqref{eq:SH_terms}:
\eq{
    \langle \psi_{\text{out}} | S_H | \psi_{\text{in}} \rangle  =
    \int d\Pi_{\psi_{\text{out}}'}
    \int d\Pi_{\psi_{\text{in}}'}
    \underbrace{\langle \psi_{\text{out}} | \Omega_{+}^{\text{as}}
    \ket{\psi_{\text{out}}'}}_{\color{darkgreen} \text{asymptotic region}}
    \underbrace{ \bra{\psi_{\text{out}}'}
    S
    \ket{\psi_{\text{in}}'}}_{\color{darkblue} \text{central region}}
    \underbrace{\bra{\psi_{\text{in}}'}
    \Omega^{\text{as}\dag}_-
    | \psi_{\text{in}} \rangle}_{\color{darkgreen} \text{asymptotic region}}
}
  We call the two outer matrix elements the {\it asymptotic region} and the part involving $\bra{\psi_{\text{out}}'} S \ket{\psi_{\text{in}}'}$ the {\it central region}. The asymptotic regions go from $0 > t > -\infty$ and $\infty > t > 0$, both backward in time. The central region calculation is just that of an ordinary $S$-matrix. A cartoon of the division is shown in Fig~\ref{fig:SH3regions}.
  In this section we establish the Feynman rules for the asymptotic regions, which are similar to those in old-fashioned, time-ordered perturbation theory with a few changes. We also give an example calculation in $\phi^3$ theory that clarifies some of the subtleties. Calculations for physical process in QED, QCD and $\cN=4$ SYM theories are given in subsequent sections.
\subsection{Asymptotic region Feynman rules \label{sec:asymrules}}
We have reduced the problem of computing matrix elements of $S_H$ to calculating matrix elements of $S$ and matrix elements of the form
\eq{
\langle \psi_{\text{out}} | \Omega_{+}^{\text{as}}
\ket{\psi_{\text{out}}'} \qquad \text{and} \qquad
\langle \psi'_{\text{in}} | \Omega^{\text{as}\dag}_-
\ket{\psi_{\text{in}}}
}
in perturbation theory.
To evaluate these matrix elements, we separate the asymptotic Hamiltonian into a free part and an interaction part
\eq{
    H_{\text{as}} = H_0 + V_{\text{as}}
}
Defining the operator $U_{+}^{\text{as}}(t)$ by the equation $\Omega_{+}^{\text{as}} = \lim_{t\to \infty} U_{+}^{\text{as}}(t)$, it satisfies the differential equation
\eq{
    -i \partial_t U_{+}^{\text{as}}(t) & = U_{+}^{\text{as}}(t) V^I_{\text{as}}(t) \\
    U_{+}^{\text{as}}(0) & = 1
}
where the superscript $I$ indicates that $V^I_{\text{as}}$ is the interaction picture potential, i.e.\ the asymptotic potential $V_{\text{as}}[\phi_0]=-\int d^3 x \mc{L}_{\text{as}} \left[\phi_0\right]$ expressed in terms of freely-evolving interaction picture fields $\phi_0$, and where $\mc{L}_{\text{as}}$ is the Lagrangian density corresponding to the asymptotic interactions. This differential equation has the solution
\eq{
    U_+^{\text{as}}(t) & = 1 + i \int\limits_0^{t} dt' V^I_{\text{as}}(t') + i^2 \int_0^t dt' \int_0^{t'} dt'' V^I_{\text{as}}(t'') V^I_{\text{as}}(t') + \ldots
    \\
    & = \overline{T} \left\{ \text{exp} \left[i\int_0^t dt' \int d^3 \vec{x}  V^I_{\text{as}}(t') \right] \right\}
}
where $\overline{T}$ denotes an  anti time-ordered product.

To see how to evaluate matrix elements of this operator, consider the following diagram in scalar $\phi^3$ theory:
\eq{
S^{+(2)}_A =
\begin{gathered}
    \begin{tikzpicture}
    \node[darkred] at (-1.5,1.5) {${}^{t=\infty}$};
    \node[darkred] at (1.5,1.5)  {${}^{t=0}$};
    \draw [<-, line width=0.5,darkred] (-2,1.7) to (2,1.7);
    \node[darkred] at (-1.5,1.9)  {${}^{\text{time}}$};
    \draw[dashed, line width=1] (-1.5,-2) to (-0.25,-1);
    \draw[dashed, line width=1] (-0.25,-1) to (1.5,-2);
    \draw[dashed, line width=1] (-0.25,-1) to (0.25,0);
    \draw[dashed, line width=1] (-1.5,1) to (0.25,0);
    \draw[dashed, line width=1] (0.25,0) to (1.5,1);
    \node at (-1.8,1)  {$p_1'$};
    \node at (-1.8,-2)  {$p_2'$};
    \node at (1.8,1)  {$p_1$};
    \node at (1.8,-2)  {$p_2$};
    \node at (0.55,0)  {$x$};
    \node at (-0.55,-1)  {$y$};
    \node at (0.3,-0.5)  {$k$};
    \end{tikzpicture}
    \end{gathered}
}
\vspace{0.5cm}
The free fields are given by
\be
\phi_0(x) = \int \frac{d^3 p}{(2\pi)^3}\frac{1}{\sqrt{2\omega_p}} \left(a_p e^{- i p x} + a^\dagger_p e^{i p x} \right)
\label{phi0def}
\ee
One-particle states in the free theory are:
\eq{
    \ket{p} = \sqrt{2 \omega_p} a^\dagger_p \ket{0}
}
\noindent
Up to renormalization, which will be discussed later, the external states are as usual taken to be free creation operators acting on the free vacuum. We therefore aim to calculate
\eq{
    S^+=& \bra{p_1 p_2} \Omega_{+}^{\text{as}}(t) \ket{p_1' p_2'}
    =
    \bra{p_1 p_2}
\overline{T} \left\{ \text{exp} \left[-i\int_0^\infty dt' \int d^3 \vec{x}  \mc{L}_{\text{as}} \left[\phi_0\right] \right] \right\}
\ket{p_1' p_2'}
}
The second order term in $g$ is:
\eq{
S^{+^{(2)}} & =
\bra{0}
\sqrt{2 \omega_{p_1}} a_{p_1}
\sqrt{2 \omega_{p_2}} a_{p_2}
\int\limits_{0}^{\infty} dt_x  \, \int\limits_{t_x}^{\infty} dt_y \, \int d^3 \vec{x} \, \int d^3 \vec{y} \,
\frac{-i g}{3!}\phi_0^3(x) \frac{-ig}{3!}\phi_0^3(y)
\sqrt{2 \omega_{p_1'}} a^\dagger_{p_1'}
\sqrt{2 \omega_{p_2'}} a^\dagger_{p_2'}
\ket{0}
}
Inserting Eq.~\eqref{phi0def} and commuting creation and annihilation operators, gives the following expression corresponding to the diagram above:
\eq{
S^{+(2)}_A & = \left(-i g\right)^2
\bra{0}
\sqrt{2 \omega_{p_1}} a_{p_1}
\sqrt{2 \omega_{p_2}} a_{p_2}
\int\limits_{0}^{\infty} dt_x  \, \int\limits_{t_x}^{\infty} dt_y \, \int d^3 \vec{x} \, \int d^3 \vec{y}
\\ & \hspace{0.2cm}
\momintFields{q_1}
a^\dagger_{q_1} e^{i q_1 x}
\momintFields{q_2}
a^\dagger_{q_2} e^{i q_2 y}
\momintFields{k} a_{k'} e^{-i k' x}
\momintFields{k'} a^\dagger_{k} e^{i k y}
\\ & \hspace{0.2cm}
\momintFields{q_2'} a_{q_2'} e^{-i q_2' y}
\momintFields{q_1'} a_{q_1'} e^{-i q_1' x}
\sqrt{2 \omega_{p_1'}} a^\dagger_{p_1'}
\sqrt{2 \omega_{p_2'}} a^\dagger_{p_2'}
\ket{0}
}
Integrating over $\vec{x}$ and $\vec{y}$ gives $\delta$-function.
Integrating over these $\delta$-functions
and the additional $\delta$-functions coming from the creation and annihilation operators reduces the expression to
\be
S^{+(2)}_A  =
 \deltaD{3}{\vec{p}_1+\vec{p}_2-\vec{p}_1{}' - \vec{p}_2{}'}
  (-i g)^2
  \frac{1}{2 \omega_k}
 \int_0^\infty d t_x \int_{t_x}^\infty d t_y
e^{i (\omega_1 -\omega'_1 -\omega_k) t_x}
e^{i (\omega_2 - \omega'_2 + \omega_k) t_y}
\label{nested}
\ee
Finally, the integrals over $t_x$ and $t_y$ give
\eq{
    S^{+(2)}_A =
 \deltaD{3}{\vec{p}_1+\vec{p}_2-\vec{p}_1{}' - \vec{p}_2{}'}(-i g)^2
    \frac{1}{2 \omega_k}
    \frac{-i}{\omega_1'+\omega_2'-\omega_1-\omega_2-i\varepsilon}
    \frac{-i}{\omega_2'-\omega_2-\omega_k-i\varepsilon}
}

More generally, the Feynman rules for the asymptotic regions are the same as those in ordinary relativistic time-ordered perturbation theory (see~\cite{Sterman:1994ce} for example) with two differences: 1) Since the outermost integral goes from $0$ to $\infty$ instead of $-\infty$ to $\infty$, the overall energy-conserving $\delta$-function; and $2\pi \delta(E_f-E_i)$ is replaced by a propagator $\frac{i}{E_f-E_i+i\varepsilon}$ 2) the evolution is backwards in time ($e^{i \Has t}$ instead of $e^{-i \Has t}$) so the whole amplitude is complex conjugated. This means $i g \to -i g$ and $\frac{i}{ E + i\varepsilon} \to \frac{-i}{ E - i\varepsilon}$.

For explicit computations and consistency checks, one has to be very careful about the $i\varepsilon$ prescription. It is important to keep in mind that the propagators $\frac{-i}{E - i \varepsilon}$ are distributions, only defined under integration. The $i \varepsilon$ comes from an integral representation of the $\theta$ function:
\begin{multline}
\int_0^\infty d t e^{-i \omega t} =\int_{-\infty}^\infty d t\theta(t) e^{-i \omega t} \\
=\int_{-\infty}^\infty dt \left[\int_{-\infty}^\infty \frac{d E}{2\pi} e^{i E t} \frac{-i}{E-i\varepsilon} \right]e^{-i \omega t}
=\int_{-\infty}^\infty d E  \delta(E-\omega) \frac{-i}{E-i\varepsilon} =  \frac{-i}{\omega - i \varepsilon}
\end{multline}
so it really should be associated with the shift $\omega \to \omega -i\varepsilon$ for any integral ending at $t=+\infty$ or $\omega \to \omega + i \varepsilon$ for any integral starting at $t=-\infty$.
When we have a nested integral, like Eq.~\eqref{nested}, we get
\be
\int_0^\infty d t_2 \int_{t_1}^\infty d t_2 e^{i \omega_1 t_x} e^{i \omega_2 t_2}  \to
\int_0^\infty d t_2 \int_{t_1}^\infty d t_2 e^{i (\omega_1 - i\varepsilon) t_1} e^{i (\omega_2 - i\varepsilon) t_2}
=\frac{-i}{\omega_2 - i \varepsilon} \frac{-i}{\omega_1+\omega_2 - 2 i \varepsilon}
\label{2es}
\ee
So each vertex gives another factor of $\varepsilon$.
An example of the importance of careful treatment of these distributions is given in Section~\ref{sec:scalar_loop}.

In summary, the Feynman rules for $\langle \psi_{\text{out}} | \Omega_{+}^{\text{as}}    \ket{\psi'_{\text{out}}}$ are as follows


\begin{itemize}
    \item Draw all relevant time-ordered diagrams between the state $  \ket{\psi_{\text{out}}}$ at $t=0$ on the right and $  \ket{\psi'_{\text{out}}}$ at $t=\infty$ on the left:
    \eq
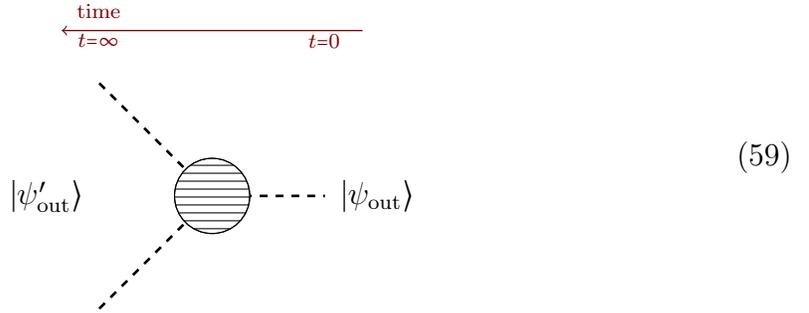
{
    \begin{tikzpicture}
    \node[darkred] at (-1.5,1.5) {${}^{t=\infty}$};
    \node[darkred] at (1.5,1.5)  {${}^{t=0}$};
    \draw [<-, line width=0.5,darkred] (-2,1.7) to (2,1.7);
    \node[darkred] at (-1.5,1.9)  {${}^{\text{time}}$};
    \draw[dashed, line width=1] (-1.5,-2) to (0,-0.5);
    \draw[dashed, line width=1] (-1.5,1) to (0,-0.5);
    \draw[dashed, line width=1] (0,-0.5) to (1.5,-0.5);
    \draw[fill=white, radius=0.5] (0,-0.5) circle;
    \draw[pattern=horizontal lines, radius=0.5] (0,-0.5) circle;
    \node at (2.2,-0.5)  {$\ket{\psi_{\text{out}}}$};
    \node at (-2.2,-0.5)  {$\ket{\psi'_{\text{out}}}$};
    \end{tikzpicture}
}
   \item Assign momenta $k_i^\mu$ to each internal line, with $k_i^0 = \omega_k = \sqrt{m^2 + \vec{k}^2_i}$ the on-shell energy.
    \item Start at the far left of the diagram ($t=\infty$), and move a vertical cut rightwards in time until a vertex is crossed. After each vertex is crossed, include a factor of
        \eqbreak{
        \frac{-i}{(E_\text{out}' - n i \varepsilon) - E_\text{cut}}
    }
    \noindent
    where $E_\text{cut} = \sum \omega_{\text{cut}}$ is the total energy of the particles in the cut, $E_\text{out}' = \sum \omega'_{\text{out}}$ is the total energy of the particles in $|\psi'_{\text{out}}\rangle$, and $n$ is the number of vertices that have already been crossed in the asymptotic region. Note that the $- i \varepsilon$ comes from a $+ i \varepsilon$ from the $t=+\infty$ region, and is then complex conjugated.
     \item For each vertex, add a factor of $(2\pi)^3 \delta^3(\sum\vec{p}_i)$ to impose 3-momentum conservation and $-i g$ for the interaction (or whatever the interaction is, just as in regular Feynman rules, complex-conjugated).
    \item Integrate over $\prod_i \int \frac{d^3k_i}{(2\pi)^3 2 \omega_i}$ for the momentum of each internal line.
\end{itemize}

The Feynman rules for $\langle \psi_{\text{in}}' | \Omega_{-}^{\text{as}}    \ket{\psi_{\text{in}}}$ are identical except that the
diagrams go from $t=-\infty$ on the right to $t=0$ on the left
\eq{
    \begin{tikzpicture}
    \node[darkred] at (-1.5,1.5) {${}^{t=0}$};
    \node[darkred] at (1.5,1.5)  {${}^{t=-\infty}$};
    \draw [<-, line width=0.5,darkred] (-2,1.7) to (2,1.7);
    \node[darkred] at (-1.5,1.9)  {${}^{\text{time}}$};
    \draw[dashed, line width=1] (1.5,-2) to (0,-0.5);
    \draw[dashed, line width=1] (1.5,1) to (0,-0.5);
    \draw[dashed, line width=1] (0,-0.5) to (-1.5,-0.5);
    \draw[fill=white, radius=0.5] (0,-0.5) circle;
    \draw[pattern=horizontal lines, radius=0.5] (0,-0.5) circle;
    \node at (2.2,-0.5)  {$\ket{\psi'_{\text{in}}}$};
    \node at (-2.2,-0.5)  {$\ket{\psi_{\text{in}}}$};
    \end{tikzpicture}
}
and the propagators are
\be
  \frac{-i}{(E_\text{in}'- i n \varepsilon) - E_\text{cut}}
\ee
where $E_\text{in}' = \sum \omega'_{\text{in}}$ is the total energy of the particles in $|\psi'_{\text{in}}\rangle$.

\subsection{Cross check in $\phi^3$ theory}
\label{sec:scalar_loop}
To validate the Feynman rules, consider the case where $\Has=H$. In this case, the hard $S$-matrix is trivial $S_H=\ \bbone$.
Perturbatively, this means that diagrams with all vertices in the central region should be exactly canceled by diagrams
involving vertices in the asymptotic regions.  Moreover, the cancellation should occur for each time-ordered diagram on its own. We can check this cancellation in any theory and any diagram, so we take $\phi^3$ theory with Lagrangian $\mc{L} = - \frac{1}{2} \phi \Box \phi +  \frac{g}{3!} \phi^3$ for simplicity and consider the diagram
\vspace{-5mm}
\setlength{\unitlength}{0.05cm}
\eq{
\begin{gathered}
\begin{tikzpicture}
 \node at (-0.05,0) {
\resizebox{40mm}{!}{
\fmfframe(0,0)(0,0){
\begin{fmfgraph*}(80,80)
\fmfleft{i1}
\fmfright{o1}
\fmf{dashes,tension=1}{i1,v1}
\fmf{dashes, left, tension=0.4}{v1,v2,v1}
\fmf{dashes,tension=1}{v2,o1}
\fmfv{label=$t_1$,label.angle=110,label.dist=0.4cm}{v1}
\fmfv{label=$t_2$,label.angle=70,label.dist=0.4cm}{v2}
\fmfv{label=$\vec{p}$,label.angle=110,label.dist=0.1cm}{i1}
\fmfv{label=$\vec{p}$,label.angle=70,label.dist=0.1cm}{o1}
\fmffreeze
\fmf{phantom, left, tension=0.4}{v1,v3,v1}
\fmf{phantom, left, tension=0.4}{v1,v4,v1}
\fmfv{label=$\vec{k}$,label.angle=60,label.dist=1.2cm}{v3}
\fmfv{label=$\vec{p}-\vec{k}$,label.angle=-60,label.dist=1.0cm}{v4}
\end{fmfgraph*}
}}};
\end{tikzpicture}
\end{gathered}
\label{to1}
}
\vspace{-5mm}

\noindent We sum over diagrams with $t_1$ and $t_2$ going from $0$ to $-\infty$ to $\infty$ and back to $0$.
Let us call the initial energy as $\omega_i=\omega_p$, the final energy $\omega_f=\omega_p$ and the energy of the intermediate state $\omega_{c}=\omega_{p-k}+\omega_{k}$.

The usual time-ordered perturbation theory loop (i.e. the contribution from $S$ to $S_H$ with all vertices in the central region) is
\setlength{\unitlength}{0.025cm}
\eq{
\mc{S}_{1} & =
\begin{gathered}
\begin{tikzpicture}
 \node at (-0.05,0) {
\resizebox{16mm}{!}{
     \fmfframe(0,0)(0,0){
\begin{fmfgraph*}(80,80)
\fmfleft{L1}
\fmfright{R1}
\fmf{dashes,tension=2}{L1,R1}
\end{fmfgraph*}
}}};
\draw[dashed,darkred,thick] (0.8,0.8) -- (0.8,-0.8);
\end{tikzpicture}
\end{gathered}
\hspace{-2mm}
\begin{gathered}
\begin{tikzpicture}
 \node at (-0.05,0) {
\resizebox{16mm}{!}{
\fmfframe(0,0)(0,0){
\begin{fmfgraph*}(80,80)
\fmfleft{i1}
\fmfright{o1}
\fmf{dashes,tension=2}{i1,v1}
\fmf{dashes, left, tension=0.4}{v1,v2,v1}
\fmf{dashes,tension=2}{v2,o1}
\end{fmfgraph*}
}}};
\draw[dashed,darkred,thick] (0.8,0.8) -- (0.8,-0.8);
\end{tikzpicture}
\end{gathered}
\hspace{-2mm}
\begin{gathered}
\begin{tikzpicture}
 \node at (-0.05,0) {
\resizebox{16mm}{!}{
\fmfframe(0,0)(0,0){
\begin{fmfgraph*}(80,80)
\fmfleft{L1}
\fmfright{R1}
\fmf{dashes,tension=2}{L1,R1}
\end{fmfgraph*}
}}};
\end{tikzpicture}
\end{gathered}
 = \frac{(ig)^2}{2} \int \frac{d^3 k }{\left(2\pi\right)^3 4\, \omega_k\, \omega_{p-k}} \frac{i}{\omega_i-\omega_c + i\varepsilon} 2 \pi \delta (\omega_i - \omega_f)
}
To see this cancel other diagrams, it is helpful to break this diagram down further, into the contribution into 3 regions: first, $-\infty < t_1 < t_2 < 0$
then $-\infty < t_1 < 0 < t_2 < \infty$ and finally $0 < t_1 < t_2 < \infty$:
\begin{multline}
\mc{S}_{1} = \frac{(ig)^2}{2} \int \frac{d^3 k }{\left(2\pi\right)^3 4\, \omega_k\, \omega_{p-k}}
\left[
\frac{i}{\omega_i - \omega_c + i\varepsilon}
\frac{i}{\omega_i - \omega_f + 2i\varepsilon}
\right.
\\
\left.
+
\frac{i}{\omega_i-\omega_c + i\varepsilon}
\frac{i}{\omega_f-\omega_c + i \varepsilon}
+
\frac{i}{\omega_f-\omega_c + i\varepsilon}
\frac{i}{\omega_f-\omega_i +2 i\varepsilon }
\right]
\end{multline}
In this decomposition, we have employed the careful treatment of the distributions discussed around Eq.~\eqref{2es}.

Contributions from the loop in the asymptotic region are given by
\eq{
\mc{S}_{2} & =
\begin{gathered}
\begin{tikzpicture}
 \node at (-0.05,0) {
\resizebox{16mm}{!}{
     \fmfframe(0,0)(0,0){
\begin{fmfgraph*}(80,80)
\fmfleft{L1}
\fmfright{R1}
\fmf{dashes,tension=2}{L1,R1}
\end{fmfgraph*}
}}};
\draw[dashed,darkred,thick] (0.8,0.8) -- (0.8,-0.8);
\end{tikzpicture}
\end{gathered}
\hspace{-2mm}
\begin{gathered}
\begin{tikzpicture}
 \node at (-0.05,0) {
\resizebox{16mm}{!}{
\fmfframe(0,0)(0,0){
\begin{fmfgraph*}(80,80)
\fmfleft{L1}
\fmfright{R1}
\fmf{dashes,tension=2}{L1,R1}
\end{fmfgraph*}
}}};
\draw[dashed,darkred,thick] (0.8,0.8) -- (0.8,-0.8);
\end{tikzpicture}
\end{gathered}
\hspace{-2mm}
\begin{gathered}
\begin{tikzpicture}
 \node at (-0.05,0) {
\resizebox{16mm}{!}{
\fmfframe(0,0)(0,0){
\begin{fmfgraph*}(80,80)
\fmfleft{i1}
\fmfright{o1}
\fmf{dashes,tension=2}{i1,v1}
\fmf{dashes, left, tension=0.4}{v1,v2,v1}
\fmf{dashes,tension=2}{v2,o1}
\end{fmfgraph*}
}}};
\end{tikzpicture}
\end{gathered}
=
\frac{(-ig)^2}{2} \int \frac{d^3 k }{\left(2\pi\right)^3 4\, \omega_k\, \omega_{p-k}} \frac{-i}{\omega_i-\omega_c- i\varepsilon}
\frac{-i}{\omega_i-\omega_f -2 i \varepsilon}
\\
\mc{S}_{3} & =
\begin{gathered}
\begin{tikzpicture}
 \node at (-0.05,0) {
\resizebox{16mm}{!}{
     \fmfframe(0,0)(0,0){
\begin{fmfgraph*}(80,80)
\fmfleft{i1}
\fmfright{o1}
\fmf{dashes,tension=2}{i1,v1}
\fmf{dashes, left, tension=0.4}{v1,v2,v1}
\fmf{dashes,tension=2}{v2,o1}
\end{fmfgraph*}
}}};
\draw[dashed,darkred,thick] (0.8,0.8) -- (0.8,-0.8);
\end{tikzpicture}
\end{gathered}
\hspace{-2mm}
\begin{gathered}
\begin{tikzpicture}
 \node at (-0.05,0) {
\resizebox{16mm}{!}{
\fmfframe(0,0)(0,0){
\begin{fmfgraph*}(80,80)
\fmfleft{L1}
\fmfright{R1}
\fmf{dashes,tension=2}{L1,R1}
\end{fmfgraph*}
}}};
\draw[dashed,darkred,thick] (0.8,0.8) -- (0.8,-0.8);
\end{tikzpicture}
\end{gathered}
\hspace{-2mm}
\begin{gathered}
\begin{tikzpicture}
 \node at (-0.05,0) {
\resizebox{16mm}{!}{
\fmfframe(0,0)(0,0){
\begin{fmfgraph*}(80,80)
\fmfleft{L1}
\fmfright{R1}
\fmf{dashes,tension=2}{L1,R1}
\end{fmfgraph*}
}}};
\end{tikzpicture}
\end{gathered}
=
\frac{(-ig)^2}{2} \int \frac{d^3 k }{\left(2\pi\right)^3 4\, \omega_k\, \omega_{p-k}}
\frac{-i}{\omega_f - \omega_c - i\varepsilon}
\frac{-i}{\omega_f-\omega_i -2 i \varepsilon}
}
and contributions from the loop divided between the two asymptotic regions is
\eq{
\mc{S}_{4} & =
\begin{gathered}
\begin{tikzpicture}
 \node at (-0.05,0) {
\resizebox{16mm}{!}{
     \fmfframe(0,0)(0,0){
\begin{fmfgraph*}(80,80)
\fmfleft{i1}
\fmfright{o1,o2}
\fmf{dashes}{i1,v1}
\fmf{dashes}{v1,o1}
\fmf{dashes}{v1,o2}
\end{fmfgraph*}
}}};
\draw[dashed,darkred,thick] (0.8,0.8) -- (0.8,-0.8);
\end{tikzpicture}
\end{gathered}
\hspace{-2mm}
\begin{gathered}
\begin{tikzpicture}
 \node at (-0.05,0) {
\resizebox{16mm}{!}{
\fmfframe(0,0)(0,0){
\begin{fmfgraph*}(80,80)
\fmfleft{i1,i2}
\fmfright{o1,o2}
\fmf{dashes}{i1,o1}
\fmf{dashes}{i2,o2}
\end{fmfgraph*}
}}};
\draw[dashed,darkred,thick] (0.8,0.8) -- (0.8,-0.8);
\end{tikzpicture}
\end{gathered}
\hspace{-2mm}
\begin{gathered}
\begin{tikzpicture}
 \node at (-0.05,0) {
\resizebox{16mm}{!}{
\fmfframe(0,0)(0,0){
\begin{fmfgraph*}(80,80)
\fmfleft{i1,i2}
\fmfright{o1}
\fmf{dashes}{i1,v1}
\fmf{dashes}{i2,v1}
\fmf{dashes}{v1,o1}
\end{fmfgraph*}
}}};
\end{tikzpicture}
\end{gathered}
= \frac{(-ig)^2}{2} \int \frac{d^3 k }{\left(2\pi\right)^3 4\, \omega_k\, \omega_{p-k}} \frac{-i}{\omega_c-\omega_i - i\varepsilon}
\frac{-i}{\omega_c-\omega_f - i\varepsilon}
}
Lastly, there are contributions from diagrams with one vertex in the asymptotic region:
\eq{
\\
\mc{S}_{5} & =
\begin{gathered}
\begin{tikzpicture}
 \node at (-0.05,0) {
\resizebox{16mm}{!}{
     \fmfframe(0,0)(0,0){
\begin{fmfgraph*}(80,80)
\fmfleft{L1}
\fmfright{R1}
\fmf{dashes,tension=2}{L1,R1}
\end{fmfgraph*}
}}};
\draw[dashed,darkred,thick] (0.8,0.8) -- (0.8,-0.8);
\end{tikzpicture}
\end{gathered}
\hspace{-2mm}
\begin{gathered}
\begin{tikzpicture}
 \node at (-0.05,0) {
\resizebox{16mm}{!}{
\fmfframe(0,0)(0,0){
\begin{fmfgraph*}(80,80)
\fmfleft{i1}
\fmfright{o1,o2}
\fmf{dashes}{i1,v1}
\fmf{dashes}{v1,o1}
\fmf{dashes}{v1,o2}
\end{fmfgraph*}
}}};
\draw[dashed,darkred,thick] (0.8,0.8) -- (0.8,-0.8);
\end{tikzpicture}
\end{gathered}
\hspace{-2mm}
\begin{gathered}
\begin{tikzpicture}
 \node at (-0.05,0) {
\resizebox{16mm}{!}{
\fmfframe(0,0)(0,0){
\begin{fmfgraph*}(80,80)
\fmfleft{i1,i2}
\fmfright{o1}
\fmf{dashes}{i1,v1}
\fmf{dashes}{i2,v1}
\fmf{dashes}{v1,o1}
\end{fmfgraph*}
}}};
\end{tikzpicture}
\end{gathered}
= \frac{(ig)(-ig)}{2} \int \frac{d^3 k }{\left(2\pi\right)^3 4\, \omega_k\, \omega_{p-k}} \frac{-i}{\omega_c-\omega_f -i\varepsilon} \left(2\pi\right) \delta(\omega_c-\omega_i)
\\
\mc{S}_{6} & =
\begin{gathered}
\begin{tikzpicture}
 \node at (-0.05,0) {
\resizebox{16mm}{!}{
     \fmfframe(0,0)(0,0){
\begin{fmfgraph*}(80,80)
\fmfleft{i1}
\fmfright{o1,o2}
\fmf{dashes}{i1,v1}
\fmf{dashes}{v1,o1}
\fmf{dashes}{v1,o2}
\end{fmfgraph*}
}}};
\draw[dashed,darkred,thick] (0.8,0.8) -- (0.8,-0.8);
\end{tikzpicture}
\end{gathered}
\hspace{-2mm}
\begin{gathered}
\begin{tikzpicture}
 \node at (-0.05,0) {
\resizebox{16mm}{!}{
\fmfframe(0,0)(0,0){
\begin{fmfgraph*}(80,80)
\fmfleft{i1,i2}
\fmfright{o1}
\fmf{dashes}{i1,v1}
\fmf{dashes}{i2,v1}
\fmf{dashes}{v1,o1}
\end{fmfgraph*}
}}};
\draw[dashed,darkred,thick] (0.8,0.8) -- (0.8,-0.8);
\end{tikzpicture}
\end{gathered}
\hspace{-2mm}
\begin{gathered}
\begin{tikzpicture}
 \node at (-0.05,0) {
\resizebox{16mm}{!}{
\fmfframe(0,0)(0,0){
\begin{fmfgraph*}(80,80)
\fmfleft{L1}
\fmfright{R1}
\fmf{dashes,tension=2}{L1,R1}
\end{fmfgraph*}
}}};
\end{tikzpicture}
\end{gathered}
= \frac{(ig)(-ig)}{2} \int \frac{d^3 k }{\left(2\pi\right)^3 4\, \omega_k\, \omega_{p-k}} \frac{-i}{\omega_c-\omega_i - i\varepsilon} \left(2\pi\right) \delta(\omega_c-\omega_f)
}
Adding these contributions up, we find
\eq{
    \sum_{i=1}^6\mc{S}_i=0
}
Similarly, all the contributions to the other time ordering of the diagram in Eq.~\eqref{to1} sum up to zero. Note that for the cancellation to occur, it was important to keep track of the distributional nature of the diagrams as encoded in the factors of $\varepsilon$.

\section{QED: Deep Inelastic Scattering}
\label{sec:DIS}
As a first real application, we consider the $e^- \gamma^\star \to e^-$ in QED with a massless electron. We call this
deep inelastic scattering (DIS) in reference to the analogous process in QCD at the parton level, although obviously there is nothing inelastic about this scattering.
We want to establish two facts about this process: that the hard $S$-matrix is IR-finite and what its value is. To compute the value for $S_H$ it is most sensible to use dimensional regularization. In dim reg, all the diagrams with interactions in the asymptotic region give scaleless integrals that formally vanish, so the bare $S_H$-matrix element is determined by the $S$-matrix element alone. However, in pure dimensional regularization, it is difficult to separate UV from IR singularities. Therefore to check the cancellation of IR divergences, we use explicit cutoffs in the asymptotic regions.

\subsection{$S_H$ using cutoffs on $\Has$}
\label{sec:cutoffs}

In this section, we look at the diagram where a photon is exchanged between the two electron legs. The Feynman diagram in Feynman-'t Hooft gauge is given by~\cite{Manohar:2003vb}
\begin{multline}
\mc{S}^{(1)}
=
\unitlength = 1mm
\begin{gathered}
\begin{tikzpicture}
 \node at (0,0) {
\parbox{40mm} {
\resizebox{40mm}{!}{
     \fmfframe(0,00)(0,0){
 \begin{fmfgraph*}(40,20)
    \fmftop{R1,R2}
    \fmfbottom{L1}
    \fmf{fermion}{v1,L1}
    \fmf{fermion,label=$\vec{p}_i$,l.s=right}{R1,v1}
    \fmf{fermion,label=$\vec{p}_f$,l.side=right}{v2,R2}
    \fmf{fermion}{L1,v2}
    \fmf{photon,tension=0,l.s=left,label=$k$}{v1,v2}
        \fmfv{d.sh=circle, d.f=30,d.si=0.1w}{L1}
\end{fmfgraph*}
}}}};
\draw[->, line width = 1] (-0.6,-1.5) to (-0.2,-1.2);
\node[above] at (-0.6,-1.4) {$q$};
\end{tikzpicture}
\end{gathered}
\label{SAform}
\\
= i \mathcal{M}_0 (2\pi)^d
        \delta^{d}(p_i+q-p_f)
        \frac{\alpha}{4 \pi}
        \left[
        \frac{1}{\epsilon_{\text{UV}}}
        -
        \frac{2}{ \epsilon_{\text{IR}}^2}
        - \frac{4}{\epsilon_{\text{IR}}}-\frac{2\ln \frac{\widetilde{\mu}^2}{Q^2}}{\epsilon_{\text{IR}}}
        -
        \ln^2 \frac{\widetilde{\mu}^2}{Q^2}
        -
        3 \ln \frac{\widetilde{\mu}^2}{Q^2}
        - 8
        + \frac{\pi^2}{6}
        \right]
\end{multline}
with $\widetilde{\mu}^2=4 \pi e^{-\gamma_E} \mu^2$ and $\mc{M}_0=-e\overline{u}_f \gamma^\alpha u_i$ the tree-level matrix element. To get a cancellation of the IR divergent terms, we need to add contributions to $S_H$ from graphs with vertices in the asymptotic regions. We would like to avoid the possible double counting of the soft and collinear degrees of freedom in $\Has$.
Working in pure dimensional regularization, the soft-collinear overlap always gives scaleless integrals that vanish. Indeed, the method-of-regions approach is to simply discount the overlap region all together. If one works with regulators that separate the UV from IR, one can explicitly remove the overlap through a zero-bin subtraction procedure~\cite{Manohar:2006nz}. In SCET, this is done by computing the soft contribution and the collinear contribution then subtracting the soft-collinear overlap through a soft-collinear power expansion at the diagram level. If one formulates SCET in terms of operators with full theory fields, as in~\cite{Feige:2014wja}, the zero-bin subtraction appears as an operator-level subtraction. In this section, we take the pragmatic approach of ~\cite{Feige:2014wja}: we exclude by hand the soft-collinear region in $\Has$. So we compute soft contributions from $\Has$ by power expanding in the soft limit, and then integrating photon momenta up to some $\omax$. We compute the collinear contributions by power expanding in the collinear limit and including only those photons with energy greater than $\omax$
that are within $\tmax$ of one of the collinear directions. Similar calculations showing IR divergence cancellations for
thrust and jet broadening can be found in~\cite{Feige:2015rea}.

To check IR divergence cancellations, we only need to look at a subset of time-ordered perturbation theory diagrams. For example, the diagrams
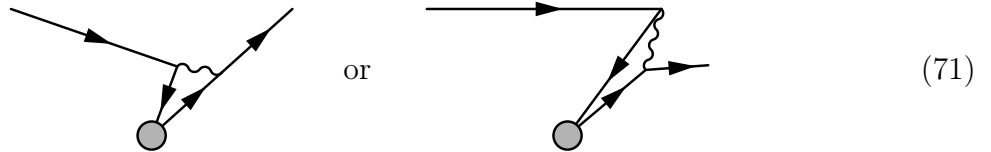
\eq{
\unitlength = 1mm
\begin{gathered}
\begin{tikzpicture}
 \node at (0,0) {
\parbox{40mm} {
\resizebox{40mm}{!}{
     \fmfframe(0,00)(0,0){
 \begin{fmfgraph*}(40,20)
    \fmfbottom{B1}
    \fmftop{L1,R1}
    \fmf{phantom}{L1,w1,w2,w3,w4,w5,R1}
    \fmf{phantom,tension=0.5}{B1,x1,v2,w4}
    \fmf{phantom,tension=0.5}{B1,v1,R1}
    \fmffreeze
    \fmf{fermion}{v1,R1}
    \fmf{fermion}{B1,v1}
    \fmf{fermion}{v2,B1}
    \fmf{fermion}{L1,v2}
    \fmf{photon}{v1,v2}
        \fmfv{d.sh=circle, d.f=30,d.si=0.1w}{B1}
\end{fmfgraph*}
}}}};
\end{tikzpicture}
\end{gathered}
\quad\text{or}\quad
\begin{gathered}
\begin{tikzpicture}
 \node at (0,0) {
\parbox{40mm} {
\resizebox{40mm}{!}{
     \fmfframe(0,00)(0,0){
 \begin{fmfgraph*}(40,20)
    \fmfbottom{B1}
    \fmftop{L1,T1}
    \fmfright{R1}
    \fmf{phantom}{L1,w1,w2,w3,w4,w5,T1}
    \fmf{fermion,tension=0}{L1,w5}
    \fmffreeze
    \fmf{fermion}{w5,B1}
    \fmf{fermion}{v1,R1}
    \fmf{fermion}{B1,v1}
    \fmf{photon}{v1,w5}
        \fmfv{d.sh=circle, d.f=30,d.si=0.1w}{B1}
\end{fmfgraph*}
}}}};
\end{tikzpicture}
\end{gathered}
}
are not IR divergent.
Although these diagrams give finite contributions to $S_H$, they do not need to be analyzed for the purposes of demonstrating IR finiteness.

 It is natural to work in the Breit or ``brick-wall'' frame, where the off-shell photon has no energy, $q^\mu=(0,0,0,Q)$ and $p_i$ and $p_f$ are back to back. Defining $\theta$ as the angle between $\vec{k}$ and $Q$, we have
\be
p_i^\mu = (\omega_i,0,0,\omega_i),\quad
p_f^\mu = (\omega_f,0,0,-\omega_f),\quad
k^\mu = (\omega_k,0, \omega_k \sin \theta, \omega_k \cos\theta)
\ee
and
\eq{
\omega_{i - k} &= \sqrt{\omega_i^2 - 2 \omega_i \omega_k \cos \theta + \omega_k^2}, \hspace{1cm}
\omega_{f - k} &= \sqrt{\omega_i^2 + 2 \omega_i \omega_k \cos \theta + \omega_k^2},
}
If we were to impose overall energy conservation, then we would also have $\omega_i = \omega_f = \frac{Q}{2}$.
However, in time-ordered perturbation theory graphs involving vertices in the asymptotic regions, energy conservation is not guaranteed, so for those diagrams we leave $\omega_i$ and $\omega_f$ more general until energy conservation can be established. With these kinematics the phase space integral becomes
\be
    \int \frac{d^{d-1} k}{\left(2\pi\right)^{d-1} } = \frac{\Omega_{d-2}}{\left(2\pi\right)^{d-1}}  \int d\omega_k \omega_k^{d-2} \int_{-1}^1 d\cos\theta(1-\cos^2\theta)^{\frac{d-4}{2}} \label{pspace}
\ee
where $\Omega_{d-2} = 2 \pi^{\frac{d-2}{2}}/ \Gamma(\frac{d-2}{2})$ is the $d-2$-dimensional solid angle.

The graph with all the vertices in the central region is
\eq{
\mc{S}_{A}
&
=
\unitlength = 1mm
\begin{gathered}
\begin{tikzpicture}
 \node at (0,0) {
\parbox{40mm} {
\resizebox{40mm}{!}{
     \fmfframe(0,00)(0,0){
 \begin{fmfgraph*}(40,20)
    \fmftop{R1,R2}
    \fmfbottom{L1}
    \fmf{fermion}{v1,L1}
    \fmf{fermion,label=$\vec{p}_i$,l.s=right}{R1,v1}
    \fmf{fermion,label=$\vec{p}_f$,l.side=right}{v2,R2}
    \fmf{fermion}{L1,v2}
    \fmf{photon,tension=0,l.s=left,label=$k$}{v1,v2}
        \fmfv{d.sh=circle, d.f=30,d.si=0.1w}{L1}
\end{fmfgraph*}
}}}};
\draw[dashed, line width = 1, darkred] (-1.1,0.8) to (-1.1,-1.1);
\draw[dashed, line width = 1, darkred] (1.1,0.8) to (1.1,-1.1);
\node[above,darkred] at (-1.1,0.9) {${}^{t=-\infty}$};
\node[above,darkred] at (1.1,0.9) {${}^{t=\infty}$};
\end{tikzpicture}
\end{gathered}
= (-ie)^3
    \, \mu^{4-d}
    \int \frac{d^{d-1} k}{\left(2\pi\right)^{d-1} }\frac{1}{2 \omega_k} \frac{1}{2\omega_{i-k}}\frac{1}{ 2 \omega_{f-k} }
\\ &
\times
   \frac{i}{\omega_i-\omega_{i-k}-\omega_k+i\epsilon}
    \frac{i}{\omega_f-\omega_{f-k}-\omega_k+i\epsilon}
    \\ & \hspace{1cm} \times
    \overline{u}_f \gamma^\mu u_{f-k} \overline{u}_{f-k} \gamma^\alpha u_{i-k} \overline{u}_{i-k} \gamma^\nu u_i \left(-g^{\mu \nu} \right)
    \deltaD{d}{p_i+q-p_f}
}
This graph is UV and IR divergent. But since this is the only IR-divergent time-ordering, we know its result must reproduce the IR divergences of the sum over all time orderings, i.e. the Feynman diagram in the full theory. So we can then read the IR divergences directly off of Eq.~\eqref{SAform}:
\be
\cS_A  =
i \mathcal{M}_0
    \deltaD{d}{p_i+q-p_f}
    \frac{\alpha}{4\pi}
    \left[
    -
    \frac{2}{\epsilon_{\text{IR}}^2} - \frac{4}{\epsilon_{\text{IR}}}-\frac{2\ln \frac{\widetilde{\mu}^2}{Q^2}}{\epsilon_{\text{IR}}} + \text{IR-finite}
    \right]
\ee

The contribution with both interactions in an asymptotic region is given by soft photon exchange alone; there are no collinear photons that couple to both the incoming and outgoing electrons since these are back-to-back.
Thus, we need to power expand the integrand in Eq.~\eqref{SAform} at small $\omega_k$ and restrict to $\omega_k < \omax$.
{\it Before} power expanding, the time-ordered perturbation theory amplitude has the form
\eq{
\mc{S}_B
&
=
\unitlength = 1mm
\begin{gathered}
\begin{tikzpicture}
 \node at (0,0) {
\parbox{40mm} {
\resizebox{40mm}{!}{
     \fmfframe(0,00)(0,0){
 \begin{fmfgraph*}(40,20)
    \fmftop{R1,R2}
    \fmfbottom{L1}
    \fmf{fermion,label=$\vec{p}_i$,l.s=right}{R1,v1}
    \fmf{fermion,label=$\vec{p}_f$,l.side=right}{v2,R2}
    \fmf{photon,tension=0,l.s=left,label=$k$}{v1,v2}
    \fmf{fermion}{v1,L1}
    \fmf{fermion}{L1,v2}
    \fmfv{d.sh=circle, d.f=30,d.si=0.1w}{L1}
\end{fmfgraph*}
}}}};
\draw[dashed, line width = 1, darkred] (-0.7,0.8) to (-0.7,-1.1);
\draw[dashed, line width = 1, darkred] (0.7,0.8) to (0.7,-1.1);
\node[above,darkred] at (-0.7,0.9) {${}^{t=-\infty}$};
\node[above,darkred] at (0.7,0.9) {${}^{t=\infty}$};
\end{tikzpicture}
\end{gathered}
= (-ie)(ie)^2
    \, \mu^{4-d}
    \int \frac{d^{d-1} k}{\left(2\pi\right)^{d-1} }\frac{1}{2 \omega_k} \frac{1}{2\omega_{i-k}}\frac{1}{2 \omega_{f-k} }  \theta(\omax - \omega_k)
    \\ &
    \times
    \frac{-i}{\omega_{i-k}+\omega_k-\omega_i-i\varepsilon}
    \frac{-i}{\omega_{f-k}+\omega_k-\omega_f-i\varepsilon}
    \\ & \hspace{1cm} \times
    \overline{u}_f \gamma^\mu u_{f-k} \overline{u}_{f-k} \gamma^\alpha u_{i-k} \overline{u}_{i-k} \gamma^\nu u_i (-g^{\mu \nu} )
    (2\pi)^d \delta^{d-1}\left(\vec{p}_i + \vec{q}-\vec{p}_f\right)
    \delta(\omega_{i-k}-\omega_{f-k} )
    \label{SBform}
}
Note that the overall energy-conserving $\delta$-function $\delta(\omega_i - \omega_f)$ from Eq.~\eqref{SAform} is replaced with  $\delta(\omega_{i-k}-\omega_{f-k} )$.
in Eq.~\eqref{SBform}, however at leading power the two $\delta$-functions agree. In the soft limit, the energies of the intermediate electrons are
\begin{align}
\omega_{i - k} &= \sqrt{\omega_i^2 - 2 \omega_i \omega_k \cos \theta + \omega_k^2} \cong \omega_{i} - \omega_k \cos\theta\\
\omega_{f - k} &= \sqrt{\omega_f^2 + 2 \omega_f \omega_k \cos \theta + \omega_k^2} \cong \omega_{f} + \omega_k \cos\theta
\end{align}
and the numerators are expanded as
\eq{
    \overline{u}_f \gamma^\mu u_{f-k} \overline{u}_{f-k} \gamma^\alpha u_{i-k} \overline{u}_{i-k} \gamma^\nu u_i \left(-g^{\mu \nu} \right)
    \cong
    -4 \, p_i \cdot p_f \, \overline{u}_f \gamma^\alpha \overline{u}_i
    =
    - 8 \, \omega_i \omega_f \, \overline{u}_f \gamma^\alpha \overline{u}_i
}
Inserting the power expansion,
the amplitude reduces to
\begin{multline}
\cS_B = - i \cM_0  (2\pi)^d \delta^{d}(p_i+q-p_f) \frac{\Omega_{d-2}}{(2\pi)^{d-1}}
        \mu^{2 \epsilon} \\ \times
        \int_0^{\omega^{\text{max}}} d\omega_k \omega_k^{1-2\epsilon} \int_{-1}^1 d x(1-x^2)^{-\epsilon}
                  \frac{1}{\omega_k (1-x) - i \varepsilon} \frac{1}{\omega_k (1+x) - i\varepsilon}
                  \label{SBsimple}
\end{multline}
where $x=\cos\theta$. Performing the integrals gives
\be
\mc{S}_B =
i \mathcal{M}_0
       (2\pi)^d \delta^{d}(p_i+q-p_f)
        \frac{\alpha}{4\pi}
        \left[
        - \frac{2}{\epsilon_{\text{IR}}^2}
        +
        \frac{2 \ln \frac{\left(2\, \omax\right)^2}{\widetilde{\mu}^2}}{\epsilon_{\text{IR}}}
        +
        \frac{\pi^2}{2}
        -
        \ln^2\frac{\left(2\,  \omax\right)^2}{\widetilde{\mu}^2}
        \right]
\ee

The remaining two graphs are
\be
\mc{S}_C
=
\unitlength = 1mm
\begin{gathered}
\begin{tikzpicture}
 \node at (0,0) {
\parbox{40mm} {
\resizebox{40mm}{!}{
     \fmfframe(0,00)(0,0){
 \begin{fmfgraph*}(40,20)
    \fmftop{R1,R2}
    \fmfbottom{L1}
    \fmf{fermion}{R1,v1}
    \fmf{fermion}{v2,R2}
    \fmf{photon,tension=0,l.s=left}{v1,v2}
    \fmf{fermion}{v1,L1}
    \fmf{fermion}{L1,v2}
    \fmfv{d.sh=circle, d.f=30,d.si=0.1w}{L1}
\end{fmfgraph*}
}}}};
\draw[dashed, line width = 1, darkred] (-0.7,0.8) to (-0.7,-1.1);
\draw[dashed, line width = 1, darkred] (1.1,0.8) to (1.1,-1.1);
\node[above,darkred] at (-0.7,0.9) {${}^{t=-\infty}$};
\node[above,darkred] at (1.1,0.9) {${}^{t=\infty}$};
\end{tikzpicture}
\end{gathered}
\quad
\text{and}
\quad
\mc{S}_D
=
\unitlength = 1mm
\begin{gathered}
\begin{tikzpicture}
 \node at (0,0) {
\parbox{40mm} {
\resizebox{40mm}{!}{
     \fmfframe(0,00)(0,0){
 \begin{fmfgraph*}(40,20)
    \fmftop{R1,R2}
    \fmfbottom{L1}
    \fmf{fermion}{R1,v1}
    \fmf{fermion}{v2,R2}
    \fmf{photon,tension=0,l.s=left}{v1,v2}
    \fmf{fermion}{v1,L1}
    \fmf{fermion}{L1,v2}
    \fmfv{d.sh=circle, d.f=30,d.si=0.1w}{L1}
\end{fmfgraph*}
}}}};
\draw[dashed, line width = 1, darkred] (-1.1,0.8) to (-1.1,-1.1);
\draw[dashed, line width = 1, darkred] (0.7,0.8) to (0.7,-1.1);
\node[above,darkred] at (-1.1,0.9) {${}^{t=-\infty}$};
\node[above,darkred] at (0.7,0.9) {${}^{t=\infty}$};
\end{tikzpicture}
\end{gathered}
\ee
These have one vertex in the asymptotic region and one in the central region.
In the first graph, the asymptotic vertex forces the exchanged photon to either be soft or collinear to the direction of the outgoing electron. In the second graph, the photon can be soft or collinear to the incoming electron. We must therefore power expand each in soft and collinear limits separately.

Before doing any expansion the first graph is
\eq{
\cS_C & = (-ie)^2(ie)
    \, \mu^{4-d}
    \int \frac{d^{d-1} k}{\left(2\pi\right)^{d-1} }\frac{1}{2 \omega_k} \frac{1}{2\omega_{i-k}}
    \frac{1}{2 \omega_{f-k}}
    \frac{-i}{\omega_{i-k}+\omega_k-\omega_i-i\varepsilon}
    \frac{i}
    {\omega_f-\omega_{f-k}-\omega_k+i\varepsilon}
    \\ & \hspace{2cm}
    \overline{u}_f \gamma^\mu u_{f-k} \overline{u}_{f-k} \gamma^\alpha u_{i-k} \overline{u}_{i-k} \gamma^\nu u_i (-g^{\mu \nu} ) \left(2\pi\right)^d \delta^{d-1}\left(\vec{p}_i + \vec{q}-\vec{p}_f\right) \delta(\omega_{i-k}+\omega_k-\omega_f)
}
In the soft limit, this reduces to the same integral as in $\cS_B$ up to a sign flip since only one vertex is anti-time ordered. $\cS_D$ is similar. So we get
\be
\mc{S}_C^\text{soft} =\mc{S}_D^\text{soft} =
i \mathcal{M}_0
      (2\pi)^d  \delta^{d}(p_i+q-p_f)\frac{\alpha}{4\pi}
        \left[
        \frac{2}{\epsilon_{\text{IR}}^2}
        -
        \frac{2 \ln \frac{\left(2\, \omax\right)^2}{\widetilde{\mu}^2}}{\epsilon_{\text{IR}}}
        -
        \frac{\pi^2}{2}
        +
        \ln^2\frac{\left(2\,  \omax\right)^2}{\widetilde{\mu}^2}
        \right]
\ee
These will cancel the double poles of $\cS_A + \cS_B$.

The graph $\mc{S}_C$ has a collinear singularity when $\theta\to0$. For the collinear graphs, as mentioned above, we consider collinear photons to be collinear but not soft, so they have energies $\omega_k > \omax$ and
angles $0<\theta<\theta^{\text{max}}$. In the collinear limit, $k \parallel p_i$, the energies expand to
\begin{align}
\omega_{i - k} &= \sqrt{\omega_i^2 - 2 \omega_i \omega_k \cos \theta + \omega_k^2} \cong \omega_{i} - \omega_k + \frac{\omega_i \omega_k}{\omega_i - \omega_k}\left(1-\cos\theta\right)\\
\omega_{f - k} &= \sqrt{\omega_f^2 + 2 \omega_f \omega_k \cos \theta + \omega_k^2} \cong \omega_{f} + \omega_k
\end{align}
Since these expansions are only valid in the regime where the electron does not recoil against the photon, i.e.\ for $\omega_k<\omega_i$, we put $\omega_i$ as an upper cutoff on the photon energy.
The spinors in the numerator are on-shell, so in the collinear limit the numerator can be approximated using $p_{i-k} \cong \frac{\omega_{i-k}}{\omega_i} p_i$ and $p_{f-k} \cong \frac{\omega_{f-k}}{\omega_f} p_f$, and hence
\eq{
    \overline{u}_f \gamma^\mu u_{f-k} \overline{u}_{f-k} \gamma^\alpha u_{i-k} \overline{u}_{i-k} \gamma^\nu u_i (-g^{\mu \nu} )
    & \cong
    -4 \, p_i \cdot p_f \frac{\omega_{i-k} \omega_{f-k}}{\omega_i \omega_f} \overline{u}_f \gamma^\alpha u_i
    \cong
    -8 \, \omega_{i-k} \omega_{f-k} \overline{u}_f \gamma^\alpha u_i
}
Then $\cS_C^{\text{coll}}$ reduces to
\begin{multline}
\mc{S}_C^\text{coll} = -i e^2 \cM_0 (2\pi)^d
        \delta^{d}(p_i + q -p_f) \frac{\Omega_{d-2}}{(2\pi)^{d-1}}
        \mu^{2 \epsilon} \\ \times
        \int_{\omax}^{\omega_i} d\omega_k \omega_k^{1-2\epsilon} \int_0^{\tmax} d \theta \sin^{1-2 \epsilon} \theta \frac{1-\frac{\omega_k}{\omega_i}}{\omega_k (1-\cos \theta) - i \varepsilon} \frac{1}{-2\omega_k + i\varepsilon}
\\
\hspace{-10cm} = i \cM_0 (2\pi)^d
        \delta^{d}(p_i + q -p_f) \\ \frac{\alpha}{4\pi}
        \left[
        \frac{2}{\epsilon_{\text{IR}}}
        +
        \frac{\ln \frac{\left(2 \omega^{\text{max}}\right)^2}{Q^2}}{ \epsilon_{\text{IR}}}
        +
        \left(2+ \ln \frac{\left(2\omega^{\text{max}}\right)^2}{Q^2} \right)
        \left(
        2-\ln \frac{\left(\theta^{\text{max}}\omega^{\text{max}}\right)^2}{\widetilde{\mu}^2}
        \right)
        +
        \frac{1}{2} \ln^2 \frac{\left(2\omega^{\text{max}}\right)^2}{Q^2}
        \right]
\end{multline}
Note that this graph has a single $\frac{1}{\epsilon}$ pole corresponding to the collinear-but-not-soft region. The amplitude $\cS_D^\text{coll}$ is the same as $\cS_C^\text{coll}$.

In summary, extracting just the IR poles
\begin{align}
\cS_A &=
i \mathcal{M}_0
    \deltaD{d}{p_i+q-p_f}
    \frac{\alpha}{4\pi}
    \left[
    -
    \frac{2}{\epsilon_{\text{IR}}^2} - \frac{4}{\epsilon_{\text{IR}}}-\frac{2\ln \frac{\widetilde{\mu}^2}{Q^2}}{\epsilon_{\text{IR}}} + \text{IR-finite}
    \right]
\\
\cS_B & = i \mathcal{M}_0
    \deltaD{d}{p_i+q-p_f}
    \frac{\alpha}{4\pi}
    \left[
        -\frac{2}{\eir^2} + \frac{2 \ln \frac{\left(2 \omax\right)^2}{\widetilde{\mu}^2}}{\eir}
        + \text{IR-finite}
        \right]
  \\
\cS_C &  = i \mathcal{M}_0 (2\pi)^d
        \delta^{d}(p_i+q-p_f) \frac{\alpha}{4\pi}
        \left[
        \frac{2}{\eir^2}
        +
        \frac{2}{\eir}
        +
        \frac{\ln \frac{\left(2 \omax\right)^2}{Q^2}}{ \eir}
        -
        \frac{2\ln \frac{\left(2 \omax\right)^2}{\widetilde{\mu}^2}}{\eir}
        + \text{IR-finite}
        \right]
         \\
\cS_D &  = i \mathcal{M}_0 (2\pi)^d
        \delta^{d}(p_i+q-p_f) \frac{\alpha}{4\pi}
        \left[
        \frac{2}{\eir^2}
        +
        \frac{2}{\eir}
        +
        \frac{\ln \frac{\left(2 \omax\right)^2}{Q^2}}{ \eir}
        -
        \frac{2\ln \frac{\left(2 \omax\right)^2}{\widetilde{\mu}^2}}{\eir}
        + \text{IR-finite}
        \right]
\end{align}
with $\mc{M}_0=-e\overline{u}_f \gamma^\alpha u_i$ the tree-level matrix element.
Summing these graphs, the IR divergences all cancel.

Note that this is a different mechanism from the way the cancellation happens in a matching calculation for the DIS Wilson coefficient in SCET~\cite{Manohar:2003vb}. There, the soft graph is subtracted from the full theory graph ($\cS_A - \cS_B$) to achieve the cancellation. Here those graphs are added, and additional graphs come in to effect the cancellation.

\subsection{$S_H$ in dimensional regularization}
Imposing cutoffs on the asymptotic Hamiltonian is useful for showing the cancellation of IR divergences. In practice, however, the calculations are much simpler using pure dimensional regularization. Dimensional regularization respects both Lorentz and gauge invariance, while explicit cutoffs do not. Moreover all  1PI graphs involving vertices in the asymptotic region are scaleless and formally vanish. This follows from simple power counting arguments: in the soft limit, we take all hard scales to infinity so there are no scales left for the amplitude to depend on. In collinear limits, only lightlike momenta in one direction are relevant and no Lorentz-invariant scale can be constructed from collinear lightlike momenta.

For an explicit example,  consider the soft graph $\cS_B$, from Eq.~\eqref{SBsimple}
\be
\mc{S}_B
=
\unitlength = 1mm
\begin{gathered}
\begin{tikzpicture}
 \node at (0,0) {
\parbox{40mm} {
\resizebox{40mm}{!}{
     \fmfframe(0,00)(0,0){
 \begin{fmfgraph*}(40,20)
    \fmftop{R1,R2}
    \fmfbottom{L1}
    \fmf{fermion,label=$\vec{p}_i$,l.s=right}{R1,v1}
    \fmf{fermion,label=$\vec{p}_f$,l.side=right}{v2,R2}
    \fmf{photon,tension=0,l.s=left,label=$k$}{v1,v2}
    \fmf{fermion}{v1,L1}
    \fmf{fermion}{L1,v2}
    \fmfv{d.sh=circle, d.f=30,d.si=0.1w}{L1}
\end{fmfgraph*}
}}}};
\draw[dashed, line width = 1, darkred] (-0.7,0.8) to (-0.7,-1.1);
\draw[dashed, line width = 1, darkred] (0.7,0.8) to (0.7,-1.1);
\node[above,darkred] at (-0.7,0.9) {${}^{t=-\infty}$};
\node[above,darkred] at (0.7,0.9) {${}^{t=\infty}$};
\end{tikzpicture}
\end{gathered}
=  - i \cM_0  (2\pi)^d \delta^{d}(p_i+q-p_f)
         \frac{\Omega_{d-2} }{(2\pi)^{d-2}} \mu^{2 \epsilon} \int_0^\infty d\omega_k \omega_k^{-1-2\epsilon} \int_{-1}^1 d x (1-x^2)^{-1-\epsilon}
\ee
The integral over $\omega_k$ is scaleless and formally vanishes in dimensional regularization. Note that there is also a IR divergence in this case in the angular, $x$, integral, so the final result has an overlapping UV/IR $\frac{1}{\epsilon_\text{UV}} \frac{1}{\epsilon_\text{IR}}$ singularity. Such singularities never occur in renormalizable theory, but they do occur in SCET. However, since when one adds up all the diagrams we know that the IR divergences cancel, the overlapping UV/IR divergences must cancel as well. These cancellations have been studied extensively in SCET (see the reviews~\cite{Becher:2014oda,stewart2013lectures}).

Thus the only non-vanishing graphs in pure dimensional regularization are those with all vertices in the central region. In the central region, hard interactions are present, and these are associated with particular scales. In $d=4-2\epsilon$ dimensions, in Feynman gauge, the result for the loop is given in Eq.~\eqref{SAform}.
For this diagram, the UV and IR divergences can be unambiguously separated since the UV divergences are known separately to be cancelled by the ordinary QED counterterms. For $S_H$ diagrams, such a separation is also possible, but much more difficult, since there can be overlapping UV and IR singularities (see~\cite{Manohar:2006nz,Hornig:2009kv,Feige:2015rea} for some discussion).

In any case, since the other diagrams contributing to $S_H$ are scaleless and since $S_H$ is IR finite, we can immediately write down the bare $S_H$ amplitude using Eq.~\eqref{SAform}. Writing, for $|\psi_\text{out}\rangle \ne |\psi_\text{in}\rangle$
\be
\langle \psi_\text{out}|S_H | \psi_\text{in} \rangle = (2\pi)^d \delta^d(p_\text{in} - p_\text{out}) i {\widehat M}
\ee
we then have
\be
{\widehat M}_\text{bare}
=\cM_0 \left[1+ \frac{\alpha(\mu)}{4 \pi}
\left(
-\frac{2}{\euv^2}-\frac{2 \ln \frac{\widetilde{\mu}^{2}}{Q^{2}}+3}{\euv}
-\ln ^{2} \frac{\widetilde{\mu}^{2}}{Q^{2}}-3 \ln \frac{\widetilde{\mu}^{2}}{Q^{2}}-8+\frac{\pi^{2}}{6} \right)
+\cO(\alpha^2) \right] \label{SHdis}
\ee

The renormalized $S_H$-matrix element is related to the bare one by operator renormalization. To remove the UV divergences, we can rescale the $S$-matrix by
\be
Z = 1+ \frac{\alpha(\mu)}{4 \pi}
\left(
-\frac{2}{\euv^2}-\frac{2 \ln \frac{\widetilde{\mu}^{2}}{Q^{2}}+3}{\euv}
 \right)
+\cO(\alpha^2)  \label{Zdis}
\ee
So that the renormalized matrix element in $\msbar$ is then
\be
{\widehat M} = \left[\frac{1}{Z_4} {\widehat M}_{\text{bare}}\right]= \cM_0 \left[ 1+ \frac{\alpha(\mu)}{4 \pi}
\left(
-\ln ^{2} \frac{\widetilde{\mu}^{2}}{Q^{2}}-3 \ln \frac{\widetilde{\mu}^{2}}{Q^{2}}-8+\frac{\pi^{2}}{6} \right)
+\cO(\alpha^2) \right]
\ee
which is UV and IR finite.

It may seem surprising that $Z$ can depend on the scale $Q$: normally $Z$-factors are just numbers. In fact, the $Q$ dependence is just shorthand for a more formal dependence of the $S_H$-matrix elements on the labels of the collinear fields. In the label formalism, the $S$-matrix for $e^-(p_1) \gamma^\star(q) \to e^- (p_2)$  can depend on its labels, which are the large components of the momenta of the collinear particles,
  $p_1^- = \bar{n}_1 \cdt p_1 \sim Q$ and $p_2^+ = \bar{n}_2 \cdt p_2 \sim Q$. These labels are non-dynamical, and so the $Z$-factor can depend on them. Thus, one could more pedantically write
  \be
Z_{p_1^- p_2^+} = 1+ \frac{\alpha(\mu)}{4 \pi}
\left(
-\frac{2}{\euv^2}-\frac{2 \ln \frac{\widetilde{\mu}^{2}}{p_1^- p_2^+}+3}{\euv}
 \right)
+\cO(\alpha^2)
\ee
and $S^{H,\text{bare}}_{p_1^- p_2^+} = Z_{p_1^- p_2^+} S^{H}_{p_1^- p_2^+}$. But writing the dependence as on $Q$ or more generally $s_{ij} = (p_i + p_j)^2$ is simpler.

It is perhaps worth commenting on why $S_H$ needs to be renormalized in the first place. The traditional $S$-matrix is also an operator, however it does not normally get an operator renormalization: its UV divergences are cancelled by rescaling the interaction strengths in the Lagrangian and the fields. The reason $S_H$ needs to be renormalized is due to diagrams that have both interactions in the asymptotic regions and hard momentum flowing through the graph due to interactions in the central region. The soft particles in $\Has$ cannot resolve the hard scales and there are no interactions in $\Has$ which could be renormalized to remove the associated UV divergences. While $S$-matrix elements are smooth, differentiable functions of momenta, the smoothness is lost in the soft power expansion generating $S_H$. Thus hard scattering, from the point of $S_H$ looks instantaneous and non-local, like a sharp, non-differentiable cusp at the hard vertex. In other words, the additional renormalization required in $S_H$ is the same as the need for renormalization associated with cusps in Wilson line matrix elements.  The non-locality of SCET (on hard length scales) and cusp renormalization is discussed more in~\cite{Becher:2014oda,stewart2013lectures}.

\section{QCD: $e^+e^- \to $ jets}
\label{sec:Zee}
To illustrate the use of $S_H$ to compute infrared-safe observables, we will explore as an example, the computation of thrust in $e^+e^-$ events to NLO in QCD.

The hard matrix element for $\gamma^\star \to \bar{q} q$ is the same as for DIS, up to a crossing. Explicitly,
\be
{\widehat M} =  \cM_0 \left[ 1+  \frac{\alpha_s(\mu)}{4 \pi}  C_F
\left(
-\ln ^{2} \frac{\widetilde{\mu}^{2}}{-Q^{2} - i \varepsilon}-3 \ln \frac{\widetilde{\mu}^{2}}{-Q^{2} - i \varepsilon}-8+\frac{\pi^{2}}{6} \right)
+\cO(\alpha^2) \right]
\label{gphase}
\ee
Due to the $\ln (-Q^2- i \varepsilon)$ term, this $S_H$-matrix element is complex. The imaginary part is the leading order expansion of the Coulomb/Glauber phase, and is present in processes with more than one charged particle in the initial or final state.

\subsection{Glauber graph}
 It is perhaps illuminating to see the origin of the imaginary part from the relevant asymptotic-region graphs. Part of the reason this question is interesting in our framework is because Glauber gluons are normally associated with purely off-shell modes, with entirely transverse momentum. In time-ordered perturbation theory one has only on-shell modes. So how is the Glauber contribution going to be reproduced?
 Before power expanding in the soft region,  the relevant time-ordered diagram is (up to some prefactors):
\eq{
\begin{gathered}
\begin{tikzpicture}
 \node at (-0.5,0.2) {
\resizebox{20mm}{!}{
\fmfframe(0,0)(0,0){
\begin{fmfgraph*}(80,80)
\fmfleft{L1}
\fmfright{R1,R2}
\fmf{dashes,tension=2}{L1,v1}
\fmf{fermion}{R1,v1}
\fmf{fermion}{v1,R2}
\fmfv{decor.shape=pentagram,decor.filled=shaded,decor.size=15}{v1}
\end{fmfgraph*}
}}};
\draw[dashed,darkred,thick] (0.6,1.2) -- (0.6,-1);
\node[above,darkred] at (0.6,1.3) {${}^{t=\infty}$};
\draw[dashed,darkred,thick] (-1,1.2) -- (-1,-1);
\node[above,darkred] at (-1,1.3) {${}^{t=-\infty}$};
\end{tikzpicture}
\end{gathered}
\hspace{-5mm}
\begin{gathered}
\begin{tikzpicture}
 \node at (0,0) {
\resizebox{20mm}{!}{
\fmfframe(0,15)(0,0){
\begin{fmfgraph*}(80,80)
\fmfleft{L1,L3,L2}
\fmfright{R1,R2}
\fmf{fermion,label=$ p_1+  k$,l.s=left}{L2,v2}
\fmf{plain}{v2,v3}
\fmf{fermion,label=$ p_1$}{v3,R2}
\fmf{fermion,label=$ p_2- k$,l.s=left}{v1,L1}
\fmf{fermion,label=$ p_2$}{R1,v1}
\fmffreeze
\fmf{gluon,label=$~\downarrow\!k$,l.s=right}{v1,v2}
\end{fmfgraph*}
}}};
\end{tikzpicture}
\end{gathered}
& \sim
    \int \frac{d^{d-1} k}{(2\pi)^{d-1}}\frac{1}{2 \omega_k} \frac{1}{2\omega_{1+k}}\frac{1}{2 \omega_{2-k} }
    \\[-10mm] & \hspace{0.5cm}
    \times    \frac{-i}{\omega_{1+k} - (\omega_1+\omega_k)-i\varepsilon}
    \frac{-i}{\omega_{1+k} + \omega_{2-k} - (\omega_1+\omega_2)-2 i\varepsilon}
\label{glauber}
}
If we were to enforce 3-momentum and energy-conservation in the central region, this would force $\omega_{1+k} = \omega_{p_2+k} = \omega_1 = \omega_2 = \frac{Q}{2}$. Then $k$ must have exactly zero energy, as expected for an off-shell mode, and the integrand appears ill-defined. The problem however is not that $k$ is off-shell, but that we have not been sufficiently careful handling the product of distributions.
 
 To properly evaluate the integral, we must be patient in enforcing the energy conservation in the central region. Recall that energy conservation comes from integrating over $-\infty <  t < \infty$. If we break the central region up into a $-\infty$ to $0$ region and a $0$ to $\infty$ region, then the hard vertex can be in only one of the regions. Let us also pretend for now that $\Has$ is the same as $H$ with the exception of the hard vertex. Then, if the hard vertex is at $t<0$, the evolution from $e^{-i H t}$ from $0$ to $\infty$ will be exactly be cancelled by the evolution from $t=\infty$ to $0$ in the asymptotic region.  That is
 \be
\hspace{-2mm}
\begin{gathered}
\begin{tikzpicture}
 \node at (-0.05,0) {
\resizebox{20mm}{!}{
\fmfframe(0,0)(0,0){
\begin{fmfgraph*}(80,80)
\fmfleft{L1}
\fmfright{R1,R2}
\fmf{dashes,tension=2}{L1,v2}
\fmf{fermion}{v1,R2}
\fmf{fermion}{v2,v1}
\fmf{fermion}{v3,v2}
\fmf{fermion}{R1,v3}
\fmffreeze
\fmf{phantom}{v2,x1,x2,x3,R2}
\fmf{phantom}{v2,y1,y2,y3,R1}
\fmffreeze
\fmf{gluon}{v1,y3}
\fmfv{decor.shape=pentagram,decor.filled=shaded,decor.size=15}{v2}
\end{fmfgraph*}
}}};
\draw[dashed,darkred,thick] (-1,0.8) -- (-1,-0.8);
\node[above,darkred] at (-1,0.8) {${}^{t=-\infty}$};
\draw[dashed,darkred,thick] (1,0.8) -- (1,-0.8);
\node[above,darkred] at (1,0.8) {${}^{t=\infty}$};
\draw[dashed,darkred,thick] (0,0.8) -- (0,-0.8);
\node[above,darkred] at (0,0.8) {${}^{t=0}$};
\end{tikzpicture}
\end{gathered}
\hspace{-2mm}
\hspace{-5mm}
\begin{gathered}
\begin{tikzpicture}
 \node at (0,0) {
\resizebox{20mm}{!}{
\fmfframe(0,15)(0,0){
\begin{fmfgraph*}(80,80)
\fmfleft{L1,L2}
\fmfright{R1,R2}
\fmf{fermion}{L2,R2}
\fmf{fermion}{R1,L1}
\end{fmfgraph*}
}}};
\end{tikzpicture}
\end{gathered}
%
+
\hspace{-2mm}
\begin{gathered}
\begin{tikzpicture}
 \node at (-0.05,0) {
\resizebox{20mm}{!}{
\fmfframe(0,0)(0,0){
\begin{fmfgraph*}(80,80)
\fmfleft{L1}
\fmfright{R1,R3,R2}
\fmf{dashes,tension=2}{L1,v2}
\fmf{fermion}{v1,R2}
\fmf{fermion}{v2,v1}
\fmf{fermion,tension=0.5}{R1,v2}
\fmffreeze
\fmf{gluon}{v1,R3}
\fmfv{decor.shape=pentagram,decor.filled=shaded,decor.size=15}{v2}
\end{fmfgraph*}
}}};
\draw[dashed,darkred,thick] (-1,0.8) -- (-1,-0.8);
\node[above,darkred] at (-1,0.8) {${}^{t=-\infty}$};
\draw[dashed,darkred,thick] (1,0.8) -- (1,-0.8);
\node[above,darkred] at (1,0.8) {${}^{t=\infty}$};
\draw[dashed,darkred,thick] (0,0.8) -- (0,-0.8);
\node[above,darkred] at (0,0.8) {${}^{t=0}$};
\end{tikzpicture}
\end{gathered}
\hspace{-2mm}
\hspace{-5mm}
\begin{gathered}
\begin{tikzpicture}
 \node at (0,0) {
\resizebox{20mm}{!}{
\fmfframe(0,15)(0,0){
\begin{fmfgraph*}(80,80)
\fmfleft{L1,L3,L2}
\fmfright{R1,R2}
\fmf{fermion}{L2,R2}
\fmf{fermion}{v1,L1}
\fmf{fermion}{R1,v1}
\fmffreeze
\fmf{gluon}{L3,v1}
\end{fmfgraph*}
}}};
\end{tikzpicture}
\end{gathered}
%
%
%
+
\begin{gathered}
\begin{tikzpicture}
 \node at (-0.5,0.2) {
\resizebox{20mm}{!}{
\fmfframe(0,0)(0,0){
\begin{fmfgraph*}(80,80)
\fmfleft{L1}
\fmfright{R1,R2}
\fmf{dashes,tension=2}{L1,v1}
\fmf{fermion}{R1,v1}
\fmf{fermion}{v1,R2}
\fmfv{decor.shape=pentagram,decor.filled=shaded,decor.size=15}{v1}
\end{fmfgraph*}
}}};
\draw[dashed,darkred,thick] (-1,1) -- (-1,-0.8);
\node[above,darkred] at (-1,0.8) {${}^{t=-\infty}$};
\draw[dashed,darkred,thick] (0,1) -- (0,-0.8);
\node[above,darkred] at (-0.2,0.8) {${}^{t=0}$};
\draw[dashed,darkred,thick] (0.5,1) -- (0.5,-0.8);
\node[above,darkred] at (0.6,0.8) {${}^{t=\infty}$};
\end{tikzpicture}
\end{gathered}
\hspace{-5mm}
\hspace{-3mm}
\begin{gathered}
\begin{tikzpicture}
 \node at (0,0) {
\resizebox{20mm}{!}{
\fmfframe(0,15)(0,0){
\begin{fmfgraph*}(80,80)
\fmfleft{L1,L3,L2}
\fmfright{R1,R2}
\fmf{fermion}{L2,v2}
\fmf{plain}{v2,v3}
\fmf{fermion,label}{v3,R2}
\fmf{fermion}{v1,L1}
\fmf{fermion}{R1,v1}
\fmffreeze
\fmf{gluon}{v1,v2}
\end{fmfgraph*}
}}};
\end{tikzpicture}
\end{gathered}
\label{ggsum0}
=0
\ee
In equations, the cancellation occurs point-by-point in phase space as
 \begin{multline}
  \left[
 \frac{i}{\omega_f - \omega_i + 2i \varepsilon} \frac{i}{ \omega_f  - \omega_c + i \varepsilon}
 -
 \frac{i}{\omega_c-\omega_i + i \varepsilon} \frac{-i}{  \omega_c -  \omega_f -  i \varepsilon}
 +
 \frac{-i}{\omega_i  - \omega_c - i \varepsilon} \frac{-i}{\omega_i -  \omega_f -  2i \varepsilon } \right]\\
 \times \frac{i}{Q- \omega_i + i \varepsilon} = 0
 \end{multline}
 where $\omega_i =  \omega_{1 + k}+ \omega_{2 - k}$, $\omega_c =  \omega_{1 + k}+ \omega_{2 - k} + \omega_k$ and $\omega_f = \omega_1 + \omega_2$.
In the real case, where $\Has$ is not exactly the same as $H$ without the hard vertex, these graphs will not sum to precisely zero, but to something that is IR finite.

 The cancellation of the graphs with the hard vertex at $t<0$ implies that the nonzero contribution of the graph in Eq.~\eqref{glauber} comes from the region where the hard vertex is at $t>0$. So we must look at
\begin{multline}
\cM_G = \begin{gathered}
\begin{tikzpicture}
 \node at (-0.5,0.2) {
\resizebox{20mm}{!}{
\fmfframe(0,0)(0,0){
\begin{fmfgraph*}(80,80)
\fmfleft{L1}
\fmfright{R1,R2}
\fmf{dashes,tension=2}{L1,v1}
\fmf{fermion}{R1,v1}
\fmf{fermion}{v1,R2}
\fmfv{decor.shape=pentagram,decor.filled=shaded,decor.size=15}{v1}
\end{fmfgraph*}
}}};
\draw[dashed,darkred,thick] (-1,1.2) -- (-1,-1);
\node[above,darkred] at (-1,1.2) {${}^{t=0}$};
\draw[dashed,darkred,thick] (0.6,1.2) -- (0.6,-1);
\node[above,darkred] at (0.5,1.2) {${}^{t=\infty}$};
\end{tikzpicture}
\end{gathered}
\hspace{-5mm}
\begin{gathered}
\begin{tikzpicture}
 \node at (0,0) {
\resizebox{20mm}{!}{
\fmfframe(0,15)(0,0){
\begin{fmfgraph*}(80,80)
\fmfleft{L1,L3,L2}
\fmfright{R1,R2}
\fmf{fermion,label=$p_1+k$,l.s=left}{L2,v2}
\fmf{plain}{v2,v3}
\fmf{fermion,label=$ p_1$}{v3,R2}
\fmf{fermion,label=$ p_2- k$,l.s=left}{v1,L1}
\fmf{fermion,label=$ p_2$}{R1,v1}
\fmffreeze
\fmf{gluon,label=$~\downarrow\!k$,l.s=right}{v1,v2}
\end{fmfgraph*}
}}};
\end{tikzpicture}
\end{gathered}
\sim
    \int \frac{d^{d-1} k}{(2\pi)^{d-1}}\frac{1}{2 \omega_k} \frac{1}{2\omega_{1+k}}\frac{1}{2  \omega_{2-k} }
    \frac{i}{ \omega_{1+k} + \omega_{2-k} - Q + i \varepsilon}
    \\[-2mm]
\times    \frac{-i}{\omega_{1+k} - (\omega_1+\omega_k)-i\varepsilon}
  \frac{-i}{\omega_{1+k} + \omega_{2-k} - (\omega_1+\omega_2)-2i\varepsilon}
\label{glauber2}
 \end{multline}
 Now  we only have 3-momentum conservation, not energy conservation. So, $\vec{p}_1 + \vec{p}_2=0$ and thus $\omega_1=\omega_2$, but nothing forces $\omega_{1} = \frac{Q}{2}$.
  Defining the angle between $\vec{k}$ and $\vec{p}_1$ as  $\theta$, in the soft limit
  $\omega_{1 + k} \cong \omega_1 + \omega_k \cos\theta$
  and $\omega_{2 - k} \cong \omega_2 + \omega_k \cos\theta$, so performing the power expansion results in
  \begin{align}
\cM_G & \sim   \frac{i}{ \omega_1 + \omega_2 -  Q + i \varepsilon}
\int \frac{d^{d-1} k}{(2\pi)^{d-1}}
      \frac{1}{\omega_k^3}
     \frac{1}{  \cos\theta - 1 - i\varepsilon}
  \frac{1}{ \cos\theta - i\varepsilon}
   \label{MGresult1}\\
 & \sim  \frac{i}{ \omega_1 + \omega_2 -  Q + i \varepsilon} \int d \omega_k \omega_k^{d-5} \left( \frac{1}{\eir} - i \pi + \cdots\right)
 \label{MGresult}
  \end{align}
The $\omega_k$ integral is scaleless, being both UV and IR divergent.  The $i \pi$ in this expression corresponds to the imaginary part in Eq.~\eqref{gphase}, and is known to exponentiate into the Coulomb/Glauber phase. The third graph in Eq.~\eqref{ggsum0} is similar, leading to the same result as in Eq.~\eqref{MGresult} with $  \frac{i}{ \omega_1 + \omega_2 -  Q + i \varepsilon} $
  replaced by $  \frac{i}{ Q- \omega_1 - \omega_2 + i \varepsilon}$. The two graphs combine to produce the expected $\delta(\omega_1 + \omega_2 - Q)$ factor.

  So we see that the Glauber phase is indeed reproduced by asymptotic diagrams with on-shell modes. Moreover, energy {\it is} conserved in this process. The key was to carefully handle the imaginary parts of the propagators and $\delta$ distributions.
  There are of course many other ways to compute this imaginary part (cf.~\cite{Schwartz:2017nmr}), but this approach clarifies the importance of carefully treating energy conservation in $S_H$ computations.

    In more complicated processes, such as $\bar{q}q \to \bar{q}q$ in QCD at 2 loops, it is known that the Glauber contribution from the full graph (the central region) is not reproduced by the eikonal approximation~\cite{Collins:1988ig}.
  Consequences of this failure include collinear-factorization violation~\cite{Forshaw:2008cq} and the emergence of super-leading logarithms~\cite{Forshaw:2006fk}. For $S_H$ this means that the IR divergences of the central region will not be canceled by an asymptotic Hamiltonian with soft and collinear gluons alone. Fortunately, it has been shown that one can add to the SCET Lagrangian a set of Glauber operators~\cite{Rothstein:2016bsq} and remedy the failure of the soft limit. A detailed discussion of when these operators are relevant and how they resolve issues such as collinear-factorization violation can be found in~\cite{Schwartz:2007ib}.
The Glauber interactions, like soft interactions, are long distance and will persist after the hard scattering. Although they violate factorization, in the sense that they are long-distance interactions that depend on multiple directions, they are still independent of the hard scattering.
  
To connect the Glauber graph $\cM_G$ to the Glauber operator, we can massage
the imaginary part of the integral in Eq.~\eqref{MGresult1} into a more familiar form. We first drop the $i\epsilon$ in the denominator $\frac{1}{\cos\theta-1-i\epsilon}$, since the endpoint singularity at $\cos\theta =1$ is regulated for $\epsilon<0$ by the $(1-\cos^2\theta)^{-\epsilon}$ factor in the measure (see Eq.~\eqref{pspace}). Rewriting the integral in terms of $k_z=\omega_k \cos \theta$ and $\vec{k}_\perp$ gives
  \begin{align}
\cM_G& \sim
\int_{-\infty}^{\infty} d k_z
\int \frac{d^{d-2} \vec{k}_\perp}{(2\pi)^{d-2}}
      \frac{1}{\sqrt{k_z^2+\vec{k}_\perp^2}}
     \frac{1}{k_z - \sqrt{k_z^2+\vec{k}_\perp^2}}
  \frac{1}{k_z - 2 i\varepsilon}
  \end{align}
  To take the imaginary part we now use $\text{Im}\left[ \frac{1}{k_z-2i \epsilon} \right] = \pi \delta (k_z)$
  and integrate over $k_z$ to get
\begin{align}
\label{eq:glaub}
\text{Im}\left[\cM_G\right] &\sim
- i \pi
\int \frac{d^{d-2}\vec{k}_\perp}{\left(2\pi\right)^{d-2}} \frac{1}{\vec{k}_\perp^2}
\end{align}
This $\frac{1}{\vec{k}_\perp^2}$ integrand is exactly what comes out of the $\bar{\xi}_{n_1} \frac{\slashed{n}_2}{2} \xi_{n_1} \frac{1}{{\mathcal P}_\perp^2} \bar{\xi}_{n_2}\frac{\slashed{n}_1}{2}  \xi_{n_2}$ Glauber operators~\cite{Rothstein:2016bsq, Schwartz:2017nmr,Schwartz:2018obd}.
In other words, tree-level exchange in the asymptotic region corresponds to the Glauber region expansion, except it has an opposite sign. Note that since $k_z =0$ the on-shell energy of the Glauber gluon is $\omega_k = |\vec{k}_\perp|$. So the $\frac{1}{\vec{k}_\perp^2}$
is not coming from an off-shell mode but rather from energy not being conserved in time-ordered perturbation theory.
Alternative ways of understanding the Glauber phase can be found in~\cite{Korchemsky:1987wg,Chien:2011wz,Laenen:2014jga,Laenen:2015jia}.

For completeness, we list the IR divergent parts of the various contributions to $S_H$ for this process cutting off the UV divergence of the soft integrals at $\omax$, as in Section~\ref{sec:cutoffs}.
Writing $S_H = i \widehat{M} (2\pi)^d \delta^d(q-p_1-p_2)$, the contributions to $\widehat{M}$ are:
\begin{align}
\hspace{-2mm}
\begin{gathered}
\begin{tikzpicture}
 \node at (-0.05,0) {
\resizebox{20mm}{!}{
\fmfframe(0,0)(0,0){
\begin{fmfgraph*}(80,80)
\fmfleft{L1}
\fmfright{R1,R2}
\fmf{dashes,tension=2}{L1,v2}
\fmf{fermion}{v1,R2}
\fmf{fermion}{v2,v1}
\fmf{fermion}{v3,v2}
\fmf{fermion}{R1,v3}
\fmffreeze
\fmf{gluon}{v1,v3}
\fmfv{decor.shape=pentagram,decor.filled=shaded,decor.size=15}{v2}
\end{fmfgraph*}
}}};
\draw[dashed,darkred,thick] (1,0.8) -- (1,-0.8);
\node[above,darkred] at (1,0.8) {${}^{t=\infty}$};
\end{tikzpicture}
\end{gathered}
\hspace{-2mm}
\hspace{-5mm}
\begin{gathered}
\begin{tikzpicture}
 \node at (0,0) {
\resizebox{20mm}{!}{
\fmfframe(0,15)(0,0){
\begin{fmfgraph*}(80,80)
\fmfleft{L1,L2}
\fmfright{R1,R2}
\fmf{fermion}{L2,R2}
\fmf{fermion}{R1,L1}
\end{fmfgraph*}
}}};
\end{tikzpicture}
\end{gathered}
& =
\mathcal{M}_0
    \frac{\alpha_s}{4\pi}C_F
    \left[
    -
    \frac{2}{\eir^2} - \frac{4}{ \eir} + \frac{2 \ln \frac{Q^2}{\widetilde{\mu}^2}}{\eir} - \frac{2 i \pi}{\eir}
    + \text{IR-finite}
    \right]
    \label{MG1}
\\
\begin{gathered}
\begin{tikzpicture}
 \node at (-0.5,0.2) {
\resizebox{20mm}{!}{
\fmfframe(0,0)(0,0){
\begin{fmfgraph*}(80,80)
\fmfleft{L1}
\fmfright{R1,R2}
\fmf{dashes,tension=2}{L1,v1}
\fmf{fermion}{R1,v1}
\fmf{fermion}{v1,R2}
\fmfv{decor.shape=pentagram,decor.filled=shaded,decor.size=15}{v1}
\end{fmfgraph*}
}}};
\draw[dashed,darkred,thick] (0.5,1) -- (0.5,-0.8);
\node[above,darkred] at (0.6,0.8) {${}^{t=\infty}$};
\end{tikzpicture}
\end{gathered}
\hspace{-5mm}
\hspace{-3mm}
\begin{gathered}
\begin{tikzpicture}
 \node at (0,0) {
\resizebox{20mm}{!}{
\fmfframe(0,15)(0,0){
\begin{fmfgraph*}(80,80)
\fmfleft{L1,L3,L2}
\fmfright{R1,R2}
\fmf{fermion}{L2,v2}
\fmf{fermion}{v2,R2}
\fmf{fermion}{v1,L1}
\fmf{fermion}{R1,v1}
\fmffreeze
\fmf{gluon}{v1,v2}
\end{fmfgraph*}
}}};
\end{tikzpicture}
\end{gathered}
& =
\mathcal{M}_0
    \frac{\alpha_s}{4\pi}C_F
    \left[
    -\frac{2}{\eir^2} + \frac{2 \ln \frac{\left(2 \omax\right)^2}{\widetilde{\mu}^2}}{\eir} + \frac{2 i \pi}{\eir}
    + \text{IR-finite}
    \right]
 \label{MG2}
 \\
\begin{gathered}
\begin{tikzpicture}
 \node at (-0.05,0) {
\resizebox{20mm}{!}{
\fmfframe(0,0)(0,0){
\begin{fmfgraph*}(80,80)
\fmfleft{L1}
\fmfright{R1,R3,R2}
\fmf{dashes,tension=2}{L1,v2}
\fmf{fermion}{v1,R2}
\fmf{fermion}{v2,v1}
\fmf{fermion,tension=0.5}{R1,v2}
\fmffreeze
\fmf{gluon}{v1,R3}
\fmfv{decor.shape=pentagram,decor.filled=shaded,decor.size=15}{v2}
\end{fmfgraph*}
}}};
\draw[dashed,darkred,thick] (1,0.8) -- (1,-0.8);
\node[above,darkred] at (1,0.8) {${}^{t=\infty}$};
\end{tikzpicture}
\end{gathered}
\hspace{-2mm}
\hspace{-5mm}
\begin{gathered}
\begin{tikzpicture}
 \node at (0,0) {
\resizebox{20mm}{!}{
\fmfframe(0,15)(0,0){
\begin{fmfgraph*}(80,80)
\fmfleft{L1,L3,L2}
\fmfright{R1,R2}
\fmf{fermion}{L2,R2}
\fmf{fermion}{v1,L1}
\fmf{fermion}{R1,v1}
\fmffreeze
\fmf{gluon}{L3,v1}
\end{fmfgraph*}
}}};
\end{tikzpicture}
\end{gathered} & =
\mathcal{M}_0
\frac{\alpha_s}{4\pi}C_F
        \left[
        \frac{2}{\eir^2}
        +
        \frac{2}{\eir}
        +
        \frac{\ln \frac{\left(2 \omax\right)^2}{Q^2}}{ \eir}
        -
        \frac{2 \ln \frac{\left(2 \omax\right)^2}{\widetilde{\mu}^2}}{\eir}
        + \text{IR-finite}
        \right]
        \label{MG3}
         \\
         \label{MG4}
\begin{gathered}
\begin{tikzpicture}
 \node at (-0.05,0) {
\resizebox{20mm}{!}{
\fmfframe(0,0)(0,0){
\begin{fmfgraph*}(80,80)
\fmfleft{L1}
\fmfright{R1,R3,R2}
\fmf{dashes,tension=2}{L1,v2}
\fmf{fermion,tension=0.5}{v2,R2}
\fmf{fermion}{v1,v2}
\fmf{fermion}{R1,v1}
\fmffreeze
\fmf{gluon}{v1,R3}
\fmfv{decor.shape=pentagram,decor.filled=shaded,decor.size=15}{v2}
\end{fmfgraph*}
}}};
\draw[dashed,darkred,thick] (1,0.8) -- (1,-0.8);
\node[above,darkred] at (1,0.8) {${}^{t=\infty}$};
\end{tikzpicture}
\end{gathered}
\hspace{-2mm}
\hspace{-5mm}
\begin{gathered}
\begin{tikzpicture}
 \node at (0,0) {
\resizebox{20mm}{!}{
\fmfframe(0,15)(0,0){
\begin{fmfgraph*}(80,80)
\fmfleft{L1,L3,L2}
\fmfright{R1,R2}
\fmf{fermion}{L2,v1}
\fmf{fermion}{v1,R2}
\fmf{fermion}{R1,L1}
\fmffreeze
\fmf{gluon}{L3,v1}
\end{fmfgraph*}
}}};
\end{tikzpicture}
\end{gathered} & = 
\mathcal{M}_0
\frac{\alpha_s}{4\pi}C_F
        \left[
        \frac{2}{\eir^2}
        +
        \frac{2}{\eir}
        +
        \frac{\ln \frac{\left(2 \omax\right)^2}{Q^2}}{ \eir}
        -
        \frac{2\ln \frac{\left(2 \omax\right)^2}{\widetilde{\mu}^2}}{\eir}
        + \text{IR-finite}
        \right]
\end{align}
with $\widetilde{\mu}^2=4 \pi e^{-\gamma_E} \mu^2$ and $\mc{M}_0 = \overline{u}_1 \gamma^\alpha v_1$ the tree level matrix element. Summing these graphs, the IR divergences all cancel.

Note that while the imaginary  part of the Glauber graph, Eq.~\eqref{MG2}, cancels against the $S$-matrix graph, Eq.~\eqref{MG1}, the real part of the Glauber graph has the same sign as the $S$-matrix graph, and the sum of the two cancels against the cut graphs. This is different from how the cancellation occurs in matching to a 2-jet operator in SCET, where a single soft graph cancels both the real and imaginary parts of the divergences of the full-theory graph.

\subsection{Thrust \label{sec:thrust}}
Next, let us use the hard $S$-matrix to compute the thrust observable~\cite{FarhiThrust}. Thrust is a particularly simple infrared-safe $e^+e^-$ observable. It is defined as
\be
T \equiv \max_{\vec{n}} \frac{\sum_{j}\left|\vec{p}_{j} \cdot \vec{n}\right|}{\sum_{j}\left|\vec{p}_{j}\right|}
\ee
It is convenient to use $\tau = 1-T$ rather than $T$. Thrust has the property that for events that consist of two highly collimated jets $\tau \ll 1$. At small $\tau$, thrust is approximated by the sum of the masses of these two jets $\tau \cong \frac{1}{Q^2} \left(m_{J1}^2 + m_{J2}^2\right)$, with $Q$ the center of mass energy. Events that are more spherical have values of $\tau \sim 0.2 - 0.5$.
 
To compute $\frac{d\sigma}{d\tau}$ in perturbation theory using $S_H$, we start at lowest order, where the hard $S$-matrix element is
\be
\widehat \cM_0 (\gamma^\star \to \bar{q} q) = \bar{u}_i(p_q) \gamma^\mu v_j (p_{\bar{q}} )
\ee
At next to leading order we need the hard matrix element for $\bar{q} q$ final states at NLO, as given in Eq. \eqref{gphase}
\be
\widehat \cM (\gamma^\star \to \bar{q} q) =  \widehat \cM_0 \left[ 1+  \frac{\alpha_s(\mu)}{4 \pi}  C_F
\left(
-\ln ^{2} \frac{\mu^{2}}{Q^{2} }-(3+2\pi i) \ln \frac{\mu^{2}}{Q^{2}}-8-3 \pi i +\frac{7\pi^{2}}{6} \right)
+\cO(\alpha_s^2) \right]
 \ee
 We also need the matrix elements for $\gamma^\star \to \bar{q} q g$ where we treat the gluon as hard. Treating it as hard, the only contribution at order $g_s$ is from diagrams with all vertices in the central region. Then this amplitude is identical to the $S$-matrix element for the same process
 \be
\widehat  \cM(\gamma^\star \to \bar q  q g) = - g_s T^a_{ij} \bar{u}_i(p_q)\left[
 \gamma^\alpha \frac{1}{\slashed{p}_q + \slashed{p}_g} \gamma^\mu
 -
 \gamma^\mu \frac{1}{\slashed{p}_{\bar q} + \slashed{p}_g} \gamma^\alpha
 \right]
 v_j (p_{\bar{q}} )
 \epsilon_\alpha^\star(p_g) \label{emit}
 \ee
To compute the observable, we must then evolve these final states to $t=+ \infty$ using $\Has$, as discussed in Section~\ref{sec:observables}. On the formal level, this additional evolution exactly cancels the entire effect of $\Has$, so the cross section predicted is identical to that using the original $S$. On the practical level, however, one can gain additional insight into the distribution by actually using the $S_H$-matrix elements we have computed, rather than simply discarding them and starting over. To this end, it is helpful to contemplate the small $\tau$ and moderate $\tau$ regions separately.
 
For small $\tau$, the gluon is necessarily soft or collinear. Thus we can disregard the hard $\widehat  \cM(\gamma^\star \to \bar q  q g) $ contribution. Instead, we should start with $\widehat \cM_0 (\gamma^\star \to \bar{q} q)$ and then
 evolve the $\bar{q} q$ final state towards a 2-jet state with nonzero $\tau$ using $\Has$. To compute the cross section,
 we need to sum the cut graphs
 \be
\begin{gathered}
\begin{tikzpicture}
 \node at (0,0) {
\resizebox{20mm}{!}{
\fmfframe(0,0)(0,0){
\begin{fmfgraph*}(80,50)
\fmfleft{L1,L2}
\fmfright{R1,R2}
\fmf{phantom}{L1,v1,w1,R1}
\fmf{phantom}{L2,v2,w2,R2}
\fmffreeze
\fmf{fermion}{L1,v1,R1}
\fmf{fermion}{R2,w2,L2}
\fmf{gluon}{v1,w2}
\end{fmfgraph*}
}}};
\draw[dashed,darkred,thick] (-0.8,0.8) -- (-0.8,-0.8);
\draw[dotted,darkgreen, line width = 1.5] (-0.5,0.8) -- (-0.5,-0.8);
\draw[dotted,darkgreen,line width = 1.5] (0,0.8) -- (0,-0.8);
\draw[dotted,darkgreen,line width = 1.5] (0.5,0.8) -- (0.5,-0.8);
\draw[dashed,darkred,thick] (0.8,0.8) -- (0.8,-0.8);
\node[above,darkred] at(-0.8,0.8) {${}^{t=0}$};
\node[above,darkred] at(0.8,0.8) {${}^{t=0}$};
\node[below,white] at(-0.8,-0.8) {${}^{t=0}$};
\node[below,white] at(0.8,-0.8) {${}^{t=0}$};
\end{tikzpicture}
\end{gathered}
+
\begin{gathered}
\begin{tikzpicture}
 \node at (0,0) {
\resizebox{20mm}{!}{
\fmfframe(0,0)(0,0){
\begin{fmfgraph*}(80,50)
\fmfleft{L1,L2}
\fmfright{R1,R2}
\fmf{phantom}{L1,v1,w1,R1}
\fmf{phantom}{L2,v2,w2,R2}
\fmffreeze
\fmf{fermion}{L1,w1,R1}
\fmf{fermion}{R2,v2,L2}
\fmf{gluon}{v2,w1}
\end{fmfgraph*}
}}};
\draw[dashed,darkred,thick] (-0.8,0.8) -- (-0.8,-0.8);
\draw[dotted,darkgreen, line width = 1.5] (-0.5,0.8) -- (-0.5,-0.8);
\draw[dotted,darkgreen,line width = 1.5] (0,0.8) -- (0,-0.8);
\draw[dotted,darkgreen,line width = 1.5] (0.5,0.8) -- (0.5,-0.8);
\draw[dashed,darkred,thick] (0.8,0.8) -- (0.8,-0.8);
\node[above,darkred] at(-0.8,0.8) {${}^{t=0}$};
\node[above,darkred] at(0.8,0.8) {${}^{t=0}$};
\node[below,white] at(-0.8,-0.8) {${}^{t=0}$};
\node[below,white] at(0.8,-0.8) {${}^{t=0}$};
\end{tikzpicture}
\end{gathered}
+
\begin{gathered}
\begin{tikzpicture}
 \node at (0,0) {
\resizebox{20mm}{!}{
\fmfframe(0,0)(0,0){
\begin{fmfgraph*}(80,50)
\fmfleft{L1,L2}
\fmfright{R1,R2}
\fmf{phantom}{L1,v1,w1,R1}
\fmf{phantom}{L2,v2,w2,R2}
\fmffreeze
\fmf{fermion}{L1,v1,w1,R1}
\fmf{fermion}{R2,L2}
\fmf{gluon,left=2}{v1,w1}
\end{fmfgraph*}
}}};
\draw[dashed,darkred,thick] (-0.8,0.8) -- (-0.8,-0.8);
\draw[dotted,darkgreen, line width = 1.5] (-0.5,0.8) -- (-0.5,-0.8);
\draw[dotted,darkgreen,line width = 1.5] (0,0.8) -- (0,-0.8);
\draw[dotted,darkgreen,line width = 1.5] (0.5,0.8) -- (0.5,-0.8);
\draw[dashed,darkred,thick] (0.8,0.8) -- (0.8,-0.8);
\node[above,darkred] at(-0.8,0.8) {${}^{t=0}$};
\node[above,darkred] at(0.8,0.8) {${}^{t=0}$};
\node[below,white] at(-0.8,-0.8) {${}^{t=0}$};
\node[below,white] at(0.8,-0.8) {${}^{t=0}$};
\end{tikzpicture}
\end{gathered}
+
\begin{gathered}
\begin{tikzpicture}
 \node at (0,0) {
\resizebox{20mm}{!}{
\fmfframe(0,0)(0,0){
\begin{fmfgraph*}(80,50)
\fmfleft{L1,L2}
\fmfright{R1,R2}
\fmf{phantom}{L1,v1,w1,R1}
\fmf{phantom}{L2,v2,w2,R2}
\fmffreeze
\fmf{fermion}{L1,R1}
\fmf{fermion}{R2,w2,v2,L2}
\fmf{gluon,left=2}{w2,v2}
\end{fmfgraph*}
}}};
\draw[dashed,darkred,thick] (-0.8,0.8) -- (-0.8,-0.8);
\draw[dotted,darkgreen, line width = 1.5] (-0.5,0.8) -- (-0.5,-0.8);
\draw[dotted,darkgreen,line width = 1.5] (0,0.8) -- (0,-0.8);
\draw[dotted,darkgreen,line width = 1.5] (0.5,0.8) -- (0.5,-0.8);
\draw[dashed,darkred,thick] (0.8,0.8) -- (0.8,-0.8);
\node[above,darkred] at(-0.8,0.8) {${}^{t=0}$};
\node[above,darkred] at(0.8,0.8) {${}^{t=0}$};
\node[below,white] at(-0.8,-0.8) {${}^{t=0}$};
\node[below,white] at(0.8,-0.8) {${}^{t=0}$};
\end{tikzpicture}
\end{gathered}
\label{Hascuts}
 \ee
 \vspace{-20pt}

 \noindent In these graphs each dotted green line represents a separate contribution where the measurement function at $t=\infty$ is inserted.
 The first two graphs have only soft contributions and the second two soft and collinear contributions (although the soft ones vanish in Feynman gauge).
  
 The middle cuts in these graphs using soft interactions in the asymptotic regions corresponds to soft real-emission processes. The amplitude for soft emission using $\Has$ is the eikonal limit of Eq.~\eqref{emit}, with an opposite sign and without the hard matrix element,
 \be
\cM_\text{soft} = g_s T^a_{ij} \left[ \frac{p_q^\alpha}{p_q \cdot p_g} -  \frac{p_{\bar q}^\alpha}{p_{\bar q} \cdot p_g}  \right]
 \epsilon_\alpha^\star(p_g)
 \ee
 Then the contribution to the differential thrust cross section at order $\alpha_s$ from these four cuts is
  \be
\left[ \frac{d \sigma}{d \tau} \right]_\text{soft,R} = \sigma_0
 \int \frac{d^3 p_g}{(2\pi)^3} \frac{1}{2 \omega_g} \left| \cM_\text{soft} \right|^2
\left[
 \delta\left(\tau -\frac{p_{\bar{q} }\cdot p_g }{Q^2}\right) \theta(\vec{p}_g \cdot \vec{p}_{\bar q} )
+
 \delta\left(\tau -\frac{p_q \cdot p_g}{Q^2}\right) \theta(\vec{p}_g \cdot \vec{p}_q)
\right]
 \ee
 In this expression, the $\theta$-functions project onto the appropriate hemisphere defined by the thrust axis (which aligns with the $\bar{q} -q$ axis at leading power).
 The first and third cuts in all the graphs, using soft interactions, give the virtual contributions. Summing all of them, the result is the same as the contribution to thrust from the thrust soft function~\cite{Schwartz:2007ib,Becher:2008cf}:
 \be
\frac{1}{\sigma_0} \left[ \frac{d \sigma}{d \tau} \right]_\text{soft,R} +
\frac{1}{\sigma_0}\left[ \frac{d \sigma}{d \tau} \right]_\text{soft,V} =
 \delta(\tau)\left[1+C_{F} \frac{\alpha_{s}}{4 \pi}\left(\frac{\pi^{2}}{3}\right)\right]-16 C_{F} \frac{\alpha_{s}}{4 \pi}\left[\frac{\ln \frac{\tau Q}{\mu}}{\tau}\right]_{+}+\mathcal{O}\left(\alpha_{s}^{2}\right)
 \ee
 Although the real and virtual contributions are separately infrared divergent, the final contribution to the cross section is not.

Similarly, the contribution from collinear graphs gives the jet functions. The net contribution is
  \be
\frac{1}{\sigma_0} \left[ \frac{d \sigma}{d \tau} \right]_\text{coll}  =
 \delta(\tau)+C_{F} \frac{\alpha_{s}}{4 \pi}\left\{\delta(\tau)\left(7-\pi^{2}\right)+\left[\frac{-3+4 \ln \frac{\tau Q^2}{\mu^{2}}}{\tau}\right]_{+}\right\}+\mathcal{O}\left(\alpha_{s}^{2}\right)
 \ee
Multiplying these by the $S_H$-matrix element squared, the sum is
\be
\frac{1}{\sigma_0} \left[ \frac{d \sigma}{d \tau} \right]_{\text{soft+coll}} =
\delta(\tau)+C_{F} \frac{\alpha_{s}}{2 \pi}\left\{\delta(\tau)\left(\frac{\pi^{2}}{3}-1\right)-3\left[\frac{1}{\tau}\right]_{+}-4\left[\frac{\ln \tau}{\tau}\right]_{+}\right\}
\ee
This agrees with the exact NLO thrust distribution at leading power (see~\cite{Schwartz:2014wha}). Note that the $\mu$ dependence of $S_H$-matrix elements exactly cancels against the $\mu$ dependence of the soft and collinear contributions in the asymptotic region.

For values of $\tau$ that are not small, one should necessarily treat the gluon as hard. The measurement function in this region is therefore only sensitive to hard particles. Since there are no asymptotic interactions between hard gluons and hard quarks, the $S_H$-matrix element in this regime is the same as in Eq.~\eqref{emit}. Integrating the square of this matrix element against the thrust measurement function gives for $\tau>0$,
\be
\frac{1}{\sigma_0} \left[\frac{d \sigma}{d\tau}\right]_\text{3-jet} =C_F\frac{\alpha_s}{2\pi}\left\{
3(1+\tau)(3 \tau-1)
+\frac{[4+6 \tau(\tau-1)] \ln (1-2 \tau)}{\tau(1-\tau)} -\frac{\left[4+6 \tau(\tau-1)\right] \ln \tau}{\tau(1-\tau)}
\right\}
\ee
Near $\tau=0$ this contribution coming from $\widehat  \cM(\gamma^\star \to \bar q  q g)$ is singular, and the phase space integral is IR divergent. However, at $\tau=0$ there is also the contribution from $\widehat  \cM(\gamma^\star \to \bar q  q)$. Although we can define the measurement function so that it is not sensitive to any gluon that couples in $\Has$, we cannot remove the soft and collinear gluons from $\Has$. These gluons still contribute to the cross section through loops, and affect the thrust distribution at
$\tau=0$. The virtual graphs are the first and third cuts in all the diagrams in Eq.~\eqref{Hascuts}. These graphs are IR divergent. If we work in $4-2\varepsilon$ dimensions,  the full phase space integral over the 3-jet contribution $|\widehat  \cM(\gamma^\star \to \bar q  q g)|^2$ generates $\frac{1}{\eir^2}$ and $\frac{1}{\eir}$ poles that exactly cancel the $\frac{1}{\eir^2}$ and $\frac{1}{\eir}$ from the virtual graphs. The result is that
\begin{multline}
\frac{1}{\sigma_0} \left[\frac{d \sigma}{d\tau}\right]_\text{3-jet}
+
\frac{1}{\sigma_0} \left[\frac{d \sigma}{d\tau}\right]_\text{2-jet}
=\delta(\tau)+C_{F} \frac{\alpha_{s}}{2 \pi}\left\{\delta(\tau)\left(\frac{\pi^{2}}{3}-1\right)\right.\\
\left.
+\left[3(1+\tau)(3 \tau-1)+\frac{[4+6 \tau(\tau-1)] \ln (1-2 \tau)}{1-\tau}\right]\left[\frac{1}{\tau}\right]_{+}-\frac{4+6 \tau(\tau-1)}{1-\tau}\left[\frac{\ln \tau}{\tau}\right]_{+}\right\}
\end{multline}
which is the exact NLO thrust distribution in QCD.

So we see that $S_H$ is capable of both reproducing distributions in fixed order QCD and, through the asymptotic expansion, reproducing just the leading-power parts of those distributions.
An advantage of leading-power approach is that one is not forced to compute the $S_H$-matrix elements and the asymptotic evolution to the same order in $\alpha_s$. Instead, one can use exponentiation properties of the soft and collinear emission to evaluate the asymptotic evolution to all orders in perturbation theory. In particular, one can perform resummation with the renormalization group, since the soft and collinear contributions are each associated with only a single scale.  Doing so in this example reproduces the resummed thrust distribution computed using SCET~\cite{Schwartz:2007ib,Becher:2008cf}.

\section{$\cN=4$ Super Yang-Mills\label{sec:Nis4}}

To further illustrate the features of $S_H$, we now consider amplitudes in $\cN=4$ super Yang-Mills  (SYM) theory. $\cN = 4$ SYM is a superconformal $SU(N_c)$ gauge theory in which scattering amplitudes have been studied quite extensively. To leading order in $\frac{1}{N_c}$, the only Feynman diagrams that contribute have planar topology and each loop order gives an additional factor of the 't Hooft coupling  $\lambda = g_s^2 N_c$. Since only one color structure is relevant at large $N_c$, is convenient to factor out the group theory factors. In addition, the amplitude for $n$-gluon scattering is totally symmetric in the permutation
of the external legs. With these observations, it is conventional to write the $L$-loop amplitude with $n$ external legs as
\be
\mathcal{A}_{n}^{(L)}=g_s^{n-2}\left[\left(4 \pi e^{-\gamma}\right)^{\epsilon} \frac{g_s^{2} N_{c}}{8 \pi^{2}}\right]^{L} \sum_\rho \operatorname{Tr}\left(T^{a_{\rho(1)}} \ldots T^{a_{\rho(n)}}\right)A_{n}^{(L)}(\rho(1), \rho(2), \ldots, \rho(n))
\ee
where the sum is over non-cyclic permutations $\rho$ of the external legs. The arguments $\rho(1)$, etc., refer to the permutation of the momenta and helicities of the legs. It is furthermore convenient to scale out the kinematic dependence of the tree-level amplitude by
defining
\be
M_{n}^{(L)}(\epsilon) \equiv \frac{ A_{n}^{(L)}(\epsilon) }{A_{n}^{(0)}(\epsilon) }\label{MArel}
\ee
In addition, we will find it useful to discuss the terms of each order in $\epsilon$ separately, so we write
\be
M_n^{(L)}(\epsilon) = \sum \epsilon^r M_n^{(L)}(\epsilon^r)
\ee
and decompose other quantities analogously.

In general, the bare $n$-leg $L$-loop amplitude is an extremely complicated function of the external momenta, even for planar maximal-helicity violating (MHV) amplitudes. What is interesting though is that there seems to be structure in the $L$-loop amplitude after the 1-loop amplitude is subtracted. More precisely, the ABDK/BDS ansatz proposes that the $L$-loop amplitude should be expressible
in terms of the 1-loop amplitude and some transcendental constants~\cite{Bern:2005iz,Anastasiou:2003kj}. More precisely, the full matrix element with $n$ legs
has the form
\be
\cM_n^{\text{BDS}} = \exp \left[ \sum_L \left(\left(4 \pi e^{-\gamma}\right)^{\epsilon}\frac{g_s^2 N_c}{8\pi^2}\right)^{L}
\Big(f^{(L)}(\epsilon) M_n^{(1)}(L \epsilon) + C^{(L)} + E_n^{(L)}(\epsilon)\Big)\right]
\label{BDS}
\ee
where $f^{(L)}(\epsilon)$ is independent of $n$ and related to the cusp anomalous dimension (explicitly $f^{(1)}(\epsilon) = 1$ and
$f^{(2)}(\epsilon) = -\zeta_2 - \zeta_3 \epsilon - \zeta_4 \epsilon^2+\cdots$. The numbers $C^{(L)}$ are also independent of $n$
and represent the part of the $L$-loop amplitude not given by the exponentiation of the first term. By explicit
computation it is known that $C^{(1)} = 0$ and $C^{(2)} = -\frac{1}{2} \zeta_2^2$. Finally,  $E_n^{(L)}(\epsilon)$ has only
positive powers of $\epsilon$, so that $E_n^{(L)}(0) = 0$.

It turns out the BDS ansatz was not quite correct: there is more structure to the amplitudes than just the numbers $C^{(L)}$ for $n>5$. Thus,
it is common to express amplitudes as ratios of the bare amplitudes and the BDS ansatz. More precisely, the remainder function is defined as
\be
R_n = \ln\left[ \frac{\cM_n}{\cM_n^{\text{BDS}}}\right] \label{BDSR}
\ee
and one can expand $R_n$ order-by-order in $g_s$.

While the remainder functions have some nice properties, such as respecting dual conformal invariance, they violate other conditions, such as the Steinmann relations~\cite{Caron-Huot:2016owq}. To preserve the Steinmann relations, the BDS ansatz is modified to the ``BDS-like'' ansatz~\cite{Alday:2009dv}. For certain amplitudes ($n=8$ for example), it has been shown that both the Steinmann relations and dual conformal invariance cannot be satisfied simultaneously~\cite{Golden:2018gtk}. That the BDS ansatz violates the Steinmann relations is due to the additional subtraction of finite, $\cO(\epsilon^0)$, terms in Eq.~\eqref{BDS} in addition to the IR divergences. A more conservative ansatz is the ``minimally-normalized'' amplitude $\cM_{n}^{\text{min}}$ defined as~\cite{Golden:2019kks}
\be
\cM_n^{\text{min}} = \exp \left[ \sum_L \left(\left(4 \pi e^{-\gamma}\right)^{\epsilon}\frac{g_s^2 N_c}{8\pi^2}\right)^{L}
\Big(f^{(L)}(\epsilon) M_n^{(1,\text{div})}(L \epsilon) + C^{(L)} \Big)\right]
\ee
where the IR divergences of  $M_n^{(1)}$ are
\be
M_n^{(1,\text{div})}(\epsilon) = -\frac{1}{2\epsilon^2}  \sum_{i=1}^n \left( \frac{\mu^2}{-s_{i,i+1}} \right)^\epsilon
\ee
The ratio $\frac{\cM_n}{\cM_n^{\text{min}}}$ of the full amplitude to the minimally normalized amplitude is IR finite, just like the BDS remainder function in Eq.~\eqref{BDSR},
but the finite parts of
$\frac{\cM_n}{\cM_n^{\text{min}}}$ and
$\frac{\cM_n}{\cM_n^{\text{BDS}}}$ are different.

In this section we relate some of these observations to the hard $S$-matrix element. We will see that the hard $S$-matrix element  computed in $\msbar$ corresponds closely to the minimally normalized amplitude.

\subsection{4-point amplitude}
We begin by discussing the MHV amplitude with 4 external legs.
The IR divergences of the 1-loop amplitude for $n=4$ are known to agree with the divergences of
\be
C_4^{(1)}(\epsilon) = -\frac{e^{\gamma \epsilon}}{\Gamma(1-\epsilon)} \frac{1}{\epsilon^2} \left[\left(\frac{\mu^2}{-s}\right)^\epsilon+ \left(\frac{\mu^2}{-t}\right)^\epsilon\right]
\ee
and the divergences of the 2-loop amplitude agree with the divergences of
\be
C_{4}^{(2)}(\epsilon)=\frac{1}{2}\left(C_{4}^{(1)}(\epsilon)\right)^{2}+C_{4}^{(1)}(\epsilon)
\left(M_{4}^{(1)}(\epsilon)-C_{4}^{(1)}(\epsilon)\right)
-\left(\zeta_{2}+\epsilon \zeta_{3}\right) \frac{e^{-\epsilon \gamma} \Gamma(1-2 \epsilon)}{\Gamma(1-\epsilon)} C_{4}^{(1)}(2 \epsilon)
\label{C42}
\ee
These formulas are due to Catani~\cite{catani1998singular} (see also~\cite{Sterman:2002qn}). Note that $C_{4}^{(2)}(\epsilon)$ depends on the complete 1-loop amplitude $M_{4}^{(1)}(\epsilon)$. Thus, although the quantity
$M_{4}^{(2)}(\epsilon) - C_{4}^{(2)}(\epsilon)$ is IR-finite, more terms are being subtracted this way than those determined by the universality of IR-divergences. These extra terms depend on quantities such as
$M_4^{(1)}(\epsilon^2)$ which are not fixed by factorization alone. Although factorization does not determine $M_4^{(1)}(\epsilon)$, its appearance in the universal formula can be understood from the point of view of effective field theory~\cite{Becher:2009cu}: it comes from a cross term between the non-universal 1-loop Wilson coefficient and the universal 1-loop divergences. An equivalent mechanism explains its appearance during the computation of $S_H$, as we now show.

With 4 legs ($n=4$), the 1-loop amplitude is
\be
{M}_{4}^{1}(\epsilon)=
-\frac{2}{\epsilon^{2}}  + \frac{1}{\epsilon} M_4^{(1)}(\epsilon^{-1}) + M_4^{(1)}(\epsilon^0)+ \cO(\epsilon) \label{M41}
\ee
where
\begin{align}
 M_4^{(1)}(\epsilon^{-2})  &= -2 \\
 M_4^{(1)}(\epsilon^{-1})  &= -\ln \frac{\mu^{2}}{-s}-\ln \frac{\mu^{2}}{-t} \\
M_4^{(1)}(\epsilon^0) &=
- \ln \frac{\mu^2}{-t}\ln\frac{\mu^2}{-s}+\frac{2\pi^{2}}{3} \\
\label{M411}
M_4^{(1)}(\epsilon^1) &=
 - \frac{\pi^2}{2} \ln \frac{-s}{u} - \frac{1}{3}\ln^3 \frac{-s}{u} + \frac{\pi^2}{12} \ln \frac{\mu^2}{-s} - \frac{1}{6} \ln^3 \frac{\mu^2}{-s}
+ \frac{\pi^2}{4} \ln \frac{\mu^2}{u}
\\ &
+ \frac{1}{2} \ln^2 \frac{-s}{u} \ln \frac{\mu^2}{u} - \frac{1}{2} \ln \frac{-s}{u} \ln \frac{-t}{u} \ln \frac{\mu^2}{u} - \ln \frac{-s}{u} \Li_2 \frac{-s}{u} + \Li_3 \frac{-s}{u} + \frac{7}{3} \zeta_3 + (s \leftrightarrow t)
 \nonumber \\
M_4^{(1)}(\epsilon^2) &=
\label{M412}
\frac{5 \pi^2}{24} \ln^2 \frac{-s}{u} + \frac{1}{8} \ln^4 \frac{-s}{u} + \frac{3}{8} \ln \frac{-s}{u} \ln \frac{-t}{u} + \frac{1}{6} \ln^3 \frac{-s}{u} \ln \frac{-t}{u}
\\ &
- \frac{1}{4} \ln^2 \frac{-s}{u} \ln^2 \frac{-t}{u} + \frac{\pi^2}{24} \ln^2 \frac{\mu^2}{-s} - \frac{1}{24} \ln^4 \frac{\mu^2}{s} - \frac{\pi^2}{2} \ln\frac{-s}{u} \ln\frac{\mu^2}{u} - \frac{1}{3} \ln^3 \frac{-s}{u} \ln \frac{\mu^2}{u} \nonumber
\\ &
+ \frac{\pi^2}{8}  \ln^2 \frac{\mu^2}{u} + \frac{1}{4} \ln^2 \frac{-s}{u} \ln^2\frac{\mu^2}{u}-\frac{1}{4} \ln \frac{-s}{u} \ln \frac{-t}{u} \ln^2 \frac{\mu^2}{u} + \frac{7}{3} \zeta_3 \ln^2 \frac{\mu^2}{-s}
+ \frac{1}{2}\ln^2 \frac{-s}{u} \Li_2 \frac{-s}{u} \nonumber
\\ &
-\ln \frac{-s}{u} \ln \frac{\mu^2}{u} \Li_2 \frac{-s}{u}
+\ln\frac{\mu^2}{u} \Li_3 \frac{-s}{u} - \ln \frac{-s}{u} \Li_3 \frac{-t}{u} - \Li_4 \frac{-s}{u} + \frac{49 \pi^4}{720}
+ (s \leftrightarrow t)
   \nonumber
\end{align}
In these expressions, $s=(p_1+p_2)^2$, $t=(p_1+p_3)^2$, $u=-t-s$ and the convention is that incoming momenta are treated as outgoing with negative energy.  Note that the $\epsilon$ are all $\eir$ since $\cN=4$ SYM is UV finite.

 At 2-loops, the amplitude can be written as
 \begin{multline}
 {M}_{4}^{(2)}=  \frac{2}{\epsilon^4} -\frac{2 }{\epsilon ^3} M_4^{(1)}(\epsilon^{-1})
   + \frac{1}{\epsilon^2} \left[\frac{\pi^2}{12} + \frac{1}{2} M_4^{(1)}(\epsilon^{-1})^2 - 2 M_4^{(1)}(\epsilon^0) \right]\\
   +\frac{1}{\epsilon} \left[ -\frac{\pi^2}{12} M_4^{(1)}(\epsilon^{-1}) + M_4^{(1)}(\epsilon^{-1}) M_4^{(1)}(\epsilon^{0})
   - 2 M_4^{(1)}(\epsilon^{1}) + \frac{\zeta_3}{2} \right] + M_4^{(2)}(\epsilon^{0}) + \cO(\epsilon)
\end{multline}
where
\be
 {M}_{4}^{(2)}(\epsilon^0) = \frac{1}{2} \left[M_4^{(1)}(\epsilon^{0}) \right]^2
 -\frac{\pi^2}{6} M_4^{(1)}(\epsilon^{0}) - \frac{\pi^4}{120} + M_4^{(1)}(\epsilon^{-1})\left[M_4^{(1)}(\epsilon^{1}) - \frac{\zeta_3}{2} \right]
 + M_4^{(1)}(\epsilon^{-2}) M_4^{(1)}(\epsilon^{2})
  \label{M42e0}
\ee
Although there is some hint of exponentiation in this expression, it is not particularly simple. That is, if one defines an IR finite
2-loop amplitude by dropping all the singular terms in $\epsilon$ and  then taking  $\epsilon\to0$ the result,  ${M}_{4}^{(2)}(\epsilon^0)$, is complicated, with all the polylogarithms from Eqs.~\eqref{M411} and \eqref{M412}.

The appearance of the  $\cO(\epsilon^1)$ and $\cO(\epsilon^2)$ terms from $\cM_{4}^{(1)}$ in the 2-loop amplitude hints
at a relationship between them. Indeed, the BDS/ABDK ansatz notes that if we subtract $C_{4}^{(2)}(\epsilon^0)$ in Eq.~\eqref{C42} from $M_{4}^{(2)}(\epsilon^0)$
the result is relatively simple
\be
M_{4}^{(2)}(\epsilon^0)-C_{4}^{(2)}(\epsilon^0)
=\frac{1}{2}\left(M_{4}^{(1)}(\epsilon^0)-C_{4}^{(1)}(\epsilon^0)\right)^{2}
-\zeta_2  \left(M_{4}^{(1)}(\epsilon^0)-C_{4}^{(1)}(\epsilon^0)\right) -  \frac{21}{8} \zeta_4
\ee
Recall that $C_4^{(2)}(\epsilon)$ is not fixed by the IR structure alone, but includes additional terms.
Although this relation works well for the 4-point amplitude, it is somewhat ad hoc and requires modification for $n>5$ legs and higher loops.

Now let us consider the hard $S$-matrix elements. We define them analogously to $S$-matrix elements, adding a hat. So $\widehat{A}_n^{(L)}$ is the color-stripped hard-$S$-matrix element for $n$ legs at $L$ loops. This amplitude is IR finite, but UV divergent. Denoting $\widehat M_{n}^{(L)}(\epsilon) \equiv \widehat A_{n}^{(L)}(\epsilon) / \widehat A_{n}^{(0)}(\epsilon)$ in analogy to Eq. \eqref{MArel}, the 1-loop bare hard matrix element is the same as Eq.~\eqref{M41} with $\eir$ replaced by $\euv$. The renormalized matrix element is then
\be
{\widehat M}_{4}^{(1)} = \left[\frac{1}{Z_4} ({\widehat \cM}_{4})_{\text{bare}}\right]_{\overset{\text{color-stripped}}{\text{1-loop}}}=
- \ln \frac{\mu^2}{-t}\ln\frac{\mu^2}{-s}+\frac{2\pi^{2}}{3}
+\cO(\epsilon)
\ee
where, with minimal subtraction ($\msbar$)
\be
Z_4^{\msbar}  =1 + \left(4 \pi e^{-\gamma}\right)^{\epsilon}\frac{g_s^2 N_c}{8\pi^2}\left[- \frac{2}{\epsilon^2} +  \frac{1}{\epsilon}\left( -\ln \frac{\mu^{2}}{-s}-\ln \frac{\mu^{2}}{-t} \right)
\right] + \cO(g_s^4N_c^2) \quad (\msbar) \label{Z4}
\ee
Note that ${\widehat M}_{4}^{(1)}$ is finite as $\epsilon \to 0$, since the IR divergences are absent in hard $S$-matrix elements and the UV divergences are removed through renormalization. There are nevertheless terms of $\cO(\epsilon)$ and $\cO(\epsilon^2)$ in the matrix elements in $d$ dimensions. These terms are the same as the $\cO(\epsilon)$ and $\cO(\epsilon^2)$ terms in $M_4^{(1)}$.
Then the 2-loop hard $S$-matrix element gets a contribution from both the 2-loop graphs, giving $M_4^{(2)}(\epsilon^0)$ after renormalization, as well as a contribution from the cross terms between the $\frac{1}{\epsilon^2}$ and $\frac{1}{\eps}$ terms in
$Z_4$ and the $\cO(\epsilon)$ and $\cO(\epsilon^2)$ terms in $\widehat M_4^{(1)}$. The result is that
\be
{\widehat M}_{4}^{(2)}  = \frac{1}{2} \left[\widehat M_4^{(1)} - \frac{\pi^2}{6} \right]^2  - \frac{\pi^4}{45} +\frac{\zeta_3}{2}
 \left(\ln \frac{\mu^{2}}{-s}+\ln \frac{\mu^{2}}{-t}\right) \qquad (\msbar)
 \label{msamp}
\ee
This matrix element is significantly simpler than $ {M}_{4}^{(2)}(\epsilon^0)$ in Eq.~\eqref{M42e0}, and does not require any ad-hoc subtractions.

\subsection{Scheme choice}
Dimensional regularization and minimal subtraction is the most widespread scheme in use in SCET. We must keep in mind, however, that due to renormalization
there is scheme dependence in $S_H$. This is not a problem {\it per se}, since $S_H$ itself is not directly observable. One expects that one $S_H$-matrix elements
are combined into observables the scheme dependence will cancel. Indeed, the cancellations that occur will be similar to the cancellations that occur in SCET. For example, Ref.~\cite{broggio2016scet} showed that physical observables agree when conventional dimensional regularization, four-dimensional helicity scheme, or dimensional reduction are used, despite the fact that the hard, jet and soft functions are different in the different schemes.  In a normal, local field theory, the counterterms are strongly constrained: they must just be numbers. In SCET the counterterms can depend on the labels for the various collinear directions which translates to dependence of hard kinematical quantities, like $s$ and $t$, as in Eq.~\eqref{Z4}. However, one cannot choose an arbitrary function of labels, as the dependence must be canceled by contributions from soft and jet functions. Roughly speaking the combination, $H \otimes J \otimes S$ must be scheme independent, where the hard function $H$ corresponds to the square of our hard $S$-matrix elements.  More discussion of these constraints can be found in~\cite{broggio2016scet}.

Let us suppose that adding a finite part to the counterterm is not problematic. More precisely, suppose we can add a finite part
$\delta_4(\epsilon)$ to the $Z_4$ renormalization constant. Then the color-stripped hard $S$-matrix element at 1-loop shifts from the $\msbar$ version by
$\delta_4^{(1)}(\epsilon)$:
\be
\widehat M_4^{\delta,(1)}  = \widehat M_4^{(1)}  - \delta_4^{(1)}(\epsilon)
\ee
At 2-loops, the shift picks up a cross term between $\delta_4^{(1)}$ and the divergent parts of the bare amplitude
$({\widehat \cM}_{4})_{\text{bare}}$:
\be
{\widehat M}_{4}^{\delta,(2)}=\widehat M_4^{(2)}
- \sum_{j=0}^{2} \widehat M_4^{\text{bare},(1)}(\epsilon^{-j}) \delta_4^{(1)}(\epsilon^{j})- \delta_4^{(2)}
\ee
so that
\begin{multline}
{\widehat M}_{4}^{\delta,(2)}  =
 \frac{1}{2} \left[\widehat M_4^{\delta,(1)}-\frac{\pi^2}{6} \right]^2
-\frac{\pi^4}{45}
-\delta_4^{(1)}(\epsilon^1) \widehat M_4^{\text{bare},(1)}(\epsilon^{-1})
-\delta_4^{(1)}(\epsilon^2) \widehat M_4^{\text{bare},(1)}(\epsilon^{-2})
\\
-\frac{\pi^2}{6} \delta_4^{(1)}(\epsilon^0)-\frac{1}{2} [\delta_4^{(1)}(\epsilon^0)]^2
+\frac{\zeta_3}{2}  \left(\ln \frac{\mu^{2}}{-s}+\ln \frac{\mu^{2}}{-t}\right)  - \delta^{(2)}_4
 \end{multline}
This motivates choosing a ``BDS'' subtraction scheme, where
\be
 \delta_4^{(1)} = - \frac{\pi^2}{6}- \frac{\zeta_3}{2}\epsilon,\qquad
 \delta_4^{(2)} = -\frac{\pi^4}{120} + \cO(\epsilon)
\ee
or equivalently
\be
Z_4^{\text{BDS}}  =1 +  \left(4 \pi e^{-\gamma}\right)^{\epsilon}\frac{g^2 N_c}{8\pi^2}\left[- \frac{2}{\epsilon^2} +  \frac{1}{\epsilon}\left( -\ln \frac{\mu^{2}}{-s}-\ln \frac{\mu^{2}}{-t} \right) - \frac{\pi^2}{6}
- \frac{\zeta_3}{2} \epsilon
\right]  + \cO(g^4) \quad (\text{BDS scheme} )
\ee
Then we get simply
 \be
{\widehat M}_{4}^{\text{BDS},(2)}  = \frac{1}{2} \left[\widehat M_4^{\text{BDS},(1)} - \frac{\pi^2}{6} \right]^2  \quad (\text{BDS scheme} )
\label{bdsscheme}
 \ee

There are two things to note about this result. First, it is nontrivial that one can pick pure numbers for $\delta_4$ to cancel the explicit $s$ and $t$ dependence in Eq.~\eqref{msamp}. This was possible only
 because the $\ln \frac{\mu^2}{-s} + \ln\frac{\mu^2}{-t}$ factor in Eq.~\eqref{msamp} is the same as in $\widehat M^{(1)}(\epsilon^{-1})$. Second it is impossible to choose $\delta_4$ to remove the $\frac{\pi^2}{6}$ in Eq.~\eqref{bdsscheme}. Thus there is a sense in which the constant term $\frac{\pi^2}{6} = \zeta_2$ of the second order amplitude is scheme independent. This term gives the constant $C_2 = \frac{1}{2} \zeta_2^2$ from Eq.~\eqref{BDS}.
 
 The BDS ansatz implies that to all orders, the  4-gluon planar amplitude exponentiates in the BDS subtraction scheme. In the language
 of the hard $S$-matrix, this means that the finite parts of the counterterms will be pure numbers to all orders.
 Indeed, for dual-conformal invariance to be respected by the 4-point amplitude, we should not be adding extra dependence on $s$ and $t$ into the counterterms. Equivalently, we can say that the dual-conformal anomaly is manifest in the BDS subtraction scheme but somewhat obscure in $\msbar$.

 \subsection{6-point amplitude}
  The amplitude with 6 external particles is more interesting because it can depend on more kinematic invariants.
 The hard MHV $S$-matrix element with 6 legs in $\msbar$ is
%
 \begin{multline} \widehat M_{6}^{(1)}(\epsilon)
 =\sum_{\text{cycles}} \left[ -\frac{1}{2} \ln^2(-s_{12}) - \ln\frac{-s_{12}}{-s_{123}} \ln\frac{-s_{23}}{-s_{123}}
 +\frac{1}{4} \ln^2 \frac{-s_{123}}{-s_{234}}\right] \qquad (\msbar)
 \\
 -\mathrm{Li}_{2}\left(1-u\right)
 -\mathrm{Li}_{2}\left(1-v\right)
 -\mathrm{Li}_{2}\left(1-w\right)+6 \zeta_{2} + \cO(\epsilon)
 \label{M6hard}
 \end{multline}
 where the 3 dual-conformal cross ratios are
 \be
 u = \frac{ s_{12} s_{45} }{ s_{123} s_{345} },\quad
v =  \frac{ s_{23} s_{56} }{ s_{234} s_{123} },\quad
w = \frac{ s_{34} s_{61} }{ s_{345} s_{234} }
 \ee
The notation here is that $s_{123} = (p_1+p_2+p_3)^2$ and sum over cycles means sum over the 6 rotations of the labels, e.g. $s_{123} \to s_{234}$ and so on. This amplitude is simply the bare 1-loop MHV amplitude ~\cite{Bern:1994zx,Dixon:2015iva} with IR divergences converted to UV divergences by the diagrams involving $\Has$ and then removed by counterterms:
\be
 Z_6  =1 + \left(4 \pi e^{-\gamma}\right)^{\epsilon}\frac{g^2 N_c}{8\pi^2}\left[- \frac{2}{\epsilon^2} -  \frac{1}{\epsilon}
 \sum_{\text{cycles}}  \left(
 \ln \frac{\mu^{2}}{-s_{12}}  \right)
\right]  + \cO(g_s^4) \qquad (\msbar)
\label{Z6ms}
\ee
The ``BDS-like'' ansatz adds to this amplitude the terms on the second line plus another cyclic sum
 \be
 Y_6 =  \mathrm{Li}_{2}\left(1-u\right) +\mathrm{Li}_{2}\left(1-v\right) + \mathrm{Li}_{2}\left(1-w\right)
 +\frac{1}{2} \left( \ln^2 u + \ln^2 v + \ln^2 w\right)
 \ee
If we are free to shift the counterterm, $Z_6^{\text{BDS-like}}  = Z_6 - Y_6$, then the matrix element has a somewhat
 simpler form
 \be
\widehat M_{6}^{(1)}(\epsilon)
 =\sum_{\text{cycles}} \left[
 - \ln (-s_{12}) \ln (-s_{23} ) + \frac{1}{2} \ln (-s_{12}) \ln(-s_{45})
 \right] + 6 \zeta_2
 \qquad (\text{BDS-like scheme})
 \ee
 In particular, it is a function of only 2-particle invariants. This means that when the amplitude is exponentiated, it cannot
 violate the Steinmann relations (these require 3 particle invariants)~\cite{steinmann1960zusammenhang,Caron-Huot:2019vjl}.
 
 Note however, that we do not know how to specify this BDS-like subtraction scheme at higher order. More importantly, we do not know if it is consistent. As mentioned above, (see~\cite{broggio2016scet}) there are constraints on the scheme from self-consistency of SCET.  Since we also do not know general constraints on the finite parts of the counterterms, it is safest to restrict to conventional dimensional regularization with minimal subtraction, where SCET at least is believed to be consistent.  In $\msbar$, the
counterterm is in Eq.~\eqref{Z6ms} and the hard matrix element is in Eq.~\eqref{M6hard}. In $\msbar$, the hard matrix elements agree with the minimally-normalized amplitudes discussed in~\cite{Golden:2019kks} up to at least 2-loops and preserve the Steinmann relations.

\label{sec:N4}

\section{Summary and Outlook}
\label{sec:summary}
The traditional $S$-matrix is only well-defined if time-evolution of a theory is well-approximated by
free evolution at early or late times. Indeed, the free Hamiltonian $H_0$ is part of the definition of
$S$ used for perturbative calculations.
When a theory has massless particles, the interactions do not
die off fast enough at asymptotic times, resulting in a poorly defined, divergent $S$-matrix.
We argue that a sensible, finite $S$-matrix is obtained by replacing $H_0$ in its definition with an asymptotic Hamiltonian $\Has$ that correctly accounts for all the asymptotic interactions. Our key principle for choosing $\Has$ is that the states should evolve before and after they scatter independently of how they scatter. That
such an $\Has$ exists and makes the $S$-matrix finite is guaranteed by theorems of hard-collinear-soft factorization. Capitalizing on these theorems, we define $\Has$
as the leading power expansion of the full Hamiltonian in soft and collinear limits,
and call the corresponding $S$-matrix the hard $S$-matrix, $S_H$. $S_H$ is finite order-by-order in perturbation theory, as we have verified through a number of explicit examples in QED, QCD and $\cN=4$ super Yang-Mills theory.

While the traditional $S$-matrix is IR divergent, it can still be used to compute IR-finite observables. This is done by
summing over a broad enough set of processes so that the sum is finite even though individual contributions are divergent. With $S_H$, the same physical predictions result using the matrix elements of a scattering operator that are finite process-by-process.

 We presented a method and Feynman rules for the perturbative calculation of $S_H$-matrix elements.
The method involves
separating $S_H$ into three parts: An asymptotic part evolving the state from $t=0$ to $t=-\infty$, the evolution from $t=-\infty$ to $t=\infty$ and an asymptotic part evolving from $t=\infty$ to $t=0$. Each asymptotic part is calculated using Feynman rules similar to those in time-ordered perturbation theory but without overall energy conservation, and the middle part consists of conventional Feynman diagrams. The three part picture is presented for calculational convenience, since it breaks up calculations into essentially usual time-ordered perturbation theory and Feynman diagrams, and bypasses the need to derive a new interaction picture with modified propagators.

The hard $S$-matrix has numerous advantages over the traditional $S$-matrix. The first advantage is the obvious one: $S_H$ exists.
Second, matrix elements of $S_H$ have a rich structure with diverse interpretations.
One can interpret the asymptotic evolution as dressing the states, so that a  initial Fock state with a finite number of particles evolves into a dressed
state with an infinite number of particles at asymptotic times. This connects
our construction to previous work on coherent states, such as by Chung~\cite{Chung:1965zza} or Faddeev and Kulish~\cite{Kulish:1970ut}.
Alternatively, $S_H$-matrix elements can be interpreted as Wilson coefficients in Soft-Collinear Effective Theory. Finally, $S_H$-matrix elements are closely related to finite remainder functions studied in the amplitude community. Indeed, much of the progress in understanding scattering amplitudes over the last few decades has comprised results about an object, the $S$-matrix, that formally does not exist. Since there is so much interest in the $S$-matrix itself (as opposed to cross sections), it is logical to try to put this object on a firmer theoretical footing. Doing so was one of the main motivations of this paper.

There are a number of new ideas contained in this paper. These include:
\begin{itemize}
\item The first explicit calculation of a finite $S$-matrix in theories with massless particles. While other authors have introduced similar concepts in QED, there are no explicit calculations in the literature of actual matrix elements. The majority of papers focuses on just the IR divergence cancellation. Issues such as regulator dependence, renormalization, subtraction schemes, phase space integrals, computation of observables, completeness of the Hilbert space, etc., are all glossed over unless one is able to do explicit computations.

\item We present a new rationale for choosing the asymptotic Hamiltonian. While others have argued that the asymptotic Hamiltonian should make the $S$-matrix IR finite, we argue that such a criterion is not restrictive enough: one could choose $\Has=H$ to satisfy that requirement.  Instead we argue that one should use that the asymptotic evolution is independent of the hard scattering. That there exists an asymptotic Hamiltonian with this property in gauge theories is non-trivial and  follows from factorization theorems.

\item We connect the literature on coherent states to that of factorization and that of scattering amplitudes. In particular, the hard $S$-matrix elements can be identified as $S$-matrix elements of coherent states, as Wilson coefficients in SCET, and as finite remainder functions in $\cN = 4$ SYM fields corresponding to BDS-inspired subtraction schemes.

\item We provide an explicit set of Feynman rules to evaluate $S_H$ elements in perturbation theory. These rules involve distributions and products of distributions that must be handled with some care.

\item We provide a number of examples of $S_H$-matrix element calculations, both using pure dimensional regularization and with explicit cutoffs on $\Has$.

\item We examine how the Glauber/Coulomb phase arises in asymptotic-region diagrams. In particular, energy non-conservation in the asymptotic regions allows the Glauber contribution to be reproduced (and cancelled) without off-shell modes.

\item We demonstrate that infrared-safe observables computed with $S_H$ will agree with those computed using the normal $S$-matrix, and, to leading power, with those computed using SCET or other factorization frameworks.  We are not aware of any paper on dressed states that makes a physical prediction using them. In our framework, one can see how the dressing occurs, but also how the states get ``undressed'' in the final asymptotic evolution before the measurement is made.

\item Although predictions using $S_H$ reduce (almost trivially) to predictions using $S$, matrix elements of $S_H$ can be studied as interesting objects on their own. These matrix elements are scheme and scale-dependent, but still have physical interpretations, just like the $\msbar$ couplings $\alpha_s(\mu)$.
\end{itemize}

These last two bullets are perhaps worth some additional discussion.
The incontrovertible truth is that cross sections computed with $S$, despite coming from IR-divergent amplitudes, are in perfect agreement with observations.  Thus, no matter how one attempts to make scattering amplitudes finite, the framework must reproduce these cross sections exactly. In other words, it is foolhardy to try to make different predictions at the cross section level with a new $S$-matrix. That being said, there are situations, in particular those with charged initial states such as $e^+e^-  + \text{photons}\to Z + \text{photons}$, where it is not entirely clear what the physical cross section is supposed to be~\cite{Frye:2018xjj}. In such situations, a finite $S_H$ may provide some clarity.

Although we cannot expect $S_H$ to revolutionize the computation of physical cross sections, having a finite $S$-matrix is still enormously beneficial for the study of scattering amplitudes themselves. Indeed, the majority of research of scattering amplitudes focuses on $S$-matrix elements themselves, not on observables. So it is this community that might benefit first from $S_H$. As an example, we showed that certain $S_H$ elements in a supersymmetric theory naturally satisfy the Steinmann relations, at least to two loops. In contrast, $S$-matrix elements are IR divergent and, depending on how the IR divergences are subtracted, the Steinmann relations may or may not be satisfied. More broadly, because $S_H$ corresponds to the matrix elements of a single unitary operator, rather than a ratio of such matrix elements, it should automatically satisfy any constraints that follow from unitarity. One might also imagine that properties stemming from analyticity would be more transparent in matrix elements of a single operator rather than a ratio.

Finally, let us briefly discuss how to think about $S_H$ non-perturbatively. In this paper, we have advocated for computing
$S_H$ in dimensional regularization with $\msbar$ subtraction. At each order in perturbation theory, one can compute $S_H$ elements this way. It may seem counterintuitive, but perturbation theory has historically been the best way to orient investigation into non-perturbative physics, and a perturbative approach could be similarly successful for $S_H$.
One can also resum $S_H$ using renormalization group techniques to examine its all-orders behavior. Alternatively, one could (in principle) compute $S_H$ numerically with hard cutoffs, but to compare to the perturbative results in dimensional regularization, one would have to convert between the cutoff scheme and $\msbar$. Through various approaches like these, it should be possible to explore the analytic structure of $S_H$. It would be interesting to look at its properties in the Borel plane, for example, or whether a renormalon-free mass scheme naturally emerges. More generally, since $S_H$ is IR finite, it resembles more closely $S$ in a theory with a mass gap than the IR-divergent $S$. Thus one might hope that when massless particles are present, the $S$-matrix bootstrap program might make more progress with $S_H$ than it has on $S$.

\section*{Acknowledgements}
The authors would like to thank T. Becher, S. Caron-Huot, J. Collins, L. Dixon, D. Neuenfeld, A. McLeod and J. Thaler for helpful conversations. HH would also like to thank Mainz Institute for Theoretical Physics (MITP) and MS would like to thank the Aspen Center for Physics, where discussions related to this work were held.  This work was supported in part by the U.S. Department of Energy under contract DE-SC0013607.

\end{fmffile}

\bibliographystyle{utphys}
\bibliography{FiniteS_Long.bib}

\providecommand{\href}[2]{#2}\begingroup\raggedright\begin{thebibliography}{10}

\bibitem{Wheeler}
J.~A. Wheeler, ``{On the Mathematical Description of Light Nuclei by the Method
  of Resonating Group Structure},''
  \href{http://dx.doi.org/10.1103/PhysRev.52.1107}{{\em Phys. Rev.} {\bfseries
  52} (Dec, 1937) 1107--1122}.
  \url{https://link.aps.org/doi/10.1103/PhysRev.52.1107}.

\bibitem{Heisenberg}
W.~Heisenberg, ``{Die ``beobachtbaren Groessen'' in der Theorie der
  Elementarteilchen},'' {\em Zeits. f. Physik} {\bfseries 120} (1943) 513.

\bibitem{Feynman}
R.~P. Feynman, ``{Relativistic Cut-Off for Quantum Electrodynamics},''
  \href{http://dx.doi.org/10.1103/PhysRev.74.1430}{{\em Phys. Rev.} {\bfseries
  74} (Nov, 1948) 1430--1438}.
  \url{https://link.aps.org/doi/10.1103/PhysRev.74.1430}.

\bibitem{Dyson1}
F.~J. Dyson, ``{The S Matrix in Quantum Electrodynamics},''
  \href{http://dx.doi.org/10.1103/PhysRev.75.1736}{{\em Phys. Rev.} {\bfseries
  75} (Jun, 1949) 1736--1755}.
  \url{https://link.aps.org/doi/10.1103/PhysRev.75.1736}.

\bibitem{nelson1981origin}
C.~A. Nelson, ``{Origin of Cancellation of Infrared Divergences in Coherent
  State Approach: Forward Process $qq \to qq$ $+$ gluon},'' {\em Nuclear
  Physics B} {\bfseries 181} no.~1, (1981) 141--156.

\bibitem{Contopanagos:1991yb}
H.~F. Contopanagos and M.~B. Einhorn, ``{Theory of the Asymptotic S matrix for
  Massless Particles},''
\href{http://dx.doi.org/10.1103/PhysRevD.45.1291}{{\em Phys. Rev.} {\bfseries
  D45} (1992) 1291--1321}.

\bibitem{Bloch:1937pw}
F.~Bloch and A.~Nordsieck, ``{Note on the Radiation Field of the Electron},''
\href{http://dx.doi.org/10.1103/PhysRev.52.54}{{\em Phys. Rev.} {\bfseries 52}
  (1937) 54--59}.

\bibitem{Yennie:1961ad}
D.~R. Yennie, S.~C. Frautschi, and H.~Suura, ``{The Infrared Divergence
  Phenomena and High-Energy Processes},''
\href{http://dx.doi.org/10.1016/0003-4916(61)90151-8}{{\em Annals Phys.}
  {\bfseries 13} (1961) 379--452}.

\bibitem{Weinberg:1965nx}
S.~Weinberg, ``{Infrared Photons and Gravitons},''
\href{http://dx.doi.org/10.1103/PhysRev.140.B516}{{\em Phys. Rev.} {\bfseries
  140} (1965) B516--B524}.

\bibitem{Grammer:1973db}
G.~Grammer, Jr. and D.~R. Yennie, ``{Improved Treatment for the Infrared
  Divergence Problem in Quantum Electrodynamics},''
\href{http://dx.doi.org/10.1103/PhysRevD.8.4332}{{\em Phys. Rev.} {\bfseries
  D8} (1973) 4332--4344}.

\bibitem{Doria:1980ak}
R.~Doria, J.~Frenkel, and J.~C. Taylor, ``{Counter Example to Nonabelian
  Bloch-Nordsieck Theorem},''
\href{http://dx.doi.org/10.1016/0550-3213(80)90278-3}{{\em Nucl. Phys.}
  {\bfseries B168} (1980) 93--110}.

\bibitem{Kinoshita:1962ur}
T.~Kinoshita, ``{Mass Singularities of Feynman Amplitudes},''
\href{http://dx.doi.org/10.1063/1.1724268}{{\em J. Math. Phys.} {\bfseries 3}
  (1962) 650--677}.

\bibitem{Lee:1964is}
T.~D. Lee and M.~Nauenberg, ``{Degenerate Systems and Mass Singularities},''
\href{http://dx.doi.org/10.1103/PhysRev.133.B1549}{{\em Phys. Rev.} {\bfseries
  133} (1964) B1549--B1562}.

\bibitem{Frye:2018xjj}
C.~Frye, H.~Hannesdottir, N.~Paul, M.~D. Schwartz, and K.~Yan, ``{Infrared
  Finiteness and Forward Scattering},''
  \href{http://dx.doi.org/10.1103/PhysRevD.99.056015}{{\em Phys. Rev.}
  {\bfseries D99} no.~5, (2019) 056015},
\href{http://arxiv.org/abs/1810.10022}{{\ttfamily arXiv:1810.10022 [hep-ph]}}.

\bibitem{Collins:1988ig}
J.~C. Collins, D.~E. Soper, and G.~F. Sterman, ``{Soft Gluons and
  Factorization},''
\href{http://dx.doi.org/10.1016/0550-3213(88)90130-7}{{\em Nucl.Phys.}
  {\bfseries B308} (1988) 833}.

\bibitem{Collins:1989gx}
J.~C. Collins, D.~E. Soper, and G.~F. Sterman, ``{Factorization of Hard
  Processes in QCD},'' {\em Adv.Ser.Direct.High Energy Phys.} {\bfseries 5}
  (1988) 1--91,
\href{http://arxiv.org/abs/hep-ph/0409313}{{\ttfamily arXiv:hep-ph/0409313
  [hep-ph]}}.

\bibitem{CATANI1989323}
S.~Catani and L.~Trentadue, ``{Resummation of the QCD Perturbative Series for
  Hard Processes},''
  \href{http://dx.doi.org/https://doi.org/10.1016/0550-3213(89)90273-3}{{\em
  Nuclear Physics B} {\bfseries 327} no.~2, (1989) 323 -- 352}.
  \url{http://www.sciencedirect.com/science/article/pii/0550321389902733}.

\bibitem{Bauer:2000ew}
C.~W. Bauer, S.~Fleming, and M.~E. Luke, ``{Summing Sudakov Logarithms in $B
  \to X_s \gamma$ in Effective Field Theory},''
  \href{http://dx.doi.org/10.1103/PhysRevD.63.014006}{{\em Phys.Rev.}
  {\bfseries D63} (2000) 014006},
\href{http://arxiv.org/abs/hep-ph/0005275}{{\ttfamily arXiv:hep-ph/0005275
  [hep-ph]}}.

\bibitem{Bauer:2000yr}
C.~W. Bauer, S.~Fleming, D.~Pirjol, and I.~W. Stewart, ``{An Effective Field
  Theory for Collinear and Soft Gluons: Heavy to Light Decays},''
  \href{http://dx.doi.org/10.1103/PhysRevD.63.114020}{{\em Phys.Rev.}
  {\bfseries D63} (2001) 114020},
\href{http://arxiv.org/abs/hep-ph/0011336}{{\ttfamily arXiv:hep-ph/0011336
  [hep-ph]}}.

\bibitem{Beneke:2002ph}
M.~Beneke, A.~Chapovsky, M.~Diehl, and T.~Feldmann, ``{Soft Collinear Effective
  Theory and Heavy to Light Currents Beyond Leading Power},''
  \href{http://dx.doi.org/10.1016/S0550-3213(02)00687-9}{{\em Nucl.Phys.}
  {\bfseries B643} (2002) 431--476},
\href{http://arxiv.org/abs/hep-ph/0206152}{{\ttfamily arXiv:hep-ph/0206152
  [hep-ph]}}.

\bibitem{Beneke:2002ni}
M.~Beneke and T.~Feldmann, ``{Multipole Expanded Soft Collinear Effective
  Theory With Non-Abelian Gauge Symmetry},''
  \href{http://dx.doi.org/10.1016/S0370-2693(02)03204-5}{{\em Phys.Lett.}
  {\bfseries B553} (2003) 267--276},
\href{http://arxiv.org/abs/hep-ph/0211358}{{\ttfamily arXiv:hep-ph/0211358
  [hep-ph]}}.

\bibitem{Feige:2013zla}
I.~Feige and M.~D. Schwartz, ``{An On-Shell Approach to Factorization},''
  \href{http://dx.doi.org/10.1103/PhysRevD.88.065021}{{\em Phys. Rev.}
  {\bfseries D88} no.~6, (2013) 065021},
\href{http://arxiv.org/abs/1306.6341}{{\ttfamily arXiv:1306.6341 [hep-th]}}.

\bibitem{Feige:2014wja}
I.~Feige and M.~D. Schwartz, ``{Hard-Soft-Collinear Factorization to All
  Orders},'' \href{http://dx.doi.org/10.1103/PhysRevD.90.105020}{{\em Phys.
  Rev.} {\bfseries D90} no.~10, (2014) 105020},
\href{http://arxiv.org/abs/1403.6472}{{\ttfamily arXiv:1403.6472 [hep-ph]}}.

\bibitem{Chung:1965zza}
V.~Chung, ``{Infrared Divergence in Quantum Electrodynamics},''
\href{http://dx.doi.org/10.1103/PhysRev.140.B1110}{{\em Phys. Rev.} {\bfseries
  140} (1965) B1110--B1122}.

\bibitem{Greco:1967zza}
M.~Greco and G.~Rossi, ``{A Note on the Infrared Divergence},''
  \href{http://dx.doi.org/10.1007/BF02820731}{{\em Nuovo Cim.} {\bfseries 50}
  (1967) 168}.

\bibitem{glauber1963coherent}
R.~J. Glauber, ``{Coherent and Incoherent States of the Radiation Field},''
  {\em Physical Review} {\bfseries 131} no.~6, (1963) 2766.

\bibitem{frantz1965compton}
L.~M. Frantz, ``{Compton Scattering of an Intense Photon Beam},'' {\em Physical
  Review} {\bfseries 139} no.~5B, (1965) B1326.

\bibitem{Ilderton:2017xbj}
A.~Ilderton and D.~Seipt, ``{Backreaction on Background Fields: A Coherent
  State Approach},'' \href{http://dx.doi.org/10.1103/PhysRevD.97.016007}{{\em
  Phys. Rev. D} {\bfseries 97} no.~1, (2018) 016007},
  \href{http://arxiv.org/abs/1709.10085}{{\ttfamily arXiv:1709.10085
  [hep-th]}}.

\bibitem{Kibble:1968sfb}
T.~W.~B. Kibble, ``{Coherent Soft-Photon States and Infrared Divergences. i.
  Classical Currents},''
\href{http://dx.doi.org/10.1063/1.1664582}{{\em J. Math. Phys.} {\bfseries 9}
  no.~2, (1968) 315--324}.

\bibitem{Kibble:1969ip}
T.~W.~B. Kibble, ``{Coherent Soft-Photon States and Infrared Divergences. ii.
  Mass-Shell Singularities of Green's Functions},''
\href{http://dx.doi.org/10.1103/PhysRev.173.1527}{{\em Phys. Rev.} {\bfseries
  173} (1968) 1527--1535}.

\bibitem{Kibble:1969ep}
T.~W.~B. Kibble, ``{Coherent Soft-Photon States and Infrared Divergences. iii.
  Asymptotic States and Reduction Formulas},''
\href{http://dx.doi.org/10.1103/PhysRev.174.1882}{{\em Phys. Rev.} {\bfseries
  174} (1968) 1882--1901}.

\bibitem{Kibble:1969kd}
T.~W.~B. Kibble, ``{Coherent Soft-Photon States and Infrared Divergences. iv.
  The Scattering Operator},''
\href{http://dx.doi.org/10.1103/PhysRev.175.1624}{{\em Phys. Rev.} {\bfseries
  175} (1968) 1624--1640}.

\bibitem{Catani:1985xt}
S.~Catani, M.~Ciafaloni, and G.~Marchesini, ``{Noncancelling Infrared
  Divergences in QCD Coherent State},''
\href{http://dx.doi.org/10.1016/0550-3213(86)90500-6}{{\em Nucl. Phys.}
  {\bfseries B264} (1986) 588--620}.

\bibitem{Gonzo:2019fai}
R.~Gonzo, T.~Mc~Loughlin, D.~Medrano, and A.~Spiering, ``{Asymptotic Charges
  and Coherent States in QCD},''
\href{http://arxiv.org/abs/1906.11763}{{\ttfamily arXiv:1906.11763 [hep-th]}}.

\bibitem{dollard1971quantum}
J.~D. Dollard, ``{Quantum-Mechanical Scattering Theory for Short-Range and
  Coulomb Interactions},'' {\em The Rocky Mountain Journal of Mathematics}
  (1971) 5--88.

\bibitem{Kulish:1970ut}
P.~P. Kulish and L.~D. Faddeev, ``{Asymptotic Conditions and Infrared
  Divergences in Quantum Electrodynamics},''
\href{http://dx.doi.org/10.1007/BF01066485}{{\em Theor. Math. Phys.} {\bfseries
  4} (1970) 745}.

\bibitem{Butler:1978rd}
D.~R. Butler and C.~A. Nelson, ``{Nonabelian Structure of {Yang-Mills} Theory
  and Infrared Finite Asymptotic States},''
\href{http://dx.doi.org/10.1103/PhysRevD.18.1196}{{\em Phys. Rev.} {\bfseries
  D18} (1978) 1196}.

\bibitem{Nelson:1980qs}
C.~A. Nelson, ``{Avoidance of Counter Example to Nonabelian Bloch-Nordsieck
  Conjecture by Using Coherent State Approach},''
\href{http://dx.doi.org/10.1016/0550-3213(81)90099-7}{{\em Nucl. Phys.}
  {\bfseries B186} (1981) 187--204}.

\bibitem{Andrasi:1980qw}
A.~Andrasi, M.~Day, R.~Doria, J.~Frenkel, and J.~C. Taylor, ``{Soft Divergences
  in Perturbative {QCD}},''
\href{http://dx.doi.org/10.1016/0550-3213(81)90460-0}{{\em Nucl. Phys.}
  {\bfseries B182} (1981) 104--124}.

\bibitem{Gatheral:1983cz}
J.~G.~M. Gatheral, ``{Exponentiation of Eikonal Cross-sections in Nonabelian
  Gauge Theories},''
\href{http://dx.doi.org/10.1016/0370-2693(83)90112-0}{{\em Phys. Lett.}
  {\bfseries 133B} (1983) 90--94}.

\bibitem{Frenkel:1984pz}
J.~Frenkel and J.~C. Taylor, ``{Nonabelian Eikonal Exponentiation},''
\href{http://dx.doi.org/10.1016/0550-3213(84)90294-3}{{\em Nucl. Phys.}
  {\bfseries B246} (1984) 231--245}.

\bibitem{Giavarini:1987ts}
G.~Giavarini and G.~Marchesini, ``{{IR} Finite S Matrix in the {QCD} Coherent
  State Basis},''
\href{http://dx.doi.org/10.1016/0550-3213(88)90031-4}{{\em Nucl. Phys.}
  {\bfseries B296} (1988) 546--556}.

\bibitem{curci1978mass}
G.~Curci and M.~Greco, ``{Mass Singularities and Coherent States in Gauge
  Theories},'' {\em Physics Letters B} {\bfseries 79} no.~4-5, (1978) 406--410.

\bibitem{Havemann:1985ra}
F.~N. Havemann,
``{Collinear Divergences and Asymptotic States},''.

\bibitem{DelDuca:1989jt}
V.~Del~Duca, L.~Magnea, and G.~F. Sterman, ``{Collinear Infrared Factorization
  and Asymptotic Evolution},''
\href{http://dx.doi.org/10.1016/0550-3213(89)90472-0}{{\em Nucl. Phys.}
  {\bfseries B324} (1989) 391--411}.

\bibitem{Forde:2003jt}
D.~A. Forde and A.~Signer, ``{Infrared Finite Amplitudes for Massless Gauge
  Theories},'' \href{http://dx.doi.org/10.1016/j.nuclphysb.2004.02.024}{{\em
  Nucl. Phys.} {\bfseries B684} (2004) 125--161},
\href{http://arxiv.org/abs/hep-ph/0311059}{{\ttfamily arXiv:hep-ph/0311059
  [hep-ph]}}.

\bibitem{Kosower:1999xi}
D.~A. Kosower, ``{All Order Collinear Behavior in Gauge Theories},''
  \href{http://dx.doi.org/10.1016/S0550-3213(99)00251-5}{{\em Nucl. Phys.}
  {\bfseries B552} (1999) 319--336},
\href{http://arxiv.org/abs/hep-ph/9901201}{{\ttfamily arXiv:hep-ph/9901201
  [hep-ph]}}.

\bibitem{Bauer:2002nz}
C.~W. Bauer, S.~Fleming, D.~Pirjol, I.~Z. Rothstein, and I.~W. Stewart, ``{Hard
  Scattering Factorization From Effective Field Theory},''
  \href{http://dx.doi.org/10.1103/PhysRevD.66.014017}{{\em Phys.Rev.}
  {\bfseries D66} (2002) 014017},
\href{http://arxiv.org/abs/hep-ph/0202088}{{\ttfamily arXiv:hep-ph/0202088
  [hep-ph]}}.

\bibitem{Bagan:1999jf}
E.~Bagan, M.~Lavelle, and D.~McMullan, ``{Charges From Dressed Matter:
  Construction},'' \href{http://dx.doi.org/10.1006/aphy.2000.6048}{{\em Annals
  Phys.} {\bfseries 282} (2000) 471--502},
\href{http://arxiv.org/abs/hep-ph/9909257}{{\ttfamily arXiv:hep-ph/9909257
  [hep-ph]}}.

\bibitem{catani1987gauge}
S.~Catani and M.~Ciafaloni, ``{Gauge Covariance of QCD Coherent States},'' {\em
  Nuclear Physics B} {\bfseries 289} (1987) 535--556.

\bibitem{Kapec:2017tkm}
D.~Kapec, M.~Perry, A.-M. Raclariu, and A.~Strominger, ``{Infrared Divergences
  in QED, Revisited},''
  \href{http://dx.doi.org/10.1103/PhysRevD.96.085002}{{\em Phys. Rev.}
  {\bfseries D96} no.~8, (2017) 085002},
\href{http://arxiv.org/abs/1705.04311}{{\ttfamily arXiv:1705.04311 [hep-th]}}.

\bibitem{Strominger:2017zoo}
A.~Strominger, ``{Lectures on the Infrared Structure of Gravity and Gauge
  Theory},''
\href{http://arxiv.org/abs/1703.05448}{{\ttfamily arXiv:1703.05448 [hep-th]}}.

\bibitem{Carney:2017jut}
D.~Carney, L.~Chaurette, D.~Neuenfeld, and G.~W. Semenoff, ``{Infrared Quantum
  Information},'' \href{http://dx.doi.org/10.1103/PhysRevLett.119.180502}{{\em
  Phys. Rev. Lett.} {\bfseries 119} no.~18, (2017) 180502},
\href{http://arxiv.org/abs/1706.03782}{{\ttfamily arXiv:1706.03782 [hep-th]}}.

\bibitem{Carney:2018ygh}
D.~Carney, L.~Chaurette, D.~Neuenfeld, and G.~Semenoff, ``{On the Need for Soft
  Dressing},'' \href{http://dx.doi.org/10.1007/JHEP09(2018)121}{{\em JHEP}
  {\bfseries 09} (2018) 121},
\href{http://arxiv.org/abs/1803.02370}{{\ttfamily arXiv:1803.02370 [hep-th]}}.

\bibitem{Gomez:2018war}
C.~Gomez, R.~Letschka, and S.~Zell, ``{The Scales of the Infrared},''
  \href{http://dx.doi.org/10.1007/JHEP09(2018)115}{{\em JHEP} {\bfseries 09}
  (2018) 115}, \href{http://arxiv.org/abs/1807.07079}{{\ttfamily
  arXiv:1807.07079 [hep-th]}}.

\bibitem{Chien:2011wz}
Y.-T. Chien, M.~D. Schwartz, D.~Simmons-Duffin, and I.~W. Stewart, ``{Jet
  Physics from Static Charges in AdS},''
  \href{http://dx.doi.org/10.1103/PhysRevD.85.045010}{{\em Phys. Rev.}
  {\bfseries D85} (2012) 045010},
\href{http://arxiv.org/abs/1109.6010}{{\ttfamily arXiv:1109.6010 [hep-th]}}.

\bibitem{Korchemsky:1987wg}
G.~P. Korchemsky and A.~V. Radyushkin, ``{Renormalization of the Wilson Loops
  Beyond the Leading Order},''
\href{http://dx.doi.org/10.1016/0550-3213(87)90277-X}{{\em Nucl. Phys.}
  {\bfseries B283} (1987) 342--364}.

\bibitem{Gardi:2013ita}
E.~Gardi, J.~M. Smillie, and C.~D. White, ``{The Non-Abelian Exponentiation
  Theorem for Multiple Wilson Lines},''
  \href{http://dx.doi.org/10.1007/JHEP06(2013)088}{{\em JHEP} {\bfseries 06}
  (2013) 088},
\href{http://arxiv.org/abs/1304.7040}{{\ttfamily arXiv:1304.7040 [hep-ph]}}.

\bibitem{Haag:1955ev}
R.~Haag, ``{On Quantum Field Theories},''
{\em Kong. Dan. Vid. Sel. Mat. Fys. Med.} {\bfseries 29N12} (1955) 1--37.

\bibitem{stewart2013lectures}
I.~W. Stewart, ``{Lectures on the Soft-Collinear Effective Theory},'' {\em
  {Lectures on the Soft-Collinear Effective Theory, MIT Open Courseware}}
  (2013) .

\bibitem{Becher:2014oda}
T.~Becher, A.~Broggio, and A.~Ferroglia, ``{Introduction to Soft-Collinear
  Effective Theory},'' \href{http://dx.doi.org/10.1007/978-3-319-14848-9}{{\em
  Lect. Notes Phys.} {\bfseries 896} (2015) pp.1--206},
\href{http://arxiv.org/abs/1410.1892}{{\ttfamily arXiv:1410.1892 [hep-ph]}}.

\bibitem{Rothstein:2016bsq}
I.~Z. Rothstein and I.~W. Stewart, ``{An Effective Field Theory for Forward
  Scattering and Factorization Violation},''
  \href{http://dx.doi.org/10.1007/JHEP08(2016)025}{{\em JHEP} {\bfseries 08}
  (2016) 025},
\href{http://arxiv.org/abs/1601.04695}{{\ttfamily arXiv:1601.04695 [hep-ph]}}.

\bibitem{Schwartz:2017nmr}
M.~D. Schwartz, K.~Yan, and H.~X. Zhu, ``{Collinear Factorization Violation and
  Effective Field Theory},''
  \href{http://dx.doi.org/10.1103/PhysRevD.96.056005}{{\em Phys. Rev.}
  {\bfseries D96} no.~5, (2017) 056005},
\href{http://arxiv.org/abs/1703.08572}{{\ttfamily arXiv:1703.08572 [hep-ph]}}.

\bibitem{Schwartz:2013pla}
M.~D. Schwartz, {\em {Quantum Field Theory and the Standard Model}}.
\newblock Cambridge University Press, 2014.
\newblock
\url{http://www.cambridge.org/us/academic/subjects/physics/theoretical-physics-and-mathematical-physics/quantum-field-theory-and-standard-model}.
\newblock

\bibitem{Beneke:1997zp}
M.~Beneke and V.~A. Smirnov, ``{Asymptotic Expansion of Feynman Integrals Near
  Threshold},'' \href{http://dx.doi.org/10.1016/S0550-3213(98)00138-2}{{\em
  Nucl. Phys.} {\bfseries B522} (1998) 321--344},
\href{http://arxiv.org/abs/hep-ph/9711391}{{\ttfamily arXiv:hep-ph/9711391
  [hep-ph]}}.

\bibitem{Bauer:2006mk}
C.~W. Bauer and M.~D. Schwartz, ``{Event Generation from Effective Field
  Theory},'' \href{http://dx.doi.org/10.1103/PhysRevD.76.074004}{{\em
  Phys.Rev.} {\bfseries D76} (2007) 074004},
\href{http://arxiv.org/abs/hep-ph/0607296}{{\ttfamily arXiv:hep-ph/0607296
  [hep-ph]}}.

\bibitem{Bauer:2006qp}
C.~W. Bauer and M.~D. Schwartz, ``{Improving Jet Distributions With Effective
  Field Theory},'' \href{http://dx.doi.org/10.1103/PhysRevLett.97.142001}{{\em
  Phys.Rev.Lett.} {\bfseries 97} (2006) 142001},
\href{http://arxiv.org/abs/hep-ph/0604065}{{\ttfamily arXiv:hep-ph/0604065
  [hep-ph]}}.

\bibitem{Becher:2009qa}
T.~Becher and M.~Neubert, ``{On the Structure of Infrared Singularities of
  Gauge-Theory Amplitudes},''
  \href{http://dx.doi.org/10.1088/1126-6708/2009/06/081,
  10.1007/JHEP11(2013)024}{{\em JHEP} {\bfseries 06} (2009) 081},
  \href{http://arxiv.org/abs/0903.1126}{{\ttfamily arXiv:0903.1126 [hep-ph]}}.
[Erratum: JHEP11,024(2013)].

\bibitem{Gardi:2009qi}
E.~Gardi and L.~Magnea, ``{Factorization Constraints for Soft Anomalous
  Dimensions in QCD Scattering Amplitudes},''
  \href{http://dx.doi.org/10.1088/1126-6708/2009/03/079}{{\em JHEP} {\bfseries
  03} (2009) 079},
\href{http://arxiv.org/abs/0901.1091}{{\ttfamily arXiv:0901.1091 [hep-ph]}}.

\bibitem{Magnus:1954zz}
W.~Magnus, ``{On the Exponential Solution of Differential Equations for a
  Linear Operator},''
\href{http://dx.doi.org/10.1002/cpa.3160070404}{{\em Commun. Pure Appl. Math.}
  {\bfseries 7} (1954) 649--673}.

\bibitem{Sterman:1994ce}
G.~F. Sterman, {\em {An Introduction to Quantum Field Theory}}.
\newblock Cambridge University Press,
1993.
\newblock

\bibitem{Manohar:2003vb}
A.~V. Manohar, ``{Deep Inelastic Scattering as $x\to 1$ Using Soft Collinear
  Effective Theory},'' \href{http://dx.doi.org/10.1103/PhysRevD.68.114019}{{\em
  Phys.Rev.} {\bfseries D68} (2003) 114019},
\href{http://arxiv.org/abs/hep-ph/0309176}{{\ttfamily arXiv:hep-ph/0309176
  [hep-ph]}}.

\bibitem{Manohar:2006nz}
A.~V. Manohar and I.~W. Stewart, ``{The Zero-Bin and Mode Factorization in
  Quantum Field Theory},''
  \href{http://dx.doi.org/10.1103/PhysRevD.76.074002}{{\em Phys.Rev.}
  {\bfseries D76} (2007) 074002},
\href{http://arxiv.org/abs/hep-ph/0605001}{{\ttfamily arXiv:hep-ph/0605001
  [hep-ph]}}.

\bibitem{Feige:2015rea}
I.~Feige, M.~D. Schwartz, and K.~Yan, ``{Removing Phase-Space Restrictions in
  Factorized Cross Sections},''
  \href{http://dx.doi.org/10.1103/PhysRevD.91.094027}{{\em Phys. Rev.}
  {\bfseries D91} (2015) 094027},
\href{http://arxiv.org/abs/1502.05411}{{\ttfamily arXiv:1502.05411 [hep-ph]}}.

\bibitem{Hornig:2009kv}
A.~Hornig, C.~Lee, and G.~Ovanesyan, ``{Infrared Safety in Factorized Hard
  Scattering Cross-Sections},''
  \href{http://dx.doi.org/10.1016/j.physletb.2009.05.039}{{\em Phys. Lett.}
  {\bfseries B677} (2009) 272--277},
\href{http://arxiv.org/abs/0901.1897}{{\ttfamily arXiv:0901.1897 [hep-ph]}}.

\bibitem{Forshaw:2008cq}
J.~R. Forshaw, A.~Kyrieleis, and M.~H. Seymour, ``{Super-Leading Logarithms in
  Non-Global Observables in QCD: Colour Basis Independent Calculation},''
  \href{http://dx.doi.org/10.1088/1126-6708/2008/09/128}{{\em JHEP} {\bfseries
  09} (2008) 128},
\href{http://arxiv.org/abs/0808.1269}{{\ttfamily arXiv:0808.1269 [hep-ph]}}.

\bibitem{Forshaw:2006fk}
J.~R. Forshaw, A.~Kyrieleis, and M.~H. Seymour, ``{Super-Leading Logarithms in
  Non-Global Observables in QCD},''
  \href{http://dx.doi.org/10.1088/1126-6708/2006/08/059}{{\em JHEP} {\bfseries
  08} (2006) 059},
\href{http://arxiv.org/abs/hep-ph/0604094}{{\ttfamily arXiv:hep-ph/0604094
  [hep-ph]}}.

\bibitem{Schwartz:2007ib}
M.~D. Schwartz, ``{Resummation and NLO Matching of Event Shapes With Effective
  Field Theory},'' \href{http://dx.doi.org/10.1103/PhysRevD.77.014026}{{\em
  Phys. Rev.} {\bfseries D77} (2008) 014026},
\href{http://arxiv.org/abs/0709.2709}{{\ttfamily arXiv:0709.2709 [hep-ph]}}.

\bibitem{Schwartz:2018obd}
M.~D. Schwartz, K.~Yan, and H.~X. Zhu, ``{Factorization Violation and Scale
  Invariance},'' \href{http://dx.doi.org/10.1103/PhysRevD.97.096017}{{\em Phys.
  Rev.} {\bfseries D97} no.~9, (2018) 096017},
\href{http://arxiv.org/abs/1801.01138}{{\ttfamily arXiv:1801.01138 [hep-ph]}}.

\bibitem{Laenen:2014jga}
E.~Laenen, K.~J. Larsen, and R.~Rietkerk, ``{Imaginary Parts and
  Discontinuities of Wilson Line Correlators},''
  \href{http://dx.doi.org/10.1103/PhysRevLett.114.181602}{{\em Phys. Rev.
  Lett.} {\bfseries 114} no.~18, (2015) 181602},
\href{http://arxiv.org/abs/1410.5681}{{\ttfamily arXiv:1410.5681 [hep-th]}}.

\bibitem{Laenen:2015jia}
E.~Laenen, K.~J. Larsen, and R.~Rietkerk, ``{Position-Space Cuts for Wilson
  Line Correlators},'' \href{http://dx.doi.org/10.1007/JHEP07(2015)083}{{\em
  JHEP} {\bfseries 07} (2015) 083},
\href{http://arxiv.org/abs/1505.02555}{{\ttfamily arXiv:1505.02555 [hep-th]}}.

\bibitem{FarhiThrust}
E.~Farhi, ``{Quantum Chromodynamics Test for Jets},''
  \href{http://dx.doi.org/10.1103/PhysRevLett.39.1587}{{\em Phys. Rev. Lett.}
  {\bfseries 39} (Dec, 1977) 1587--1588}.
  \url{https://link.aps.org/doi/10.1103/PhysRevLett.39.1587}.

\bibitem{Becher:2008cf}
T.~Becher and M.~D. Schwartz, ``{A Precise Determination of $\alpha_s$ From LEP
  Thrust Data Using Effective Field Theory},''
  \href{http://dx.doi.org/10.1088/1126-6708/2008/07/034}{{\em JHEP} {\bfseries
  07} (2008) 034},
\href{http://arxiv.org/abs/0803.0342}{{\ttfamily arXiv:0803.0342 [hep-ph]}}.

\bibitem{Schwartz:2014wha}
M.~D. Schwartz and H.~X. Zhu, ``{Nonglobal Logarithms at Three Loops, Four
  Loops, Five Loops, and beyond},''
  \href{http://dx.doi.org/10.1103/PhysRevD.90.065004}{{\em Phys. Rev.}
  {\bfseries D90} no.~6, (2014) 065004},
\href{http://arxiv.org/abs/1403.4949}{{\ttfamily arXiv:1403.4949 [hep-ph]}}.

\bibitem{Bern:2005iz}
Z.~Bern, L.~J. Dixon, and V.~A. Smirnov, ``{Iteration of Planar Amplitudes in
  Maximally Supersymmetric Yang-Mills Theory at Three Loops and Beyond},''
  \href{http://dx.doi.org/10.1103/PhysRevD.72.085001}{{\em Phys. Rev.}
  {\bfseries D72} (2005) 085001},
\href{http://arxiv.org/abs/hep-th/0505205}{{\ttfamily arXiv:hep-th/0505205
  [hep-th]}}.

\bibitem{Anastasiou:2003kj}
C.~Anastasiou, Z.~Bern, L.~J. Dixon, and D.~A. Kosower, ``{Planar Amplitudes in
  Maximally Supersymmetric Yang-Mills Theory},''
  \href{http://dx.doi.org/10.1103/PhysRevLett.91.251602}{{\em Phys. Rev. Lett.}
  {\bfseries 91} (2003) 251602},
\href{http://arxiv.org/abs/hep-th/0309040}{{\ttfamily arXiv:hep-th/0309040
  [hep-th]}}.

\bibitem{Caron-Huot:2016owq}
S.~Caron-Huot, L.~J. Dixon, A.~McLeod, and M.~von Hippel, ``{Bootstrapping a
  Five-Loop Amplitude Using Steinmann Relations},''
  \href{http://dx.doi.org/10.1103/PhysRevLett.117.241601}{{\em Phys. Rev.
  Lett.} {\bfseries 117} no.~24, (2016) 241601},
\href{http://arxiv.org/abs/1609.00669}{{\ttfamily arXiv:1609.00669 [hep-th]}}.

\bibitem{Alday:2009dv}
L.~F. Alday, D.~Gaiotto, and J.~Maldacena, ``{Thermodynamic Bubble Ansatz},''
  \href{http://dx.doi.org/10.1007/JHEP09(2011)032}{{\em JHEP} {\bfseries 09}
  (2011) 032},
\href{http://arxiv.org/abs/0911.4708}{{\ttfamily arXiv:0911.4708 [hep-th]}}.

\bibitem{Golden:2018gtk}
J.~Golden and A.~J. Mcleod, ``{Cluster Algebras and the Subalgebra
  Constructibility of the Seven-Particle Remainder Function},''
  \href{http://dx.doi.org/10.1007/JHEP01(2019)017}{{\em JHEP} {\bfseries 01}
  (2019) 017},
\href{http://arxiv.org/abs/1810.12181}{{\ttfamily arXiv:1810.12181 [hep-th]}}.

\bibitem{Golden:2019kks}
J.~Golden, A.~J. McLeod, M.~Spradlin, and A.~Volovich, ``{The Sklyanin Bracket
  and Cluster Adjacency at All Multiplicity},''
  \href{http://dx.doi.org/10.1007/JHEP03(2019)195}{{\em JHEP} {\bfseries 03}
  (2019) 195},
\href{http://arxiv.org/abs/1902.11286}{{\ttfamily arXiv:1902.11286 [hep-th]}}.

\bibitem{catani1998singular}
S.~Catani, ``{The Singular Behaviour of QCD Amplitudes at Two-Loop Order},''
  {\em Physics Letters B} {\bfseries 427} no.~1-2, (1998) 161--171.

\bibitem{Sterman:2002qn}
G.~F. Sterman and M.~E. Tejeda-Yeomans, ``{Multiloop Amplitudes and
  Resummation},'' \href{http://dx.doi.org/10.1016/S0370-2693(02)03100-3}{{\em
  Phys. Lett.} {\bfseries B552} (2003) 48--56},
\href{http://arxiv.org/abs/hep-ph/0210130}{{\ttfamily arXiv:hep-ph/0210130
  [hep-ph]}}.

\bibitem{Becher:2009cu}
T.~Becher and M.~Neubert, ``{Infrared Singularities of Scattering Amplitudes in
  Perturbative QCD},'' \href{http://dx.doi.org/10.1103/PhysRevLett.102.162001,
  10.1103/PhysRevLett.111.199905}{{\em Phys. Rev. Lett.} {\bfseries 102} (2009)
  162001}, \href{http://arxiv.org/abs/0901.0722}{{\ttfamily arXiv:0901.0722
  [hep-ph]}}.
[Erratum: Phys. Rev. Lett.111,no.19,199905(2013)].

\bibitem{broggio2016scet}
A.~Broggio, C.~Gnendiger, A.~Signer, D.~St{\"o}ckinger, and A.~Visconti,
  ``{SCET Approach to Regularization-Scheme Dependence of QCD Amplitudes},''
  {\em Journal of High Energy Physics} {\bfseries 2016} no.~1, (2016) 78.

\bibitem{Bern:1994zx}
Z.~Bern, L.~J. Dixon, D.~C. Dunbar, and D.~A. Kosower, ``{One Loop n Point
  Gauge Theory Amplitudes, Unitarity and Collinear Limits},''
  \href{http://dx.doi.org/10.1016/0550-3213(94)90179-1}{{\em Nucl. Phys.}
  {\bfseries B425} (1994) 217--260},
\href{http://arxiv.org/abs/hep-ph/9403226}{{\ttfamily arXiv:hep-ph/9403226
  [hep-ph]}}.

\bibitem{Dixon:2015iva}
L.~J. Dixon, M.~von Hippel, and A.~J. McLeod, ``{The Four-Loop Six-Gluon NMHV
  Ratio Function},'' \href{http://dx.doi.org/10.1007/JHEP01(2016)053}{{\em
  JHEP} {\bfseries 01} (2016) 053},
\href{http://arxiv.org/abs/1509.08127}{{\ttfamily arXiv:1509.08127 [hep-th]}}.

\bibitem{steinmann1960zusammenhang}
O.~Steinmann, {\em About the Relationship Between the Wightman Functions and
  the Retarded Commutators}.
\newblock PhD thesis, ETH Zurich, 1960.

\bibitem{Caron-Huot:2019vjl}
S.~Caron-Huot, L.~J. Dixon, F.~Dulat, M.~von Hippel, A.~J. McLeod, and
  G.~Papathanasiou, ``{Six-Gluon Amplitudes in Planar $ \mathcal{N} $ = 4
  Super-Yang-Mills Theory at Six and Seven Loops},''
  \href{http://dx.doi.org/10.1007/JHEP08(2019)016}{{\em JHEP} {\bfseries 08}
  (2019) 016},
\href{http://arxiv.org/abs/1903.10890}{{\ttfamily arXiv:1903.10890 [hep-th]}}.

\end{thebibliography}\endgroup

\end{document}